\documentclass[fleqn,10pt]{wlscirep}
\usepackage[utf8]{inputenc}
\usepackage[T1]{fontenc}

\usepackage{amsmath,amsfonts}
\usepackage{algorithmic}
\usepackage{graphicx}
\usepackage{textcomp}
\usepackage{xcolor}
\usepackage{multirow}
\usepackage{xspace}
\usepackage{listings}
\usepackage{caption}
\usepackage{xcolor}
\usepackage{hyperref}

\usepackage{amsmath,amsfonts,bm}

\def\eqref#1{equation~\ref{#1}}

\def\1{\bm{1}}

\def\vx{{\bm{x}}}

\DeclareMathAlphabet{\mathsfit}{\encodingdefault}{\sfdefault}{m}{sl}
\SetMathAlphabet{\mathsfit}{bold}{\encodingdefault}{\sfdefault}{bx}{n}

\DeclareMathOperator*{\argmax}{arg\,max}

\usepackage{graphicx}
\usepackage{float}
\usepackage{amsfonts,amsmath,amssymb,amsthm}
\usepackage{cleveref}
\usepackage{multicol, multirow}
\usepackage{makecell}
\usepackage{caption}
\usepackage{subcaption}
\usepackage{adjustbox}
\usepackage{enumitem}
\usepackage{wrapfig}
\usepackage{algorithm}
\usepackage{algorithmic}

\newcommand{\ie}{\em{i.e.}}
\newcommand{\eg}{\em{e.g.}}
\newcommand{\model}{ChatDrug}

\newcommand{\TotalTaskNumber}{39}
\newcommand{\SmallMoleculeTaskNumber}{28}
\newcommand{\PeptideTaskNumber}{9}
\newcommand{\ProteinTaskNumber}{2}
\usepackage{cleveref}

\usepackage{xcolor}
\definecolor{SmallMoleculeBlue}{RGB}{216,217,250}
\definecolor{SmallMoleculeRed}{RGB}{249,216,213}
\definecolor{SmallMoleculeGreen}{RGB}{213,255,213}
\definecolor{ProteinSecondaryStructureRed}{RGB}{149,1,1}
\definecolor{ProteinSecondaryStructureYellow}{RGB}{204,204,22}
\definecolor{ProteinSecondaryStructureBlue}{RGB}{69,124,198}

\begin{document}

\title{ChatGPT-powered Conversational Drug Editing\\Using Retrieval and Domain Feedback}

\author[1,2 *]{Shengchao Liu}
\author[3 *]{Jiongxiao Wang}
\author[3]{Yijin Yang}
\author[4]{Chengpeng Wang}
\author[5]{Ling Liu}
\author[6 $\dagger$]{~ Hongyu Guo}
\author[3 $\dagger$]{Chaowei Xiao}
\affil[1]{Mila-Québec Artificial Intelligence Institute, Montréal, QC H2S 3H1, Canada}
\affil[2]{Université de Montréal, Montréal, QC H3T 1J4, Canada}
\affil[3]{Arizona State University, Tempe, AZ 85281, United States}
\affil[4]{Princeton University, Princeton, NJ 08544, United States}
\affil[5]{University of Illinois Urbana-Champaign, Champaign, IL 61801, United States}
\affil[6]{National Research Council Canada, Ottawa, ON K1A 0R6, Canada}
\def\thefootnote{*}\footnotetext{Equal contribution}\def\thefootnote{\arabic{footnote}}
\def\thefootnote{$\dagger$}\footnotetext{Equal advising}\def\thefootnote{\arabic{footnote}}

\begin{abstract}
Recent advancements in conversational large language models (LLMs), such as ChatGPT, have demonstrated remarkable promise in various domains, including drug discovery. However, existing works mainly focus on investigating the capabilities of conversational LLMs on chemical reaction and retrosynthesis. While drug editing, a critical task in the drug discovery pipeline, remains largely unexplored. To bridge this gap, we propose ChatDrug, a framework to facilitate the systematic investigation of drug editing using LLMs. ChatDrug jointly leverages a prompt module, a retrieval and domain feedback (ReDF) module, and a conversation module to streamline effective drug editing. We empirically show that ChatDrug reaches the best performance on 33 out of 39 drug editing tasks, encompassing small molecules, peptides, and proteins. We further demonstrate, through 10 case studies, that ChatDrug can successfully identify the key substructures (\textit{e.g.}, the molecule functional groups, peptide motifs, and protein structures) for manipulation, generating diverse and valid suggestions for drug editing. Promisingly, we also show that ChatDrug can offer insightful explanations from a domain-specific perspective, enhancing interpretability and enabling informed decision-making. This research sheds light on the potential of ChatGPT and conversational LLMs for drug editing. It paves the way for a more efficient and collaborative drug discovery pipeline, contributing to the advancement of pharmaceutical research and development. The source codes can be found in \href{https://github.com/chao1224/ChatDrug}{this GitHub repository}.
\end{abstract}

\maketitle

\section{Introduction} \label{sec:intro}
In recent years, artificial intelligence (AI) tools have made remarkable strides in revolutionizing the field of drug discovery, offering tremendous potential for accelerating and enhancing various stages of the process~\cite{sullivan2019tough}, including but not limited to virtual screening~\cite{doi:10.1021/ci8002649,liu2018practical}, lead optimization~\cite{jin2020hierarchical,irwin2022chemformer,wang2022retrieval,liu2022graphcg}, reaction and retrosynthesis~\cite{gottipati2020learning,ShiXG0T20,BiWSC0G21}, protein folding and inverse folding~\cite{jumper2021highly,hsu2022learning}. However, much of the existing research has predominantly focused on the drug structure information, solely considering the inherent chemical structure of the drugs as a single modality. In contrast, the drug discovery pipeline involves iterative refining processes that entail conversations with domain experts to incorporate their feedback, ultimately achieving the desired outcome. On the other hand, significant advancements have been made in large language models (LLMs)~\cite{brown2020language,devlin2018bert,yang2019xlnet}, showcasing exceptional capabilities in understanding human knowledge and exhibiting promising reasoning abilities~\cite{huang2022large,zhou2022large,kojima2022large}. Such observations inspire us to investigate the potential of leveraging LLMs' conversation and reasoning abilities for AI-assisted drug discovery in a multi-modality fashion.

\textbf{Potential of Conversational LLMs for Drug Discovery and Editing.} Conversational LLMs exhibit three compelling factors that make them highly promising for drug discovery. Firstly, these models, such as ChatGPT, are pretrained on a comprehensive knowledge base, enabling their application across various fields, including drug discovery. This extensive ``world-level'' knowledge serves as a robust foundation for drug-related tasks. Second, conversational LLMs possess outstanding abilities in fast adaptation and generalization. For example, by leveraging few-shot demonstrations, these models can effectively activate the relevant concepts learned during pretraining, enabling them to deliver more accurate and desired answers~\cite{xie2021incontext}. This adaptability and generalization capacity holds immense potential for addressing complex drug discovery challenges and generating valuable insights. Lastly, interactive communication is a vital characteristic of conversational LLMs. This feature allows for a dynamic exchange of information, enabling users to incorporate feedback from prior knowledge or domain experts into the model. This bidirectional flow of information facilitates self-calibration of the answers, leading to improved accuracy and relevance in drug discovery tasks. To sum up, these factors collectively highlight the untapped potential of conversational LLMs for drug discovery tasks. Noticeably, there exists an important and challenging task: \textbf{drug editing} (AKA \textit{lead optimization} or \textit{protein design}). This is a routine task in pharmaceutical companies, and it aims at updating the drug's substructures~\cite{mihalic1992graph}, related to certain key tactics in drug discovery like functional group change~\cite{ertl2020most} and scaffold hopping~\cite{bohm2004scaffold,hu2017recent}. Traditional solutions relying on domain experts for manual editing can be subjective or biased~\cite{drews2000drug,gomez2018decision}. Recent works~\cite{liu2023text,liu2022multi} have started to explore text-guided drug editing in a multi-modal manner. However, they do not possess conversational potentials like ChatGPT.\looseness=-1

\begin{figure}[t!]
\centering
\includegraphics[width=\linewidth]{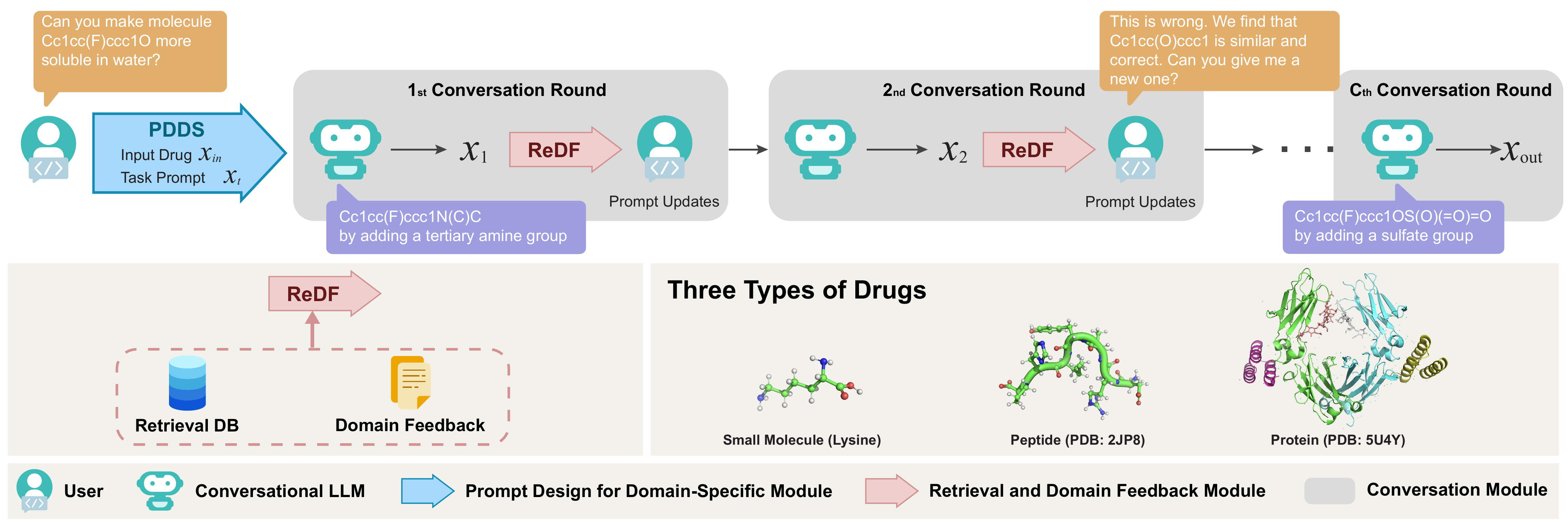}
\vspace{-4ex}
\caption{\small The pipeline for \model{} with 3 modules. PDDS generates drug editing prompts. ReDF updates the prompts using retrieved information and domain feedback. Finally, \model{} adopts the conversational module for interactive refinement. Further, we demonstrate 3 drug types: small molecules, peptides, and proteins.
}
\label{fig:pipeline_chatdrug}
\end{figure}

\textbf{Our Approach: \model{}.}
Motivated by the aforementioned factors and challenges, we propose \model{}, a framework aiming to unlock new possibilities and enhance drug editing using contrastive LLMs like ChatGPT. As shown in \Cref{fig:pipeline_chatdrug}, in the \model{} framework for drug editing, users can activate the conversation with LLMs involving domain knowledge and inject such retrieved information into the conversation. Specifically, \model{} includes the following modules for conversational drug editing. First, \model{} adopts a PDDS (prompt design for domain-specific) module, enabling strong prompt engineering capability from LLMs. Second, \model{} integrates a ReDF (retrieval and domain feedback) module. By leveraging the vast domain knowledge available, such a ReDF module serves as guidance for prompt updates and augments the model's performance in generating accurate outputs. Third, \model{} adopts a conversation-based approach, aligning with the iterative refinement nature of the drug discovery pipeline. Such interactive schema enables a dynamic and collaborative process, effectively incorporating feedback from domain experts to achieve desired outcomes.\looseness=-1

Through our design, \model{} demonstrates two appealing properties for drug editing tasks: (1) \model{} exhibits an open vocabulary property, allowing for exploring novel drug concepts beyond a fixed set of pre-defined annotations. The model can generalize to new drug-related concepts due to the unrestricted nature of natural language. (2) \model{} possesses the compositional property. It can decompose complex concepts, such as multi-objective lead optimization, into simpler attributes like binding to a new protein and high permeability, enabling handling complicated drug editing tasks.
\footnote{\small Note that \model{} aims to inspire domain experts rather than replace them. While \model{} can propose optimized drugs or novel attributes, its primary role is to serve as a tool for knowledge exploration. The generated outputs can provide valuable insights and spark inspiration for domain experts in the drug discovery process.\looseness=-1}

Then to fully verify the effectiveness of \model{}, we need to design a benchmark for a wide range of drug editing tasks. Before going into details, we would like to claim two criteria for the task design: (1) The tasks should involve indeterministic answers, as they can serve as a source of inspiration for domain experts. (2) The tasks should be able to evaluate computationally since the lab experiment can be quite laborious and is beyond the discussion of this paper. Following these criteria, we introduce 39 editing tasks over three common drugs: \SmallMoleculeTaskNumber{} for small molecules, \PeptideTaskNumber{} for peptides, and \ProteinTaskNumber{} for proteins.

Last but not least, we offer empirical evidence substantiating the capability of \model{} for a wide range of drug editing tasks. Quantitatively, \model{} can reach the best performance on 33 out of 39 drug editing tasks compared to seven baselines. Qualitatively, we further provide 10 case studies (more qualitative results in~\Cref{sec:visual_analysis}), illustrating that \model{} can successfully identify the important substructures for each type of drug, as follows. (1) For small molecules, \model{} is able to detect the key scaffold for molecule editing, such as changing polar or hydrophobic functional groups for tuning properties like solubility in water and permeability. (2) For peptides, \model{} accurately identifies the protein-specific binding motifs of the peptide sequences. (3) For proteins, \model{} can modify sequences with more $\alpha$-helix or $\beta$-strand structures after folding~\cite{jumper2021highly,lin2022language}. We additionally illustrate that \model{} can provide insightful explanations, serving as a knowledge extraction tool.\looseness=-1

\section{Preliminaries}
\vspace{-1ex}

\textbf{Data Structure of Drugs.}
Drugs~\cite{wishart2008drugbank,drug_definition} refer to certain specific substances that can be adopted to prevent, diagnose, treat, or relieve symptoms of a disease or abnormal condition. In this paper, we would like to explore the three most common drugs: small molecules~\cite{jayatunga2022ai}, proteins~\cite{frokjaer2005protein}, and peptides~\cite{craik2013future}. Small molecules are sets of atoms connected together through the covalent bonds. Commonly-used data structures include SMILES (simplified molecular-input line-entry system) strings~\cite{weininger1988smiles} and molecular graphs~\cite{duvenaud2015convolutional,kearnes2016molecular,liu2019n}. In \model{}, we consider using the SMILES strings. Proteins are complex macromolecules, and they are composed of 20 amino acids, where each amino acid is a small molecule. Regarding the protein data structure, we adopt the amino acid sequence ({\ie}, amino acid string), and the mapping between 20 alphabets and amino acids can be found in~\Cref{sec:data_specification}. Peptides are short chains of amino acids and can be viewed as a special type of protein. The demonstration of three data structures can be found in~\Cref{fig:pipeline_chatdrug}.

\textbf{Drug Editing and Problem Formulation.}
In this paper, we focus on the drug editing task. Drug editing is also known as \textit{lead optimization} or \textit{protein design}, an important drug discovery task. From the machine learning perspective, drug editing is a \textit{conditional generation} problem and can be formulated as follows. Suppose the input drug (SMILES string or amino acid sequence) is $\vx_{\text{in}}$, and a target or desired property in the textual description is also known as the \textit{text prompt} $\vx_t$ in literature~\cite{raffel2020exploring,liu2023pre}. Then condition on such text prompt, the goal is to obtain an optimized drug as:
\begin{equation} \label{eq:objective}
\small
\begin{aligned}
\vx_{\text{out}} = \text{\model{}}(\vx_{\text{in}}, \vx_t).
\end{aligned}
\end{equation}
Then an evaluation metric $E(\vx_{\text{in}}, \vx_{\text{out}}; \vx_t) \in \{\text{True}, \text{False}\}$ is to check if the edited drugs can satisfy the desired properties compared to the input drugs, and we will average this over each corresponding task to get the \textit{hit ratio}. Note that $E(\cdot,\cdot;\cdot)$ is task-specific, as will be discussed in~\Cref{sec:experiment}.

\section{Method: \model{} Framework} \label{sec:method}
\textbf{Overview.} Our framework is shown in~\Cref{fig:pipeline_chatdrug}. \model{} consists of three components: (1)Prompt Design for Domain-Specific (PDDS) module, (2) Retrieval and Domain Feedback (ReDF) module, and (3) conversation module. Given a task prompt and input drug, PDDS aims to generate the domain-specific prompt and concatenate it with the input drug to request ChatGPT for answers. One problem for current LLMs is that it does not fully utilize the prior domain knowledge. Thus, we design the ReDF module aiming to (1) guide the LLMs to solve this task by retrieving structurally similar examples from the database and adding examples into the prompt as demonstrations and (2) verify the correctness of the output by using a domain feedback function. If the output drug is incorrect after ReDF, we then adopt the conversation module to ask LLMs to generate a new drug iteratively. Note that \model{} is a parameter-free scheme and does not require any learning procedure.\looseness=-1

\subsection{PDDS Module} \label{sec:prompt_design}
\model{} is proposed to solve a challenging problem: generalization of a universally (w.r.t. data type and data source) well-trained LLM to solving scientific tasks. In natural language processing (NLP), prompt design or prompt engineering~\cite{liu2023pre} has proven to be an effective paradigm for generalizing well-trained LLMs to various NLP downstream tasks, including but not limited to sentiment classification~\cite{han2022ptr,hu2021knowledgeable}, textual entailment~\cite{webson2021prompt,shin2020autoprompt}, text summarization~\cite{he2020ctrlsum,schick2020few,dou2020gsum}.\looseness=-1

But the explorations of adapting ChatGPT for drug editing tasks have been lagging behind. In this paper, we are interested in investigating this problem on the three most common types of drugs: small molecules, protein-binding peptides, and proteins. Recall that the goal of \model{} is (as in \Cref{eq:objective}): $\vx_{\text{out}} = \text{\model{}}(\vx_{\text{in}}, \vx_t)$. Here the text prompts $\vx_t$ should be specifically designed to enable the generalization for domain-specific tasks with computationally feasible metrics. Additionally, we want to highlight that the objectives for drug editing (in $\vx_t$) should be about the \textit{high-level property} instead of \textit{exact substructure replacement}. There are two main reasons, as follows. (1) As discussed in~\Cref{sec:more_discussion}, \model{} suits better for fuzzy matching like edited drugs with desired properties. In contrast, exact substructure replacement can be easily and precisely performed by domain experts, and such replacement may lack the creative inspiration for humans. (2) Property-related questions have an ambiguous nature, leading to dispersed answers that spark inspiration for domain experts in the drug discovery process.

Then concretely on the prompt design, for small molecules, we consider properties like solubility, drug-likeness, permeability, and the number of acceptors/donors. For peptides, we consider the properties of peptide-MHC binding. For proteins, we consider the secondary structure. The text prompts are to explicitly depict the desired properties to be either higher or lower, and corresponding task prompts will be briefly explained in~\Cref{sec:experiment}. One concrete example for molecule editing is \textit{``Can you make molecule [$\vx_{\text{in}}$] more soluble in water.''}, and more details can be found in~\Cref{sec:task_specification}.

\subsection{ReDF Module} \label{sec:retrieval_with_domain_feedback}
To better utilize the domain knowledge, we propose an important module: the ReDF (retrieval and domain feedback) module. The intuition is that there exists rich domain knowledge in the form of a retrieval database (DB), and ReDF will retrieve the useful information and inject it into the text prompt, adopting the fascinating language understanding ability of conversational LLMs.

Specifically, for each input drug $\vx_{\text{in}}$ and prompt $\vx_t$, we have a candidate drug $\tilde \vx$, which does not satisfy the desired property change in $\vx_t$. The candidate drug has multiple data resources, depending on the problem setup; in \model{}, it is the output drug with the negative result at each conversation round (will be introduced in~\Cref{sec:conversational_model}). Based on these, ReDF will return a drug $\vx_R$ satisfying:
\begin{equation} \label{eq:retrieval_with_domain_feedback}
\small
\vx_R = \text{ReDF}(\vx_{\text{in}}, \tilde \vx; \vx_t) = \argmax_{\vx_R' \in \text{Retrieval DB}} \langle \tilde \vx, \vx_R' \rangle \wedge D(\vx_{\text{in}}, \vx_R'; \vx_t),
\end{equation} 
where $D(\cdot, \cdot; \cdot) \in \{\text{True}, \text{False}\}$ is the domain feedback function, and $\langle \tilde \vx, \vx_R' \rangle$ is the similarity function. We use Tanimoto similarity~\cite{bajusz2015tanimoto} for small molecules and Levenshtein distance for peptides and proteins. Notice that here we take $D(\cdot, \cdot; \cdot)$ the same as evaluation metric $E(\cdot, \cdot; \cdot)$, while there is some critical difference on the task-specific thresholds, as will be discussed in the ablation study in~\Cref{sec:ablation_study_feedback_condition}. Then the ReDF module injects $\vx_R$ into a new prompt, {\eg}, the updated prompt for a molecule task is \textit{``Your provided sequence [$\tilde \vx$] is not correct. We find a sequence [$\vx_R$] which is correct and similar to the molecule you provided. Can you give me a new molecule?''}

We also want to highlight that the domain feedback injection in ReDF is similar to the \textit{in-context learning (ICL)} paradigm~\cite{dong2022survey}. Such knowledge injection can result in performance gain~\cite{min2022rethinking} not only because of the mapping between ground truth data-label pairs, but also the format or demonstration of the in-distribution data and label space. In~\Cref{sec:experiment}, we will conduct an ablation study on ICL.

\subsection{Conversation Module} \label{sec:conversational_model}
Another appealing attribute of conversational LLMs (like ChatGPT) is their interactive capability. This enables the LLMs to iteratively update the results by injecting prior knowledge. Inspired by this, we also consider adapting the conversational strategy for \model{}, which can naturally fit the ReDF module as described in in~\Cref{sec:retrieval_with_domain_feedback}. Then concretely on this conversational strategy in \model{}, first suppose there are $C$ conversation rounds, and we have an edited drug $\vx_c$ for the conversation round $c$. If $\vx_c$ satisfies our condition in the task prompt, then \model{} will exit. Otherwise, users will tell \model{} that $\vx_c$ is wrong, and we need to retrieve another similar but correct drug from the retrieval DB using ReDF: $\vx_R = \text{ReDF}(\vx_{\text{in}}, \vx_c)$, with $\tilde \vx = \vx_c$ in~\Cref{eq:retrieval_with_domain_feedback}.

To sum up, for each conversation round, we request a drug $\vx_R$ similar to $\vx_c$, which will be updated at each conversation round. The $\vx_c$ and $\vx_R$ serve as two in-context pairs to feed into \model{}, {\ie}, \textit{"The output drug at round [c] is [$\vx_c$], which is wrong. We find a sequence [$\vx_R$] which is correct and similar. Can you help improve the edited results?"} An illustration of this conversation is in~\Cref{fig:pipeline_chatdrug}.

\section{Experiment} \label{sec:experiment}
\textbf{Specifications for \model{}.} In this section, we verify the effectiveness of \model{} for drug editing on three types of drugs: small molecules, peptides, and proteins. Here we select GPT-3.5 in our experiment. We introduce three types of drugs and five categories of tasks accordingly: task 1xx and 2xx are single- and multi-objective tasks for small molecules (each task further includes 2 subtasks w.r.t. two thresholds as will be discussed next), task 3xx and 4xx are single- and multi-objective editing tasks for peptides, and task 5xx is for single-objective protein editing. Due to the space limitation, please check~\Cref{sec:task_specification} for the full list. Details of implementation and hyperparameters are in~\Cref{sec:implementation_hyperparameters}.\looseness=-1

\subsection{Text-guided Molecule Property Editing}
The first experiment is text-guided molecule editing or molecule optimization. We adopt 16 single-objective tasks and 12 multi-objective editing tasks from MoleculeSTM~\cite{liu2022multi}. These tasks are about the high-level properties of small molecules, like solubility in water and permeability.

\begin{table}[t!]
\centering
\caption{
\small Results on 16 single-objective small molecule editing, and the evaluation is the hit ratio of the property change. For \model{}, we report the mean and std of five random seeds. The best results are marked in \textbf{bold}.
}
\label{tab:results_small_molecule_editing_single_objective}
\vspace{-2ex}
\begin{adjustbox}{max width=\linewidth}
\begin{tabular}{l l rrrrrrr}
\toprule
Single Target Property & $\Delta$ & Random & PCA & High Variance & GS-Mutate & \makecell{MoleculeSTM\\(SMILES)} & \makecell{MoleculeSTM\\(Graph)} & \makecell{\model\\(Ours)}\\
\midrule
\multirow{2}{*}{101 \textit{more soluble in water}} & 0 & 35.33 $\pm$ 1.31 & 33.80 $\pm$ 3.63 & 33.52 $\pm$ 3.75 & 52.00 $\pm$ 0.41 & 61.87 $\pm$ 2.67 & 67.86 $\pm$ 3.46 & \textbf{94.13$\pm$1.04}\\
 & 0.5 & 11.04 $\pm$ 2.40 & 10.66 $\pm$ 3.24 & 10.86 $\pm$ 2.56 & 14.67 $\pm$ 0.62 & 49.02 $\pm$ 1.84 & 54.44 $\pm$ 3.99 & \textbf{88.67$\pm$0.95}\\
\midrule
\multirow{2}{*}{102 \textit{less soluble in water}} & 0 & 43.36 $\pm$ 3.06 & 39.36 $\pm$ 2.55 & 42.89 $\pm$ 2.36 & 47.50 $\pm$ 0.41 & 52.71 $\pm$ 1.67 & 64.79 $\pm$ 2.76 & \textbf{96.86$\pm$1.10}\\
 & 0.5 & 19.75 $\pm$ 1.56 & 15.12 $\pm$ 2.93 & 18.22 $\pm$ 0.33 & 12.50 $\pm$ 0.82 & 30.47 $\pm$ 3.26 & 47.09 $\pm$ 3.42 & \textbf{70.08$\pm$3.44}\\
\midrule
\multirow{2}{*}{103 \textit{more like a drug}} & 0 & 38.06 $\pm$ 2.57 & 33.99 $\pm$ 3.72 & 36.20 $\pm$ 4.34 & 28.00 $\pm$ 0.71 & 36.52 $\pm$ 2.46 & 39.97 $\pm$ 4.32 & \textbf{48.65$\pm$3.39}\\
 & 0.1 & 5.27 $\pm$ 0.24 & 3.97 $\pm$ 0.10 & 4.44 $\pm$ 0.58 & 6.33 $\pm$ 2.09 & 8.81 $\pm$ 0.82 & 14.06 $\pm$ 3.18 & \textbf{19.37$\pm$5.54}\\
\midrule
\multirow{2}{*}{104 \textit{less like a drug}} & 0 & 36.96 $\pm$ 2.25 & 35.17 $\pm$ 2.61 & 39.99 $\pm$ 0.57 & 71.33 $\pm$ 0.85 & 58.59 $\pm$ 1.01 & \textbf{77.62 $\pm$ 2.80} & 70.75$\pm$2.92\\
 & 0.1 & 6.16 $\pm$ 1.87 & 5.26 $\pm$ 0.95 & 7.56 $\pm$ 0.29 & 27.67 $\pm$ 3.79 & 37.56 $\pm$ 1.76 & \textbf{54.22 $\pm$ 3.12} & 30.99$\pm$2.66\\
\midrule
\multirow{2}{*}{105 \textit{higher permeability}} & 0 & 25.23 $\pm$ 2.13 & 21.36 $\pm$ 0.79 & 21.98 $\pm$ 3.77 & 22.00 $\pm$ 0.82 & 57.74 $\pm$ 0.60 & \textbf{59.84 $\pm$ 0.78} & 56.56$\pm$1.84\\
 & 10 & 17.41 $\pm$ 1.43 & 14.52 $\pm$ 0.80 & 14.66 $\pm$ 2.13 & 6.17 $\pm$ 0.62 & 47.51 $\pm$ 1.88 & \textbf{50.42 $\pm$ 2.73} & 43.08$\pm$2.95\\
\midrule
\multirow{2}{*}{106 \textit{lower permeability}} & 0 & 16.79 $\pm$ 2.54 & 15.48 $\pm$ 2.40 & 17.10 $\pm$ 1.14 & 28.83 $\pm$ 1.25 & 34.13 $\pm$ 0.59 & 31.76 $\pm$ 0.97 & \textbf{77.35$\pm$1.98}\\
 & 10 & 11.02 $\pm$ 0.71 & 10.62 $\pm$ 1.86 & 12.01 $\pm$ 1.01 & 15.17 $\pm$ 1.03 & 26.48 $\pm$ 0.97 & 19.76 $\pm$ 1.31 & \textbf{66.69$\pm$2.74}\\
\midrule
\multirow{2}{*}{107 \makecell[l]{\textit{more hydrogen bond acceptors}}} & 0 & 12.64 $\pm$ 1.64 & 10.85 $\pm$ 2.29 & 11.78 $\pm$ 0.15 & 21.17 $\pm$ 3.09 & 54.01 $\pm$ 5.26 & 37.35 $\pm$ 0.79 & \textbf{95.35$\pm$0.62}\\
 & 1 & 0.69 $\pm$ 0.01 & 0.90 $\pm$ 0.84 & 0.67 $\pm$ 0.01 & 1.83 $\pm$ 0.47 & 27.33 $\pm$ 2.62 & 16.13 $\pm$ 2.87 & \textbf{72.60$\pm$2.51}\\
\midrule
\multirow{2}{*}{108 \makecell[l]{\textit{more hydrogen bond donors}}} & 0 & 2.97 $\pm$ 0.61 & 3.97 $\pm$ 0.55 & 6.23 $\pm$ 0.66 & 19.50 $\pm$ 2.86 & 28.55 $\pm$ 0.76 & 60.97 $\pm$ 5.09 & \textbf{96.54$\pm$1.31}\\
 & 1 & 0.00 $\pm$ 0.00& 0.00 $\pm$ 0.00& 0.00 $\pm$ 0.00& 1.33 $\pm$ 0.24 & 7.69 $\pm$ 0.56 & 32.35 $\pm$ 2.57 & \textbf{76.43$\pm$3.32}\\
\bottomrule
\end{tabular}
\end{adjustbox}
\end{table}

\begin{table}[tb!]
\centering
\caption{
\small
Results on 12 multi-objective small molecule editing, and the evaluation is the hit ratio of the property change. For \model{}, we report the mean and std of five random seeds. The best results are marked in \textbf{bold}.
}
\label{tab:results_small_molecule_editing_multiple_objective}
\vspace{-2ex}
\begin{adjustbox}{max width=\linewidth}
\begin{tabular}{l l rrrrrrr}
\toprule
Two Target Properties & $\Delta$ & Random & PCA & High Variance & GS-Mutate & \makecell{MoleculeSTM\\(SMILES)} & \makecell{MoleculeSTM\\(Graph)} & \makecell{\model\\(Ours)}\\
\midrule
\multirow{2}{*}{\makecell[l]{201 \textit{more soluble in water} and\\\textit{more hydrogen bond acceptors}}} & 0 -- 0 & 9.88 $\pm$ 1.03 & 8.64 $\pm$ 2.06 & 9.09 $\pm$ 1.25 & 14.00 $\pm$ 2.48 & 27.87 $\pm$ 3.86 & 27.43 $\pm$ 3.41 & \textbf{79.62$\pm$0.64}\\
 & 0.5 -- 1 & 0.23 $\pm$ 0.33 & 0.45 $\pm$ 0.64 & 0.22 $\pm$ 0.31 & 0.67 $\pm$ 0.62 & 8.80 $\pm$ 0.04 & 11.10 $\pm$ 1.80 & \textbf{49.64$\pm$2.66}\\
\midrule
\multirow{2}{*}{\makecell[l]{202 \textit{less soluble in water} and\\\textit{more hydrogen bond acceptors}}} & 0 -- 0 & 2.99 $\pm$ 0.38 & 2.00 $\pm$ 0.58 & 2.45 $\pm$ 0.67 & 7.17 $\pm$ 0.85 & 8.55 $\pm$ 2.75 & 8.21 $\pm$ 0.81 & \textbf{51.59$\pm$3.79}\\
 & 0.5 -- 1 & 0.45 $\pm$ 0.32 & 0.00 $\pm$ 0.00& 0.22 $\pm$ 0.31 & 0.17 $\pm$ 0.24 & 2.93 $\pm$ 0.30 & 0.00 $\pm$ 0.00& \textbf{24.92$\pm$4.85}\\
\midrule
\multirow{2}{*}{\makecell[l]{203 \textit{more soluble in water} and\\\textit{more hydrogen bond donors}}} & 0 -- 0 & 2.28 $\pm$ 1.15 & 2.23 $\pm$ 1.16 & 4.44 $\pm$ 0.58 & 13.83 $\pm$ 2.95 & 33.51 $\pm$ 4.08 & 49.23 $\pm$ 1.71 & \textbf{89.34$\pm$0.96}\\
 & 0.5 -- 1 & 0.00 $\pm$ 0.00& 0.00 $\pm$ 0.00& 0.00 $\pm$ 0.00& 0.00 $\pm$ 0.00& 9.98 $\pm$ 1.03 & 23.94 $\pm$ 1.09 & \textbf{53.64$\pm$5.81}\\
\midrule
\multirow{2}{*}{\makecell[l]{204 \textit{less insoluble in water} and\\\textit{more hydrogen bond donors}}} & 0 -- 0 & 0.69 $\pm$ 0.58 & 1.96 $\pm$ 0.87 & 1.79 $\pm$ 0.66 & 5.67 $\pm$ 0.62 & 17.03 $\pm$ 2.75 & 14.42 $\pm$ 3.43 & \textbf{39.90$\pm$3.86}\\
 & 0.5 -- 1 & 0.00 $\pm$ 0.00& 0.00 $\pm$ 0.00& 0.00 $\pm$ 0.00& 0.00 $\pm$ 0.00& 2.59 $\pm$ 1.14 & 3.84 $\pm$ 0.71 & \textbf{24.19$\pm$2.19}\\
\midrule
\multirow{2}{*}{\makecell[l]{205 \textit{more soluble in water} and\\\textit{higher permeability}}} & 0 -- 0 & 5.06 $\pm$ 1.21 & 3.53 $\pm$ 0.38 & 4.88 $\pm$ 2.21 & 8.17 $\pm$ 1.03 & 35.69 $\pm$ 3.19 & \textbf{39.74 $\pm$ 2.26} & 12.85$\pm$2.68\\
 & 0.5 -- 10 & 1.16 $\pm$ 0.68 & 0.67 $\pm$ 0.55 & 0.66 $\pm$ 0.54 & 0.00 $\pm$ 0.00& 19.15 $\pm$ 0.73 & \textbf{22.66 $\pm$ 1.90} & 10.44$\pm$5.75\\
\midrule
\multirow{2}{*}{\makecell[l]{206 \textit{more soluble in water} and\\\textit{lower permeability}}} & 0 -- 0 & 12.17 $\pm$ 1.05 & 10.43 $\pm$ 2.88 & 13.08 $\pm$ 2.28 & 19.83 $\pm$ 2.46 & 44.35 $\pm$ 0.68 & 30.87 $\pm$ 0.62 & \textbf{65.33$\pm$2.16}\\
 & 0.5 -- 10 & 6.20 $\pm$ 0.64 & 6.23 $\pm$ 2.31 & 6.67 $\pm$ 0.53 & 4.83 $\pm$ 0.85 & 28.67 $\pm$ 2.22 & 20.06 $\pm$ 1.26 & \textbf{52.9$\pm$2.23}\\ 
\bottomrule
\end{tabular}
\end{adjustbox}
\end{table}

\textbf{Data}: Both the input molecules and retrieval DB are sampled from ZINC~\cite{irwin2012zinc}: we sample 200 and 10K molecules (with SMILES strings) from ZINC as input molecules and retrieval DB, respectively.
\textbf{Prompt}: The text prompt is \textit{"Can you make molecule [SMILES placeholder] [task requirement]? The output molecule should be similar to the input molecule"}. The [task requirement] is the textual description for each specific task, {\eg}, \textit{more soluble in water} and \textit{with higher permeability}.
\textbf{Evaluation}. We take the hit ratio to measure the success ratio of edited molecules, {\ie}, the percentage of edited molecules that can reach the desired properties compared to the input molecules. All the properties for small molecules considered here can be calculated deterministically using RDKit~\cite{landrum2013rdkit}. Another important argument is the threshold $\Delta$: it is a successful hit if the difference between input and output properties is above the threshold.
\textbf{Baselines}: The baselines are from~\cite{liu2022multi}, based on MegaMolBART~\cite{irwin2022chemformer}, a pretrained auto-regressive model. Baselines include Random, PCA, High-Variance, GS-Mutate, and MoleculeSTM with SMILES or Graph as the molecule representation.

\textbf{Observation.}
We illustrate the descriptions and the single- and multi-objective editing results in~\Cref{tab:results_small_molecule_editing_single_objective,tab:results_small_molecule_editing_multiple_objective}, respectively.
The threshold $\Delta$ for each specific task is specified in~\Cref{tab:results_small_molecule_editing_single_objective}; for multi-objective editing tasks in~\Cref{tab:results_small_molecule_editing_multiple_objective}, the threshold $\Delta$ has two values corresponding to the two tasks. We further conduct an ablation study on the thresholds of ReDF in~\Cref{sec:ablation_study_feedback_condition}.
We can observe that \model{} can reach the best performance on 22 out of \SmallMoleculeTaskNumber{} tasks, 20 of which possess over 20\% hit ratio than the second-best method.
\Cref{tab:small_molecule_analysis} visualizes examples of 6 molecule editing tasks where \model{} successfully generates output molecules $\vx_{\text{out}}$ with desirable property change, while the output of the first conversation round $\vx_1$ fail. For example, in~\Cref{tab:small_molecule_analysis}a, $\vx_1$ converts a methyl group to a propyl which incorrectly yields a less soluble molecule. Through conversational guidance, \model{} changes its output $\vx_{\text{out}}$ to an aminoethyl group, successfully fulfilling the task. In~\Cref{tab:small_molecule_analysis}f, $\vx_1$ installs a phenyl urea to the molecule, which brings lower permeability as requested but makes the molecule less soluble. In contrast, \model{} is able to replace the hydrophobic aromatic substituent with a hydrophilic amide in $\vx_{\text{out}}$, consistent with the requirement of higher solubility in water.

\begin{table}[t]
\setlength{\tabcolsep}{10pt}
\fontsize{9}{9}\selectfont
\centering
\caption{\small
Visualization of six small molecule editing tasks. The \colorbox{SmallMoleculeBlue}{blue regions}, \colorbox{SmallMoleculeRed}{red regions}, and \colorbox{SmallMoleculeGreen}{green regions} correspond to the edited substructures in the input molecule $\vx_{\text{in}}$, intermediate molecule $\vx_1$ for the 1st conversation round, and the output molecule $\vx_{\text{out}}$, respectively.
}
\label{tab:small_molecule_analysis}
\vspace{-2ex}
    \begin{adjustbox}{max width=\linewidth}
    \small
    \begin{tabular}{cccccc}
    \toprule
    \multicolumn{3}{c}{(a) Prompt for 101 : more soluble in water} &  \multicolumn{3}{c}{(b) Prompt  for 102: less soluble in water}\\
    \cmidrule(lr){1-3}\cmidrule(lr){4-6}
    {Input Molecule $\vx_{\text{in}}$} & {Intermediate Molecule $\vx_{1}$} & {Output Molecule $\vx_{\text{out}}$} & {Input Molecule $\vx_{\text{in}}$} & {Intermediate Molecule $\vx_{1}$} & {Output Molecule $\vx_{\text{out}}$}\\
    \cmidrule(lr){1-1}\cmidrule(lr){2-2}\cmidrule(lr){3-3}\cmidrule(lr){4-4}\cmidrule(lr){5-5}\cmidrule(lr){6-6}
    \adjustbox{valign=c}{\includegraphics[width=0.25\linewidth]{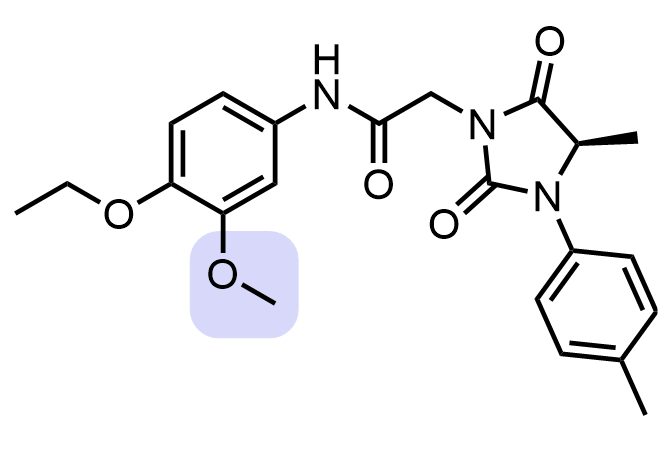}} &
    \adjustbox{valign=c}{\includegraphics[width=0.25\linewidth]{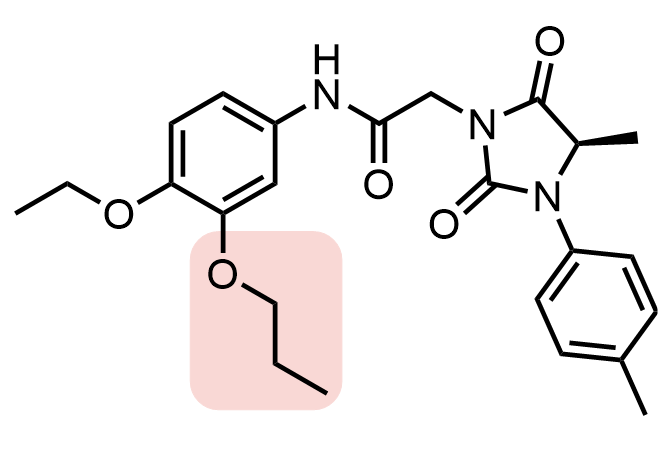}} &
    \adjustbox{valign=c}{\includegraphics[width=0.25\linewidth]{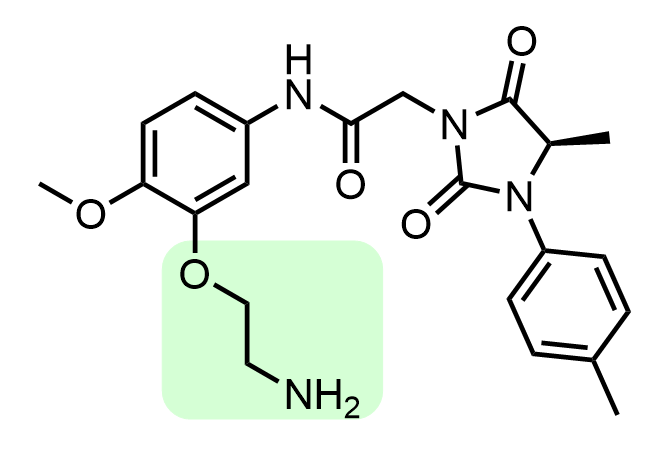}} &
    \adjustbox{valign=c}{\includegraphics[width=0.25\linewidth]{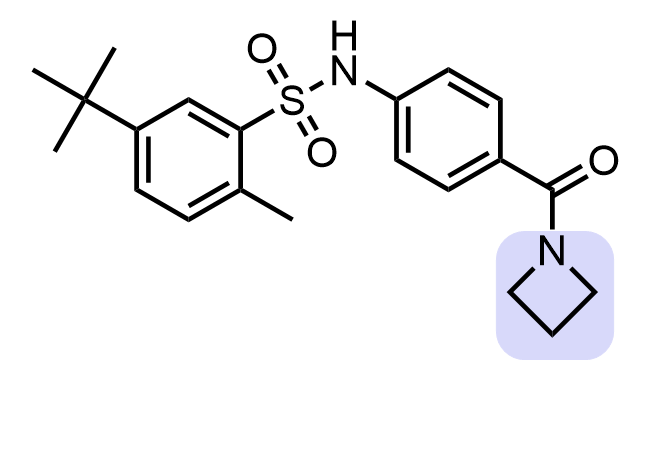}} &
    \adjustbox{valign=c}{\includegraphics[width=0.25\linewidth]{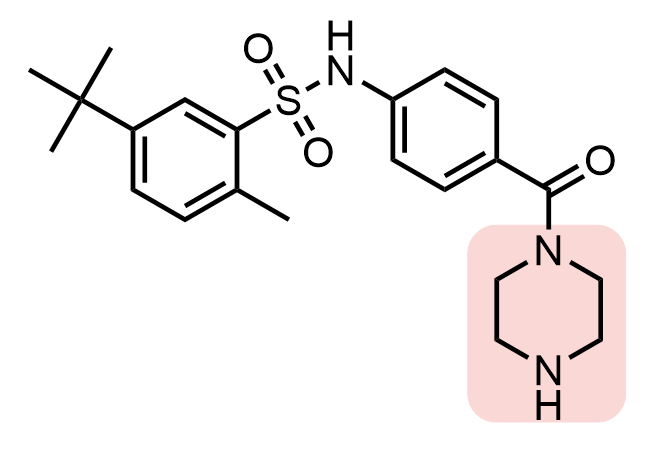}} &
    \adjustbox{valign=c}{\includegraphics[width=0.25\linewidth]{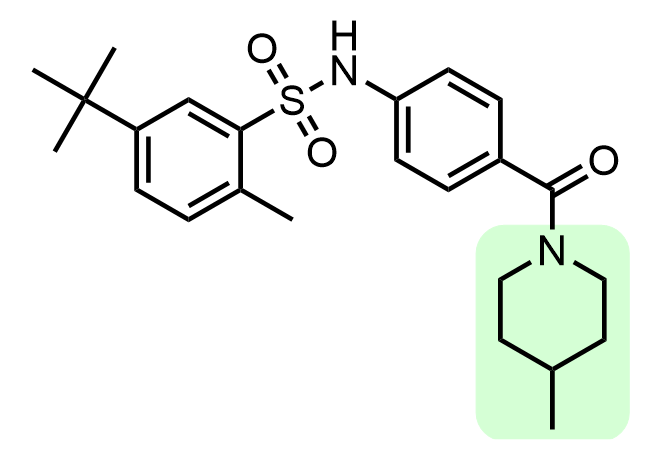}}
    \\
    {\small LogP: 1.46} & {\small LogP: 2.24} & {\small LogP: 0.40} & {\small LogP: 3.29} & {\small LogP: 2.49} & {\small LogP: 4.31}\\[5pt]
    \toprule
    \multicolumn{3}{c}{(c) Prompt for 105: higher permeability} &  \multicolumn{3}{c}{(d) Prompt for 106: lower permeability}\\
    \cmidrule(lr){1-3}\cmidrule(lr){4-6}
    {Input Molecule $\vx_{\text{in}}$} & {Intermediate Molecule $\vx_{1}$} & {Output Molecule $\vx_{\text{out}}$} & {Input Molecule $\vx_{\text{in}}$} & {Intermediate Molecule $\vx_{1}$} & {Output Molecule $\vx_{\text{out}}$}\\
    \cmidrule(lr){1-1}\cmidrule(lr){2-2}\cmidrule(lr){3-3}\cmidrule(lr){4-4}\cmidrule(lr){5-5}\cmidrule(lr){6-6}
    \adjustbox{valign=c}{\includegraphics[width=0.25\linewidth]{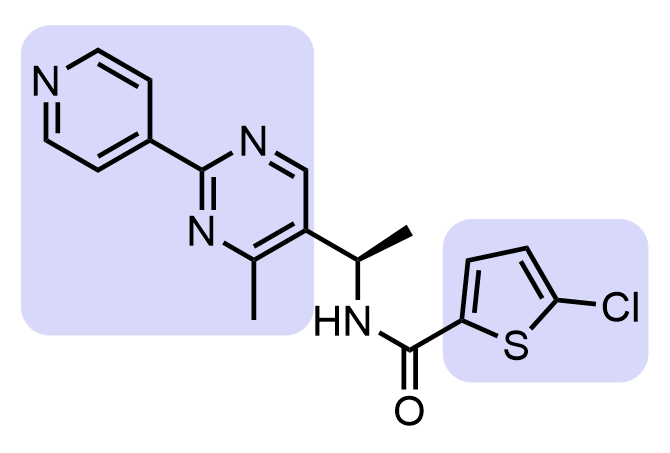}} &
    \adjustbox{valign=c}{\includegraphics[width=0.25\linewidth]{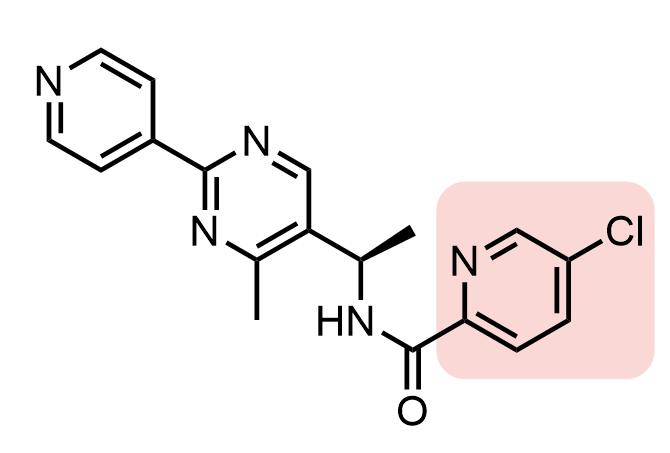}} &
    \adjustbox{valign=c}{\includegraphics[width=0.25\linewidth]{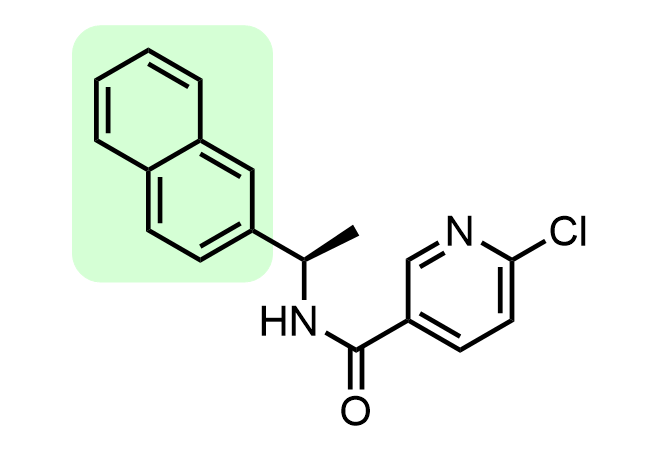}} &
    \adjustbox{valign=c}{\includegraphics[width=0.25\linewidth]{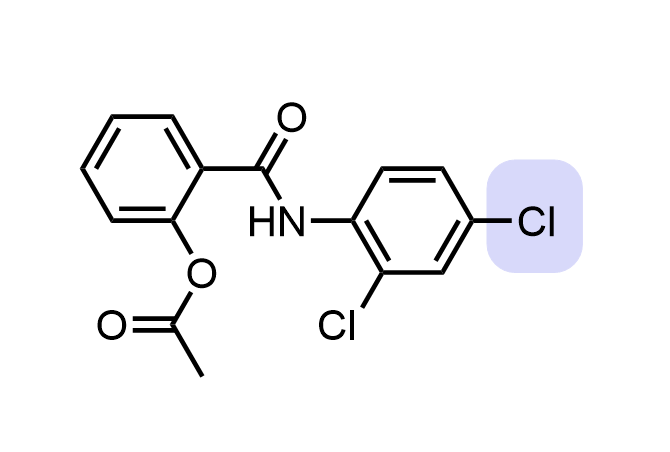}} &
    \adjustbox{valign=c}{\includegraphics[width=0.25\linewidth]{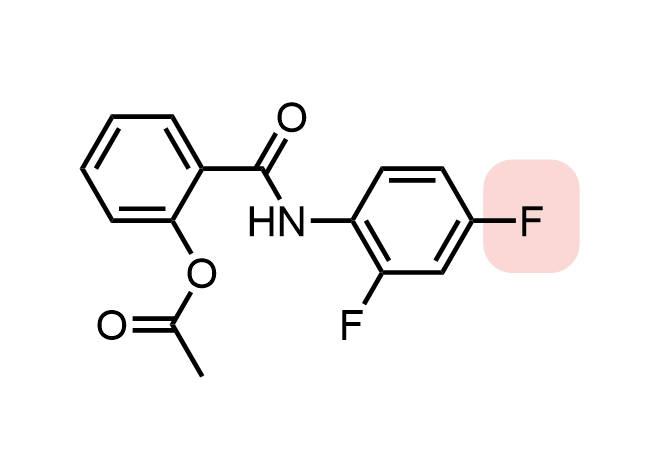}} &
    \adjustbox{valign=c}{\includegraphics[width=0.25\linewidth]{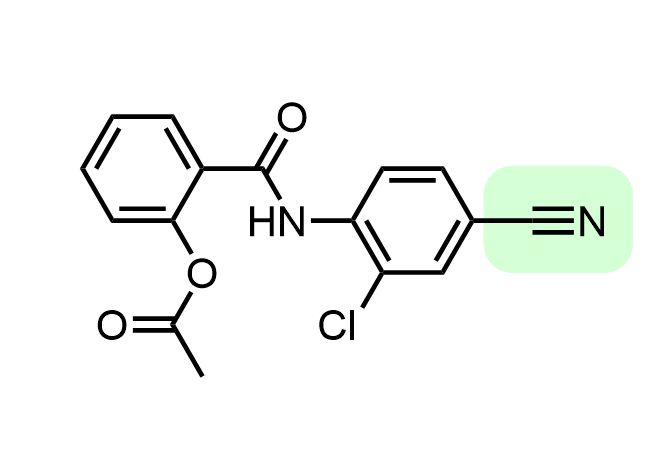}}
    \\
    {\small tPSA: 68} & {\small tPSA: 81} & {\small tPSA: 42} & {\small tPSA: 55} & {\small tPSA: 55} & {\small tPSA: 79}\\[5pt]
    \toprule
    \multicolumn{3}{c}{(e) Prompt for 205: more soluble in water and higher permeability} &  \multicolumn{3}{c}{(f) Prompt for 206: more soluble in water and lower permeability}\\
    \cmidrule(lr){1-3}\cmidrule(lr){4-6}
    {Input Molecule $\vx_{\text{in}}$} & {Intermediate Molecule $\vx_{1}$} & {Output Molecule $\vx_{\text{out}}$} & {Input Molecule $\vx_{\text{in}}$} & {Intermediate Molecule $\vx_{1}$} & {Output Molecule $\vx_{\text{out}}$}\\
    \cmidrule(lr){1-1}\cmidrule(lr){2-2}\cmidrule(lr){3-3}\cmidrule(lr){4-4}\cmidrule(lr){5-5}\cmidrule(lr){6-6}
    \adjustbox{valign=c}{\includegraphics[width=0.25\linewidth]{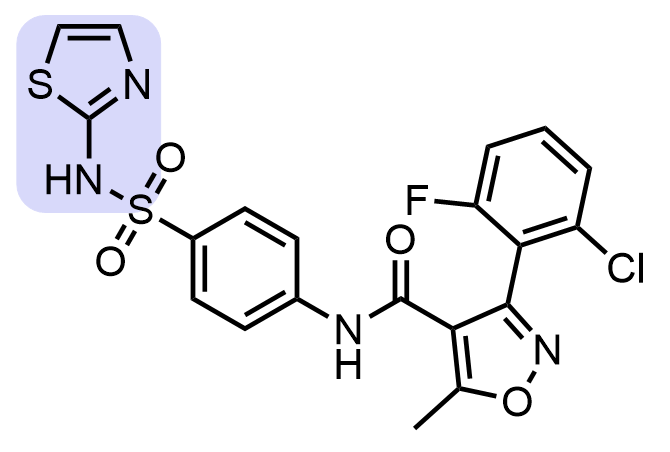}} &
    \adjustbox{valign=c}{\includegraphics[width=0.25\linewidth]{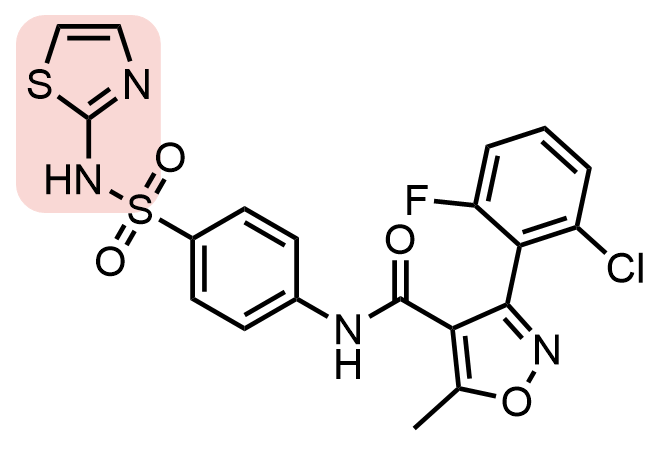}} &
    \adjustbox{valign=c}{\includegraphics[width=0.25\linewidth]{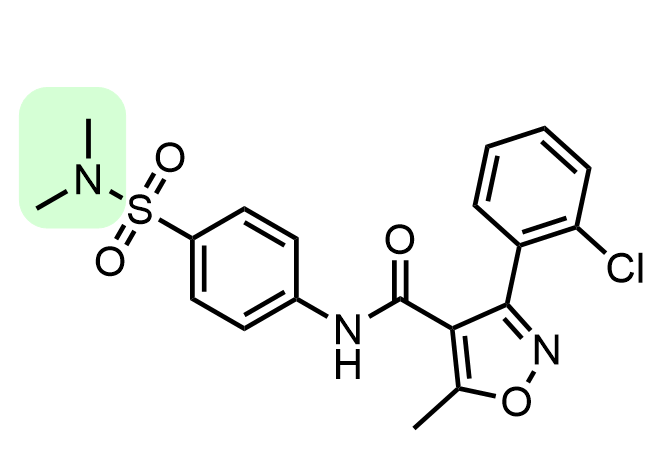}} &
    \adjustbox{valign=c}{\includegraphics[width=0.25\linewidth]{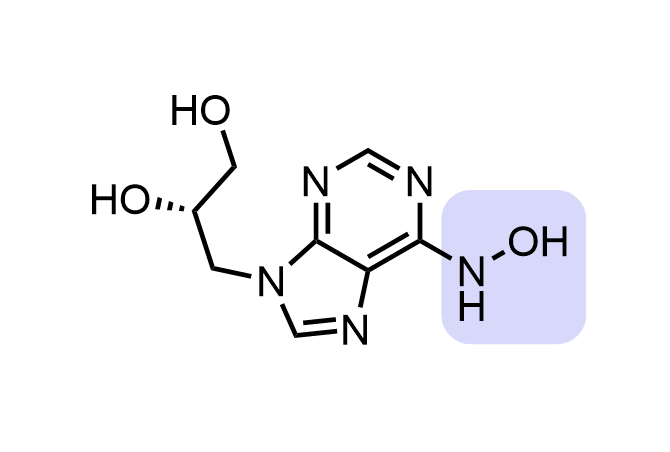}} &
    \adjustbox{valign=c}{\includegraphics[width=0.25\linewidth]{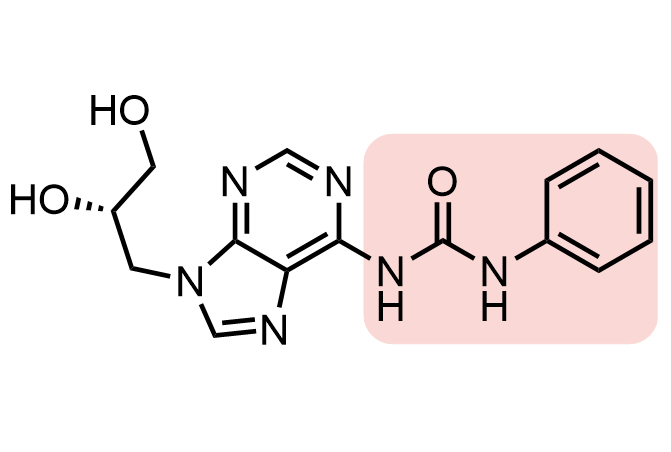}} &
    \adjustbox{valign=c}{\includegraphics[width=0.25\linewidth]{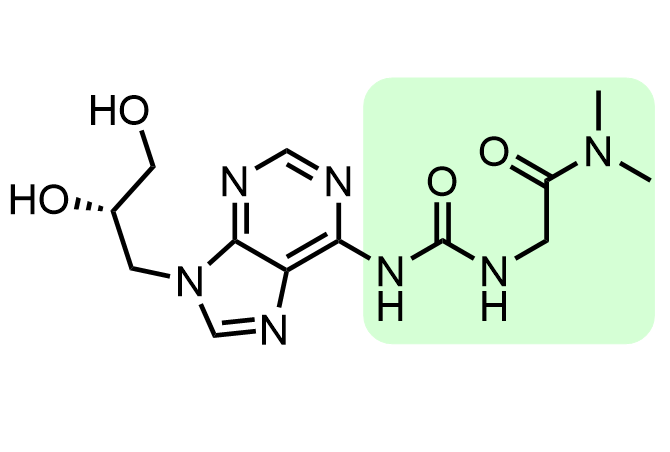}}
    \\
    {\small LogP: 3.59, tPSA: 114} & {\small LogP: 3.59, tPSA: 114} & {\small LogP: 2.83, tPSA: 93} & {\small LogP: -1.83, tPSA: 116} & {\small LogP: -0.37, tPSA: 125} & {\small LogP: -2.42, tPSA: 146}\\[5pt]
    \bottomrule
    \end{tabular}
    \end{adjustbox}
\end{table}

\subsection{Text-guided Immunogenic Binding Peptide Editing}
The second task is text-guided immunogenic binding peptide editing. Immunogenic peptides are promising therapeutic targets for the personalized vaccine, which triggers a person’s immune system, {\eg}, CD8+ T cells, to fight diseases~\cite{craiu1997two,hennecke2001t}. Immunogenic  peptides are typically degraded from intracellular antigens. To activate CD8+ T cell immune responses, these peptides must first bind to Major Histocompatibility Complex (MHC) proteins, forming peptide-MHC complexes which are then presented on the surface of infected or malignant cells to interact with the T cells. Although the peptide-MHC binding process is critical for immune response, it is highly specific, making editing known peptides to improve their binding affinity to specific MHC proteins a challenging yet crucial task for peptide engineering and discovery. Recall that peptides are typically short protein chains, with most peptides having less than 16 amino acids.

\textbf{Data}: In this experiment, we use the experimental dataset of peptide-MHC binding affinities~\cite{o2020mhcflurry}. This dataset contains 149 human MHC Class I proteins (alleles) and 309K peptides. We follow existing works~\cite{chen2023binding} on using the 30 common MHC proteins (alleles) and we randomly pick one as the source allele and one or more alleles as the target alleles. Notice that for single-allele tasks, 30 MHC proteins can be further divided into 3 categories: HLA-A, HLA-B, and HLA-C; we make sure that the sampled source and target alleles are from different categories. Then we sample 500 peptides from the source allele types. For the retrieval DB, the experimental data of the target allele(s) are adopted. The sampled MHC types are further specified in~\Cref{sec:task_specification}.
\textbf{Prompt}: We expect the edited peptides can bind to the target MHC protein(s), so the prompt template is \textit{We want a peptide that binds to [target allele]. We have a peptide [peptide sequence] that binds to [source allele], can you help modify it? The output peptide should be similar to the input peptide."}
\textbf{Evaluation}: The actual bindings require wet-lab experiments, which are expensive and prohibited for large scaled evaluation. Following existing works~\cite{Chen2020RankingbasedCN,chen2023binding}, we leverage the MHCflurry2.0~\cite{o2020mhcflurry} as a pseudo-oracle to predict the peptide-MHC binding affinity. MHCflurry2.0 is the state-of-the-art method enabling accurate estimating of the binding affinity of peptides with MHC proteins. The success of the peptide editing needs to satisfy two conditions: (1) The output peptide should have a higher binding affinity with the target allele compared to the input peptide; (2) The binding affinity of the output peptide and target allele should be above a certain threshold. Here we take the threshold as one-half of the average binding affinity of experimental data on the target allele.
\textbf{Baselines}: Since there is no existing approach for text-guided binding peptide editing, we use random mutation as the baseline, {\ie}, conducting random mutation on the amino acid sequence of the input peptides.

\begin{table}[tb!]
\centering
\caption{\small
Results on six single-objective and three multi-objective peptide editing tasks. Random Mutation-$R$ for $R$ mutated positions. The evaluation is the hit ratio of the increased binding affinity score. The best results are marked in \textbf{bold}. Due to the space limitation, please check~\Cref{sec:task_specification} for the text prompt of each task.
}
\label{tab:peptide_editing}
\vspace{-2ex}
\setlength{\tabcolsep}{10pt}
\begin{adjustbox}{max width=\linewidth}
\begin{tabular}{l rrrrrr rrr}
\toprule
& \multicolumn{6}{c}{single-objective editing} & \multicolumn{3}{c}{multi-objective editing}\\
\cmidrule(lr){2-7} \cmidrule(lr){8-10}
& 301 & 302 & 303 & 304 & 305 & 306 & 401 & 402 & 403\\
\midrule
Random Mutation-1 & 1.80 & 14.40 & 1.80 & 1.80 & 12.00 & 5.60 & 3.20 & 0.80 & 0.40\\
Random Mutation-2 & 1.80 & 13.40 & 2.80 & 3.00 & 8.40 & 4.40 & 2.20 & 0.60 & 1.20\\
Random Mutation-3 & 1.80 & 9.40 & 2.40 & 4.20 & 9.00 & 3.80 & 3.00 & 0.60 & 0.80\\
\model{} & \textbf{58.60} & \textbf{69.34} & \textbf{58.52} & \textbf{55.11} & \textbf{64.40} & \textbf{62.73} & \textbf{53.71} & \textbf{41.45} & \textbf{54.71}\\
\bottomrule
\end{tabular}
\end{adjustbox}
\end{table}

\begin{figure}[tb!]
\centering
\begin{subfigure}[b]{0.32\linewidth}
    \centering
    \includegraphics[width=\linewidth]{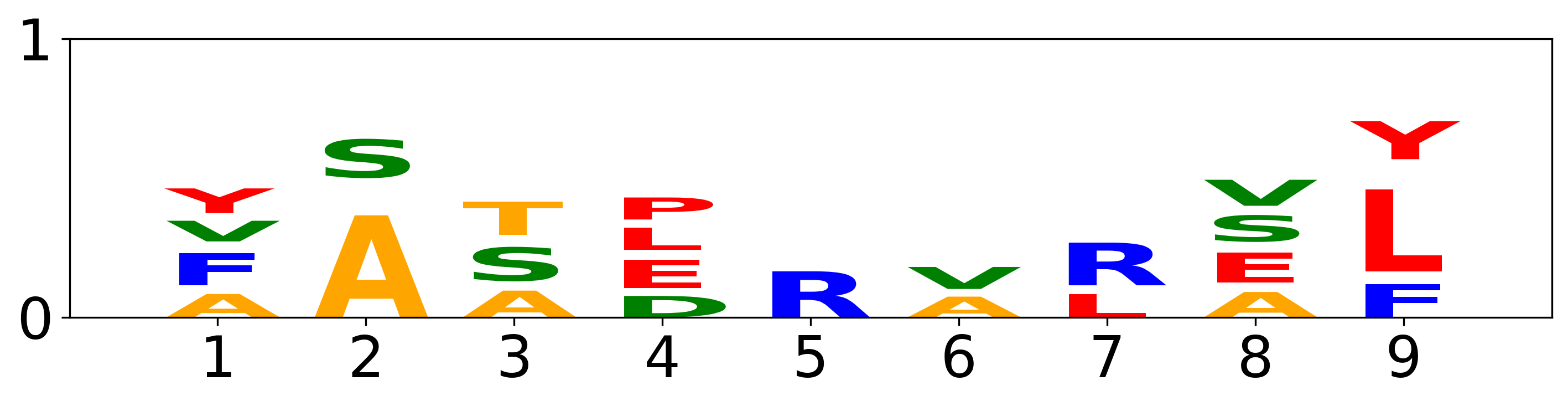}
    \vspace{-3ex}
    \caption{\fontsize{7.4}{6}\selectfont Motifs of input peptides for 301.}
\end{subfigure}
\hfill
\begin{subfigure}[b]{0.32\linewidth}
    \centering
    \includegraphics[width=\linewidth]{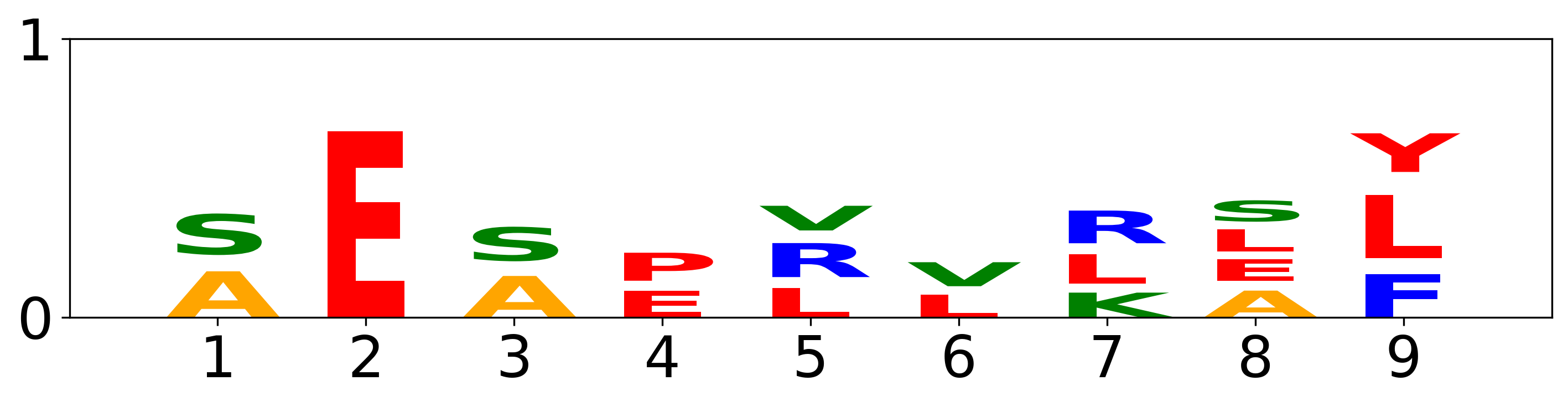}
    \vspace{-3ex}
    \caption{\fontsize{7.4}{6}\selectfont Motifs of edited peptides for 301.}
\end{subfigure}
\hfill
\begin{subfigure}[b]{0.32\linewidth}
    \centering
    \includegraphics[width=\linewidth]{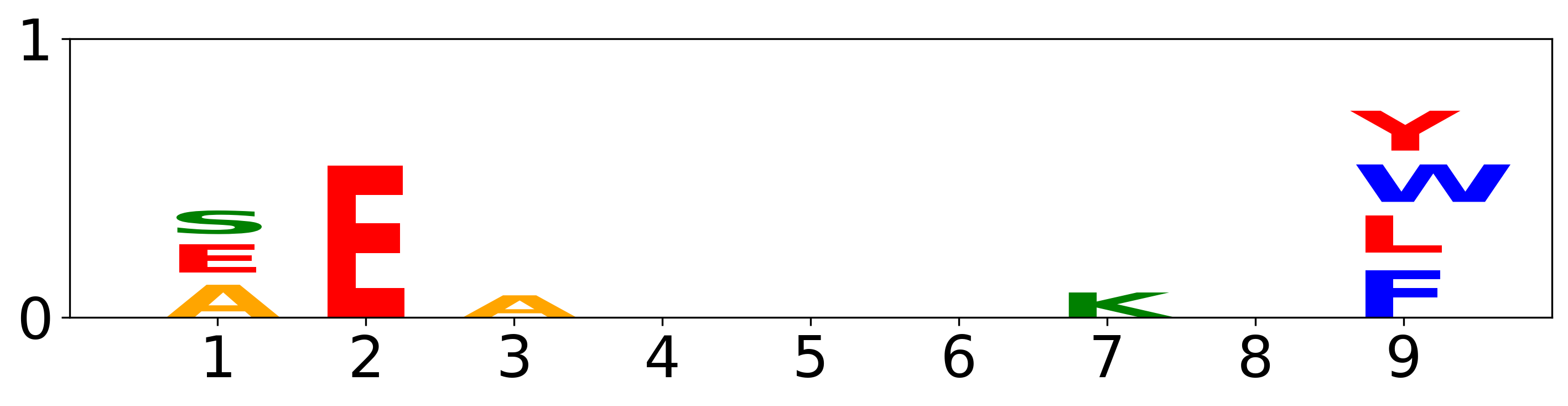}
    \vspace{-3ex}
    \caption{\fontsize{7.4}{6}\selectfont Motifs of experimental peptides for 301.}
\end{subfigure}
\hfill
\begin{subfigure}[b]{0.32\linewidth}
    \centering
    \includegraphics[width=\linewidth]{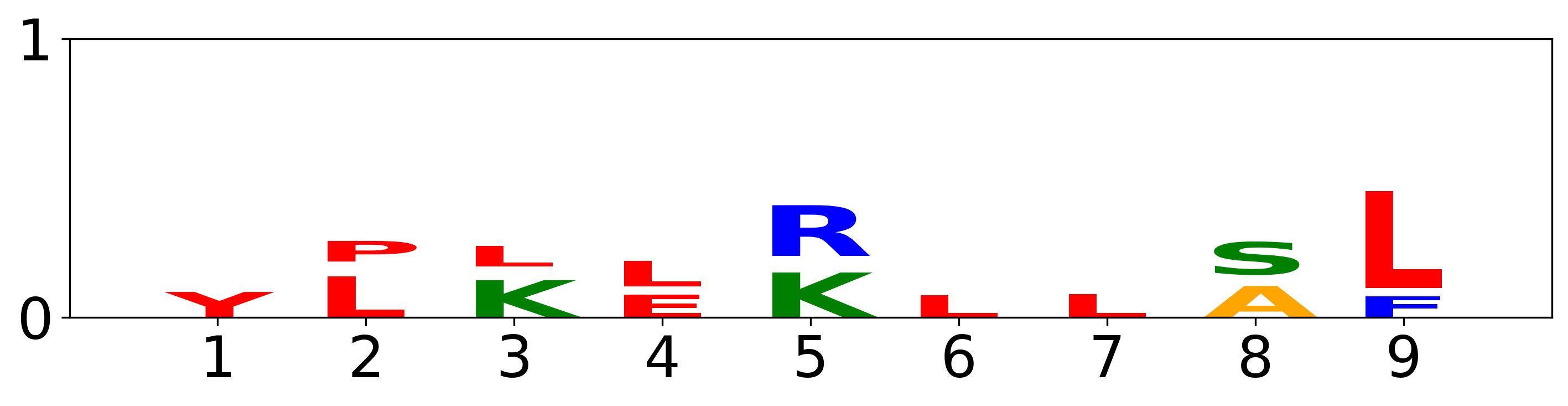}
    \vspace{-3ex}
    \caption{\fontsize{7.4}{6}\selectfont Motifs of input peptides for 302.}
\end{subfigure}
\hfill
\begin{subfigure}[b]{0.32\linewidth}
    \centering
    \includegraphics[width=\linewidth]{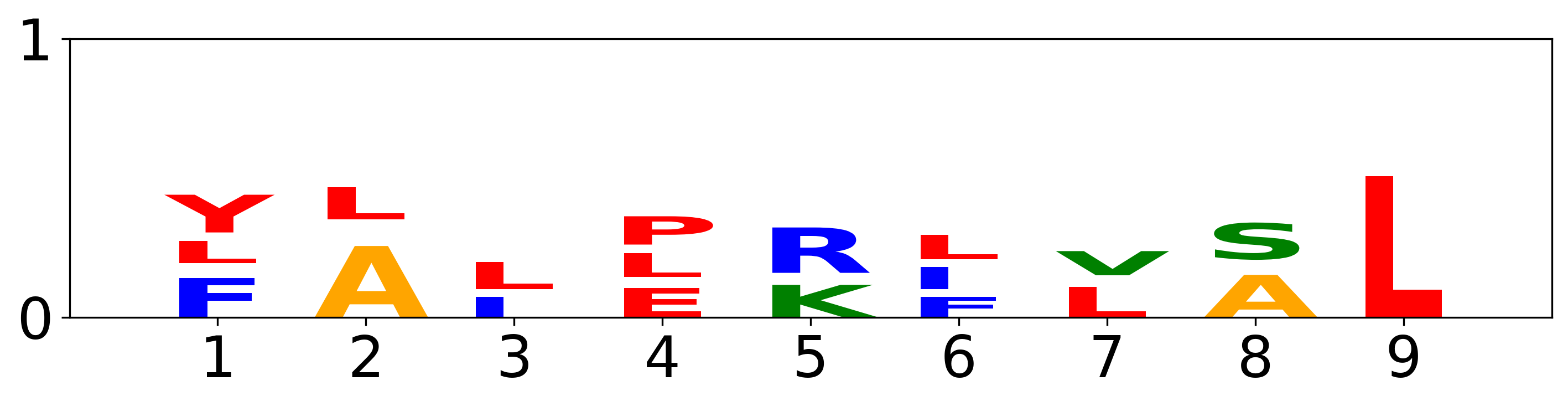}
    \vspace{-3ex}
    \caption{\fontsize{7.4}{6}\selectfont Motifs of edited peptides for 302.}
\end{subfigure}
\hfill
\begin{subfigure}[b]{0.32\linewidth}
    \centering
    \includegraphics[width=\linewidth]{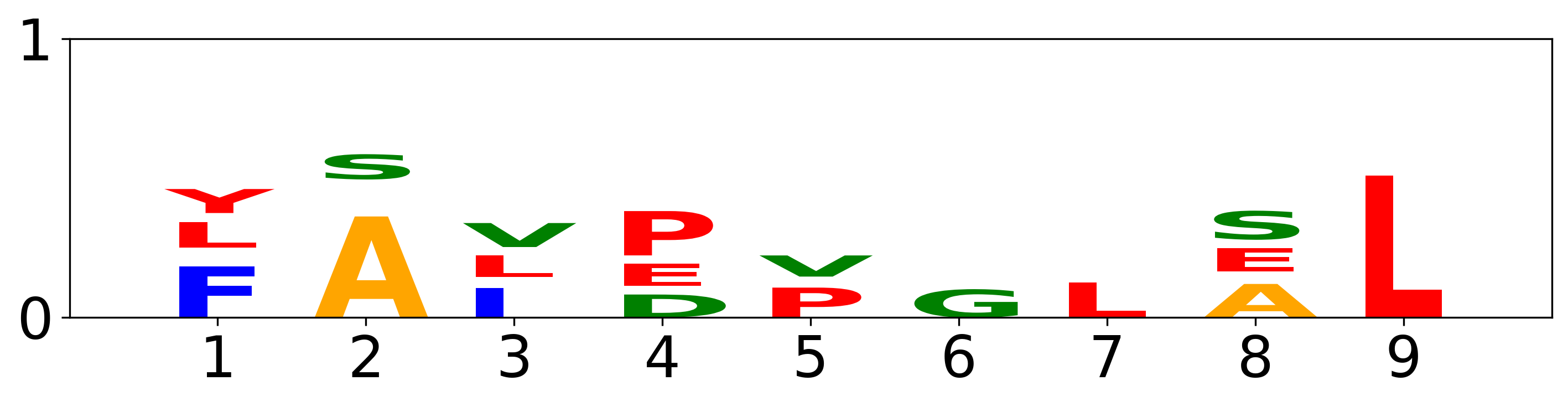}
    \vspace{-3ex}
    \caption{\fontsize{7.4}{6}\selectfont Motifs of experimental peptides for 302.}
\end{subfigure}
\vspace{-1ex}
\caption{\small
Visualization of two peptide editing tasks using PWM. The x-axis corresponds to the position index, while the y-axis corresponds to the distribution of each amino acid (in alphabets) at each position.
}
\label{fig:peptide_visualization_PWM}
\end{figure}

\textbf{Observation.}
We illustrate the single- and multi-objective editing results in~\Cref{tab:peptide_editing}. We can observe that \model{} reaches the best performance over all 9 tasks compared to the random mutation baselines. We further visualize peptides using position weight matrices (PWMs) in~\Cref{fig:peptide_visualization_PWM}. PWM has been widely used for the visualization of protein motifs (patterns), and it plots the distribution of each amino acid at the corresponding position. Thus, more important motifs with higher probabilities will be marked in higher alphabets. According to \Cref{fig:peptide_visualization_PWM}, the edited or optimized peptides follow similar patterns to the experimental data presented. For instance, for task 301, the edited peptides can successfully upweight the alphabet E (glutamic acid) at position 2; similarly, for alphabet A at position 2 and L at position 9 for task 302. These results indicate that the binding motifs of the edited peptides are highly correlated with the real binding motifs derived from experimental data. 

\newpage
\subsection{Text-guided Protein Secondary Structure Editing}
\begin{wraptable}[10]{r}{0.53\textwidth}
\setlength{\tabcolsep}{10pt}
\vspace{-2ex}
\centering
\caption{\small 
Results on two protein editing tasks. Random Mutation-$R$ for $R$ mutated positions. The evaluation is the hit ratio of increased secondary structures accordingly. The best results are marked in \textbf{bold}.
}
\label{tab:protein_editing}
\vspace{-2ex}
\begin{adjustbox}{max width=0.53\textwidth}
\begin{tabular}{l rr}
\toprule
& 501 more helix & 502 more strand\\
\midrule
Random Mutation-1 & 18.32 & 17.35\\
Random Mutation-2 & 24.95 & 19.69\\
Random Mutation-3 & 26.90 & 21.44\\
\model{} & \textbf{34.79} & \textbf{51.38} \\
\bottomrule
\end{tabular}
\end{adjustbox}
\end{wraptable}
Last but not least, we consider text-guided protein secondary structure editing (PSSE)~\cite{klausen2019netsurfp}. For protein 1D sequence, it can fold into the 3D structure, as shown in~\Cref{fig:pipeline_chatdrug}. Specifically, proteins possess four levels of structures, and secondary structures are fundamental building blocks, which are local folding patterns stabilized by hydrogen bonds. Typical secondary structures include $\alpha$-helix and $\beta$-sheet, consisting of $\beta$-strands. Here we are interested in two PSSE tasks, {\ie}, using \model{} to edit protein sequences with more helix or strand structures after folding~\cite{jumper2021highly,lin2022language}.

\textbf{Data}: TAPE~\cite{rao2019evaluating} is a benchmark for protein sequence property prediction, including the secondary structure prediction task. We take the test dataset and training dataset as the input proteins and retrieval DB, respectively.
\textbf{Prompt}: \textit{For an input protein sequence [protein sequence], can you modify it with more helix/strand structures?}
\textbf{Baselines}: Same with peptide editing, we adopt random mutation as baselines.
\textbf{Evaluation.} For evaluation, we adopt the state-of-the-art pretrained secondary structure prediction model, {\ie}, ProteinCLAP-EBM-NCE model from ProteinDT~\cite{liu2023text}. The hit condition is if the output protein sequences have more secondary structures than the input sequences.

\begin{figure}[tb!]
\centering
\begin{subfigure}[b]{0.47\linewidth}
    \centering
    \includegraphics[width=\linewidth]{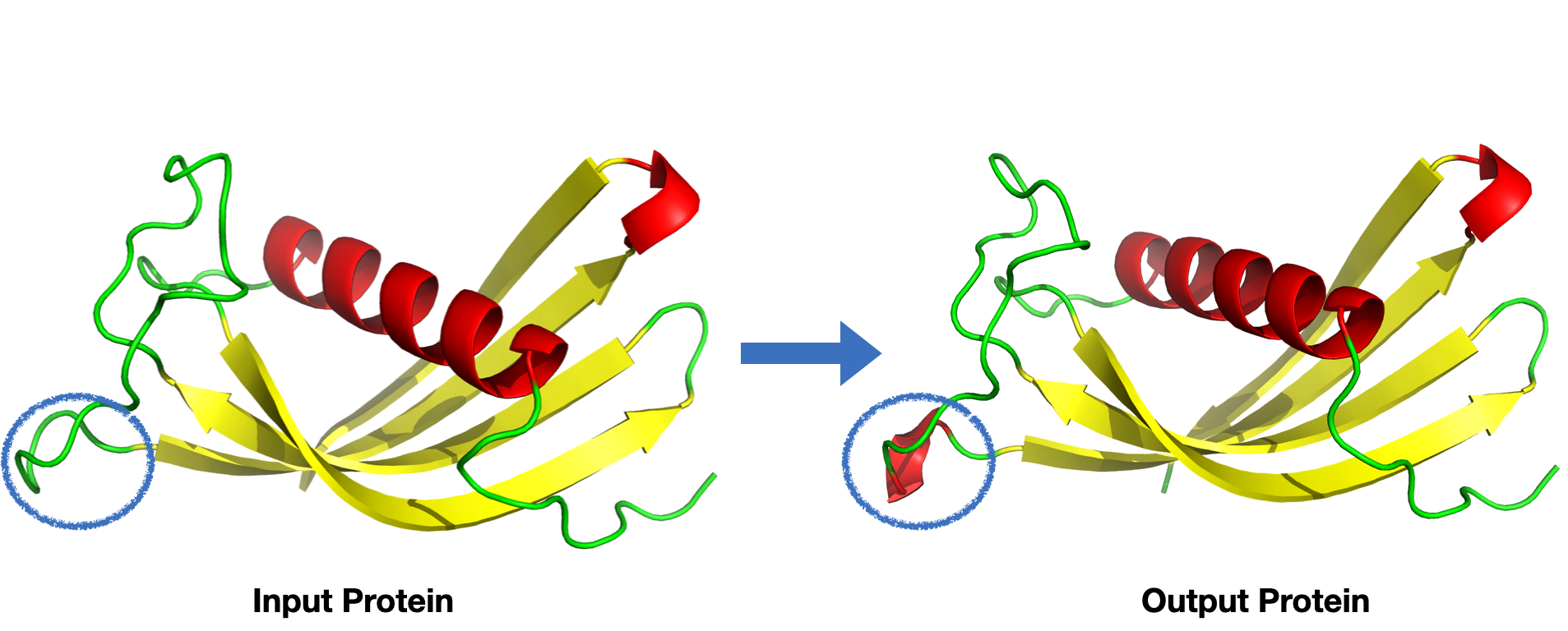}
    \caption{\small Protein editing with more helix structures.}
    \label{fig:protein_visual_701}
\end{subfigure}
\hfill
\begin{subfigure}[b]{0.47\linewidth}
    \centering
    \includegraphics[width=\linewidth]{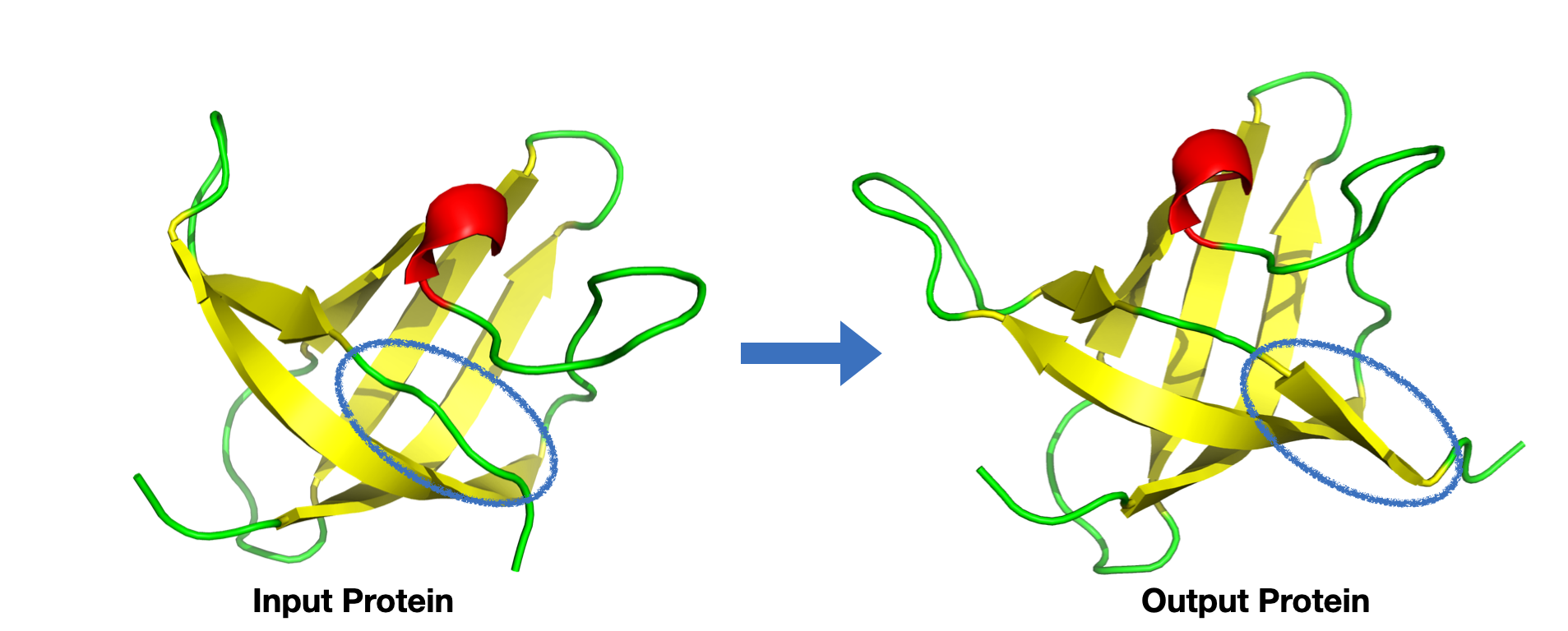}
    \caption{\small Protein editing with more strand structures.}
    \label{fig:protein_visual_702}
\end{subfigure}
\vspace{-1ex}
\caption{
\small Visualization of two protein editing tasks.
For the protein secondary structures, the $\alpha$-helix is marked in \textcolor{ProteinSecondaryStructureRed}{red}, and $\beta$-sheet is marked in \textcolor{ProteinSecondaryStructureYellow}{yellow}. The edited regions before and after \model{} are marked in \textcolor{ProteinSecondaryStructureBlue}{blue} circles.
}
\label{fig:protein_visualization}
\end{figure}

\textbf{Observation.} Because we only consider two types of secondary structures in PSSE, the tasks are single-objective tasks. As shown in~\Cref{tab:protein_editing}, we can tell the large performance gain by \model{}. We further visualize cases on how \model{} successfully edits the proteins with more helix/strand structures. We adopt pretrained ESMFold~\cite{lin2022language} for protein folding (protein sequence to protein structure prediction) and then plot the protein structures using PyMOL~\cite{pymol}. We show two examples in \Cref{fig:protein_visualization}. As circled in the \textcolor{ProteinSecondaryStructureBlue}{blue regions} in~\Cref{fig:protein_visual_701,fig:protein_visual_702}, the edited proteins possess more \textcolor{ProteinSecondaryStructureRed}{helix structures} and \textcolor{ProteinSecondaryStructureYellow}{strand structures}, respectively. More visualization can be found in~\Cref{sec:visual_analysis}.

\subsection{Ablation Study on Comparison with Zero-shot and In-context Learning} \label{sec:ablation_study_zero_shot_in_context_learning}
There are two important modules for \model{}: conversation for result refinement and the ReDF for knowledge retrieval. Thus in this ablation study, we would like to explore the effect of such two modules. The first case is zero-shot prediction. It is indeed \model{} with $c=0$, {\ie}, without conversation or ReDF. On the other hand, in-context learning (ICL) can be treated as \model{} equipped with the ReDF module but without any conversational round. Concretely, the retrieved drug is $\vx_R = \text{ReDF}(\vx_{\text{in}}, \vx_{\text{in}})$, with $\tilde \vx = \vx_{\text{in}}$ in~\Cref{eq:retrieval_with_domain_feedback}. The text prompt for zero-shot and ICL are \textit{``Can you edit the molecule [$\vx_{\text{in}}$] to be more soluble in water?''} and `\textit{`We know that [$\vx_R$] is similar to [$\vx_{\text{in}}$] and is more soluble in water. Can you edit the molecule [$\vx_{\text{in}}$] to be more soluble in water?''} The results can be found in~\Cref{tab:small_molecule_ablation_on_ICL_and_conversation_rounds}. As we can see, both \model{} and ICL are better than the zero-shot prediction, and conversational refinement performs best on all 14 tasks.

\subsection{Ablation Study on the Number of Conversation Rounds in \model{}}
In \model{}, the number of conversation rounds is an important hyperparameter. Here we conduct an ablation study on small molecules to test its effectiveness. The results are in~\Cref{tab:small_molecule_ablation_on_ICL_and_conversation_rounds}. For molecule editing tasks tested here, the performance of \model{} tends to converge after $C=2$ conversation rounds. This motives us taking $C=2$ for the main results in~\Cref{tab:results_small_molecule_editing_single_objective,tab:results_small_molecule_editing_multiple_objective,tab:peptide_editing,tab:protein_editing}.\looseness=-1

\begin{table}[tb!]
\centering
\caption{\small Ablation studies on comparison with in-context learning (ICL) and conversation rounds on molecule editing. The threshold is the loose threshold with $\Delta=0$, and the random seed is 0.}
\label{tab:small_molecule_ablation_on_ICL_and_conversation_rounds}
\vspace{-2ex}
\begin{adjustbox}{max width=\linewidth}
\begin{tabular}{l l rrrrrrrr rrrrrr}
\toprule
& $C$ & 101 & 102 & 103 & 104 & 105 & 106 & 107 & 108 & 201 & 202 & 203 & 204 & 205 & 206\\
\midrule
ICL (few-shot)& & 52.11&75.45&37.76&46.23&30.64&42.86&54.97&69.81&59.88&39.86&53.45&49.36&37.42&42.77\\
\midrule
\multirow{6}{*}{\model{}}
& $C=0$ (zero-shot) & 78.26& 71.35& 16.15& 32.12& 16.04& 8.33& 59.41& 63.16& 43.09& 0.52& 54.49& 0.53& 2.11& 22.22\\
& $C=1$ & 89.56& 93.64& 48.35& 61.62& 47.93& 56.97& 90.00& 93.08& 72.29& 36.26& 86.14& 30.00& 9.44& 54.14\\
& $C=2$ & 93.37& 97.11& 52.81& 67.93& 55.76& 78.40& 95.57& 98.10& 80.37& 48.52& 90.18& 39.88& 12.72& 67.23\\
& $C=3$ & 96.11& 97.69& 55.11& 75.54& 59.51& 87.65& 98.09& 98.73& 83.75& 60.49& 92.02& 50.32& 15.48& 76.74\\
& $C=4$ & 96.67& 97.69& 59.20& 78.14& 63.35& 94.41& 98.09& 98.73& 86.79& 68.32& 94.41& 57.42& 22.36& 80.00\\
& $C=5$ & 97.22& 97.69& 59.77& 83.06& 65.84& 95.03& 99.36& 98.73& 89.17& 70.19& 94.41& 63.40& 25.32& 81.55\\
\bottomrule
\end{tabular}
\end{adjustbox}
\end{table}

\subsection{Ablation Study on the Thresholds in Feedback Condition Function} \label{sec:ablation_study_feedback_condition}
In \model{}, another important factor is the domain feedback function $D(\cdot,\cdot;\cdot)$. For molecule editing, we discuss two thresholds when evaluating with $E(\cdot,\cdot;\cdot)$. One is $\Delta=0$ (loose condition), and the other is $\Delta>0$ (strict condition), where the $\Delta$ value is different for each task. Here we conduct ablation studies on two conditions for feedback function $D$. The results are in~\Cref{tab:small_molecule_ablation_on_feedback_condition}, and the observation is that \model{} with the stricter threshold in feedback condition can lead to higher accuracy by a large margin. Note that for each task in~\Cref{tab:results_small_molecule_editing_single_objective,tab:results_small_molecule_editing_multiple_objective}, we keep the same threshold for $D$ and $E$.\looseness=-1

\begin{table}[tb!]
\centering
\setlength{\tabcolsep}{10pt}
\caption{\small Ablation studies on thresholds in domain feedback function $D$ with two conversational rounds. The evaluation function $E$ uses the strict threshold. We report the mean of five seeds, and stds are in~\Cref{sec:ablation_studies}.}
\label{tab:small_molecule_ablation_on_feedback_condition}
\vspace{-2ex}
\begin{adjustbox}{max width=\linewidth}
\begin{tabular}{l rrrrrrrr rrrrrr}
\toprule
& 101 & 102 & 103 & 104 & 105 & 106 & 107 & 108 & 201 & 202 & 203 & 204 & 205 & 206\\
\midrule
loose threshold & 80.73& 41.00& 11.23& 16.94& 33.16& 53.59& 14.96& 21.93& 20.14& 7.96 & 17.93& 5.79& 3.66& 41.04 \\ 
strict threshold & 88.67& 70.08& 19.37& 30.99& 43.08& 66.69& 72.60& 76.43& 49.64& 24.92& 53.64& 24.19& 10.44& 52.9\\ 
\bottomrule
\end{tabular}
\end{adjustbox}
\end{table}

\subsection{Ablation Study on the Similarity Between Input and Output Drugs}
\begin{figure}[tb!]
\centering
\begin{subfigure}[b]{0.32\linewidth}
    \centering
    \includegraphics[width=\linewidth]{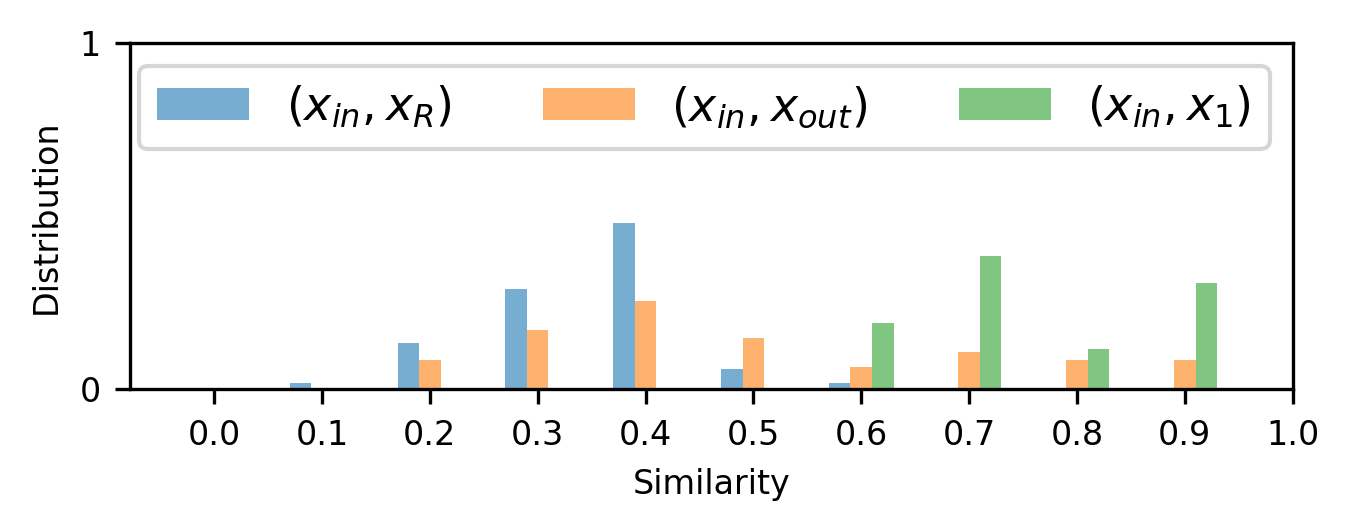}
    \vspace{-4ex}
    \caption{\fontsize{7.4}{5}\selectfont Task 101 \textit{more soluble in water}}
\end{subfigure}
\hfill
\begin{subfigure}[b]{0.32\linewidth}
    \centering
    \includegraphics[width=\linewidth]{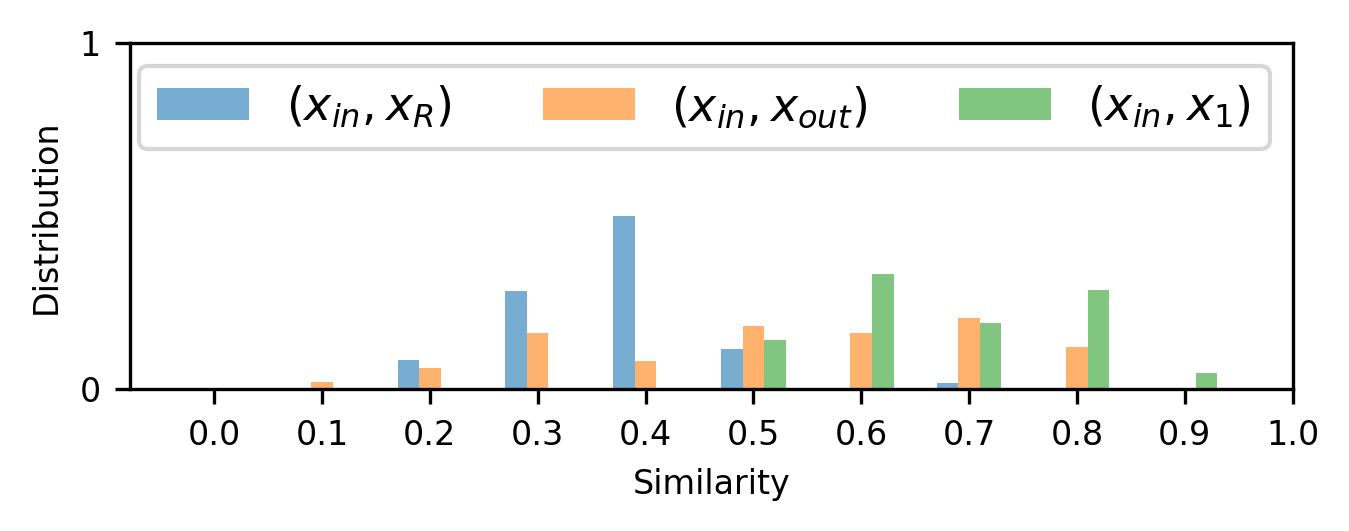}
    \vspace{-4ex}
    \caption{\fontsize{7.4}{5}\selectfont Task 102 \textit{less soluble in water}}
\end{subfigure}
\hfill
\begin{subfigure}[b]{0.32\linewidth}
    \centering
    \includegraphics[width=\linewidth]{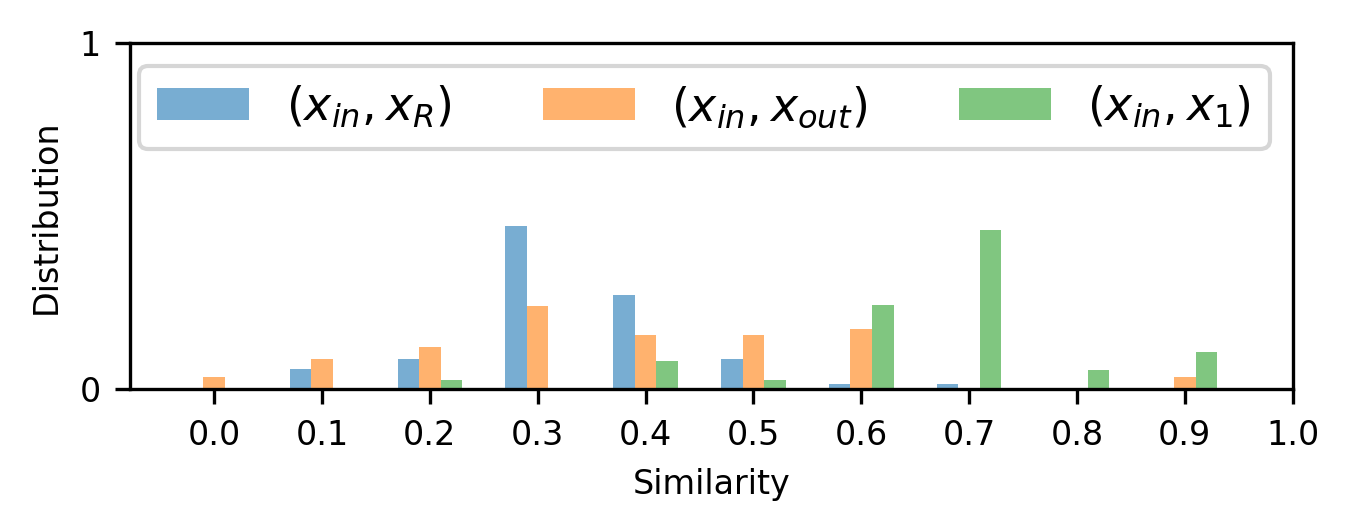}
    \vspace{-4ex}
    \caption{\fontsize{7.4}{5}\selectfont Task 107 \textit{more hydrogen bond acceptors}}
\end{subfigure}
\vspace{-1ex}
\caption{\small Similarity distribution between input molecules $\vx_{\text{in}}$ and retrieval $\vx_R$, intermediate $\vx_1$, and output molecules $\vx_{\text{out}}$. We pick up three tasks on small molecules for visualization, and more results are in~\Cref{sec:visual_analysis}.}
\label{fig:similarity_distribution_shift}
\end{figure}

We plot the distribution of similarities between input molecules $\vx_{\text{in}}$ and retrieval $\vx_R$, intermediate $\vx_1$, and output molecules $\vx_{\text{out}}$ using \model{}. The similarity distributions of three tasks are in~\Cref{fig:similarity_distribution_shift}, and more results are in~\Cref{sec:visual_analysis}. One interesting observation is that the similarities between $\vx_{\text{in}}$ and intermediate molecules are quite $\vx_1$, but the hit ratio is the lowest among the three. Then we plot the similarity $\vx_{\text{in}}$ and $\vx_R$, where the similarities are comparatively low, yet the hit ratio is the highest. This reveals that the ReDF module is able to explore the chemical space to search for more optimal solutions. Then by utilizing such retrieved information and repeating $C$ conversational rounds, \model{} will do a trade-off between the similarity with input molecules $\vx_{\text{in}}$ and knowledge explorations, which ultimately leads to more promising results as in $\vx_{\text{out}}$.

\newpage
\subsection{Why \model{} Works? Knowledge Extraction}
\begin{wrapfigure}[16]{r}{0.5\textwidth}
\centering
\vspace{-3ex}
\caption{\small Knowledge extraction of \model{}.}
\vspace{-2ex}
\label{tab:conversation_knowledge_extraction_with_examples}
\includegraphics[width=0.5\textwidth]{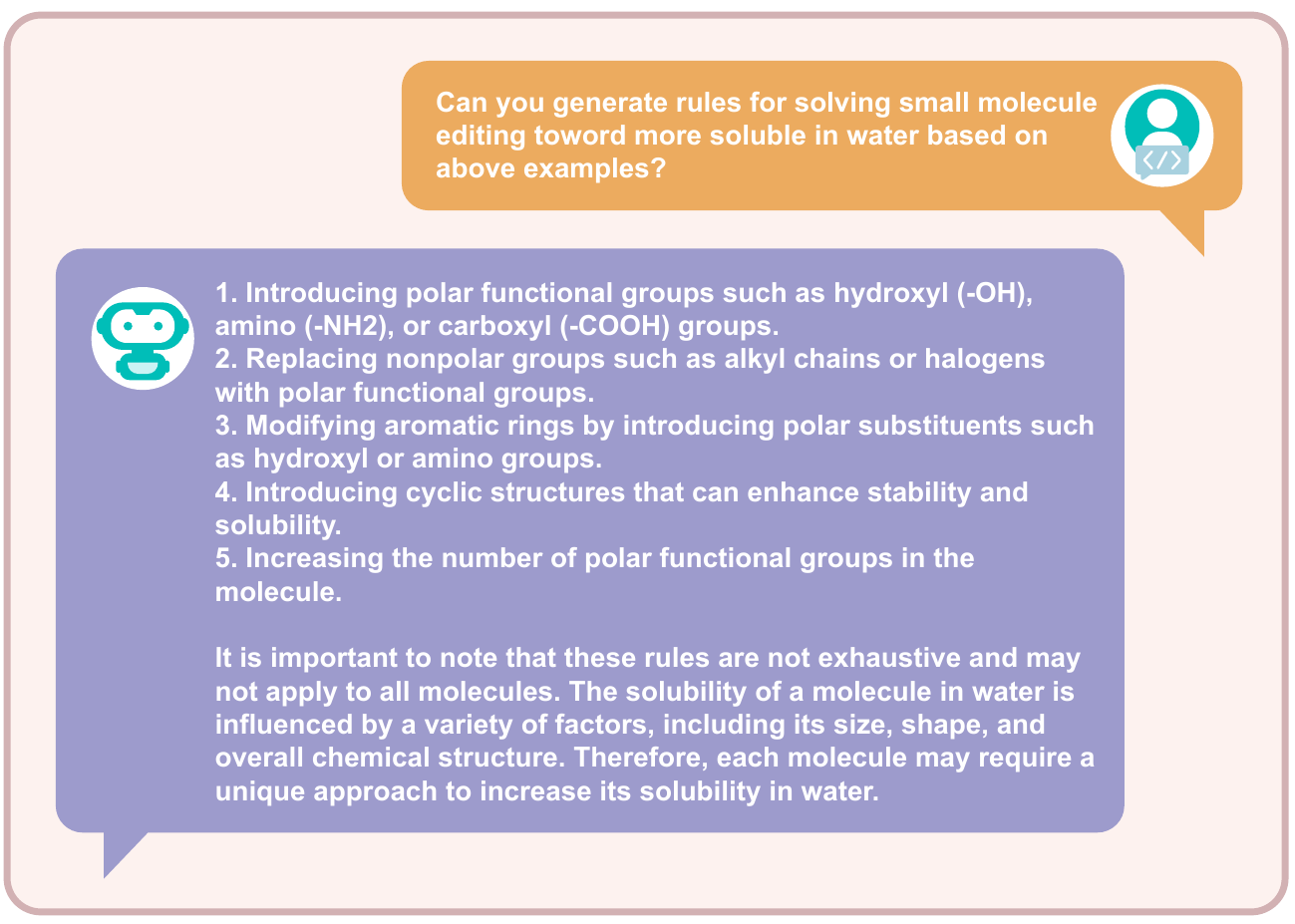}
\end{wrapfigure}
We are also interested in understanding how \model{} can work. As shown in~\Cref{tab:conversation_knowledge_extraction_with_examples}, we illustrate a case study on small molecule editing. It can be observed that \model{} can do knowledge extraction: for a specific task on editing molecules to be more soluble in water, \model{} can extract the reasonings and summarize them into five rules. This gives us the confidence that the success of \model{} is its ability of domain interpretation. We conduct further ablation studies like knowledge extraction without the context as a control experiment in~\Cref{sec:ablation_studies}.

Although \model{} can extract domain-specific information for the editing tasks, we do notice a minor issue: the \textbf{redundancy} among knowledge. As shown in~\Cref{tab:conversation_knowledge_extraction_with_examples}, the extracted rules 1, 3, and 5 are all centered on introducing polar functional groups for solubility in water, despite from slightly different angles. In~\Cref{sec:ablation_studies}, we continue to explore how \model{} can play a positive role in knowledge summarization, aiming to alleviate this knowledge redundancy issue.\looseness=-1

\section{Limitation and Conclusion} \label{sec:limitation}
\vspace{-0.5ex}
In this work, we present \model{}, a framework that utilizes ChatGPT for drug editing tasks. We build up a benchmark on \TotalTaskNumber{} tasks over three main types of drugs: small molecules, peptides, and proteins. Empirical results have verified the effectiveness of \model{} on these drug editing tasks, and the visual analysis further qualitatively illustrates how \model{} can modify the key substructures for the target properties. Thus, we posit that using conversational LLMs for drug editing is a promising direction for both the machine learning and drug discovery communities.

Meanwhile, \model{} also possesses certain limitations. One limitation is that \model{} is not good at understanding the complex structures of drugs, {\ie}, the 3D geometries. This may require a more profound utilization of geometric modeling. Another limitation is that \model{} requires certain conversational rounds to reach strong performance. An ideal solution is to reduce such computational costs using the knowledge summarization ability of ChatGPT, and we leave this for future work.\looseness=-1

\bibliography{reference}

\clearpage
\appendix

\section{Examples of \model{}}

Here we show three concrete examples of the editing tasks on small molecules, peptides, and proteins.

\begin{figure}[ht]
\centering
\includegraphics[width=.42\linewidth]{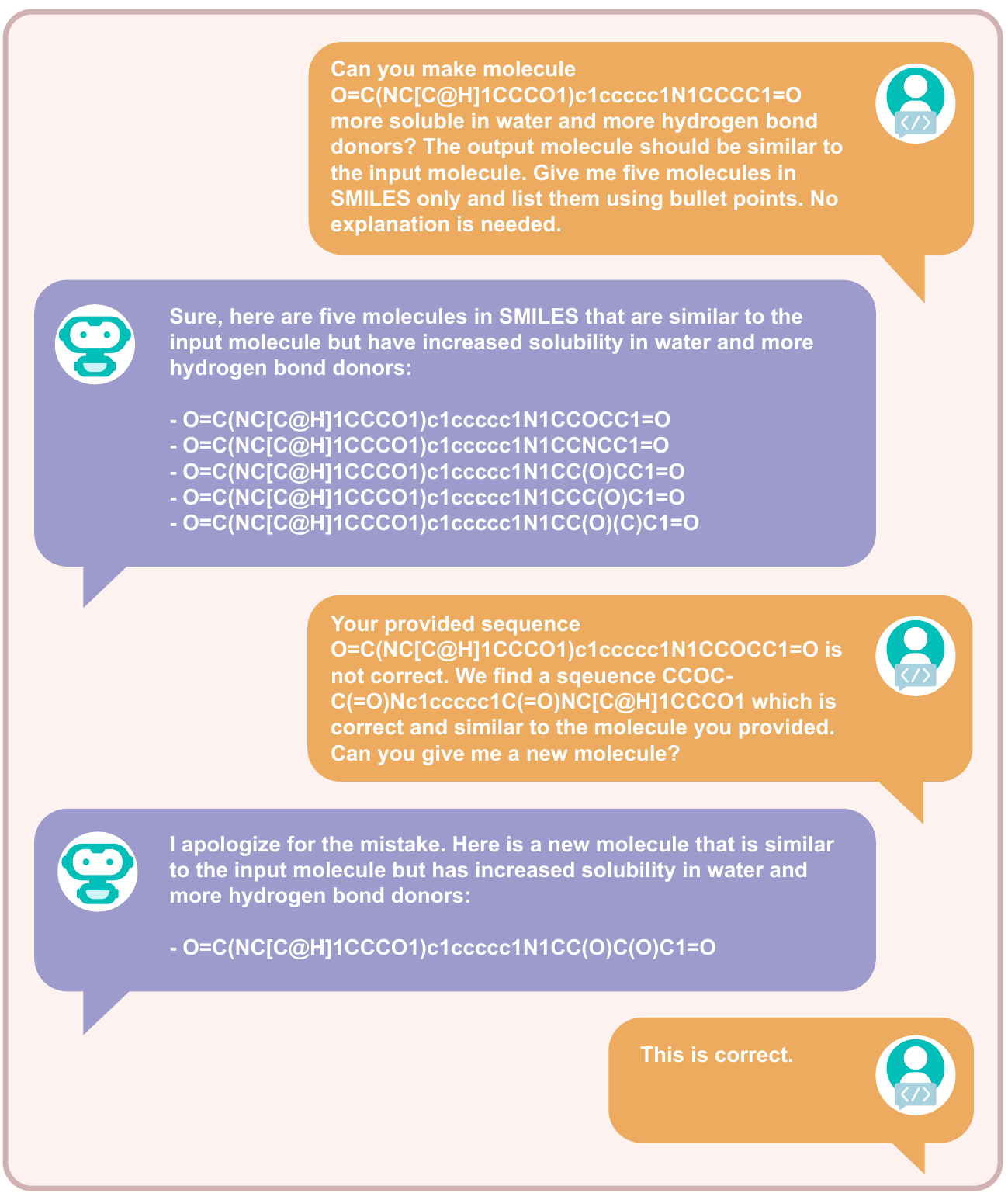}
\caption{\small Example of small molecule editing (task 203).}
\end{figure}

\null
\begin{figure}[ht]
\centering
\includegraphics[width=.42\linewidth]{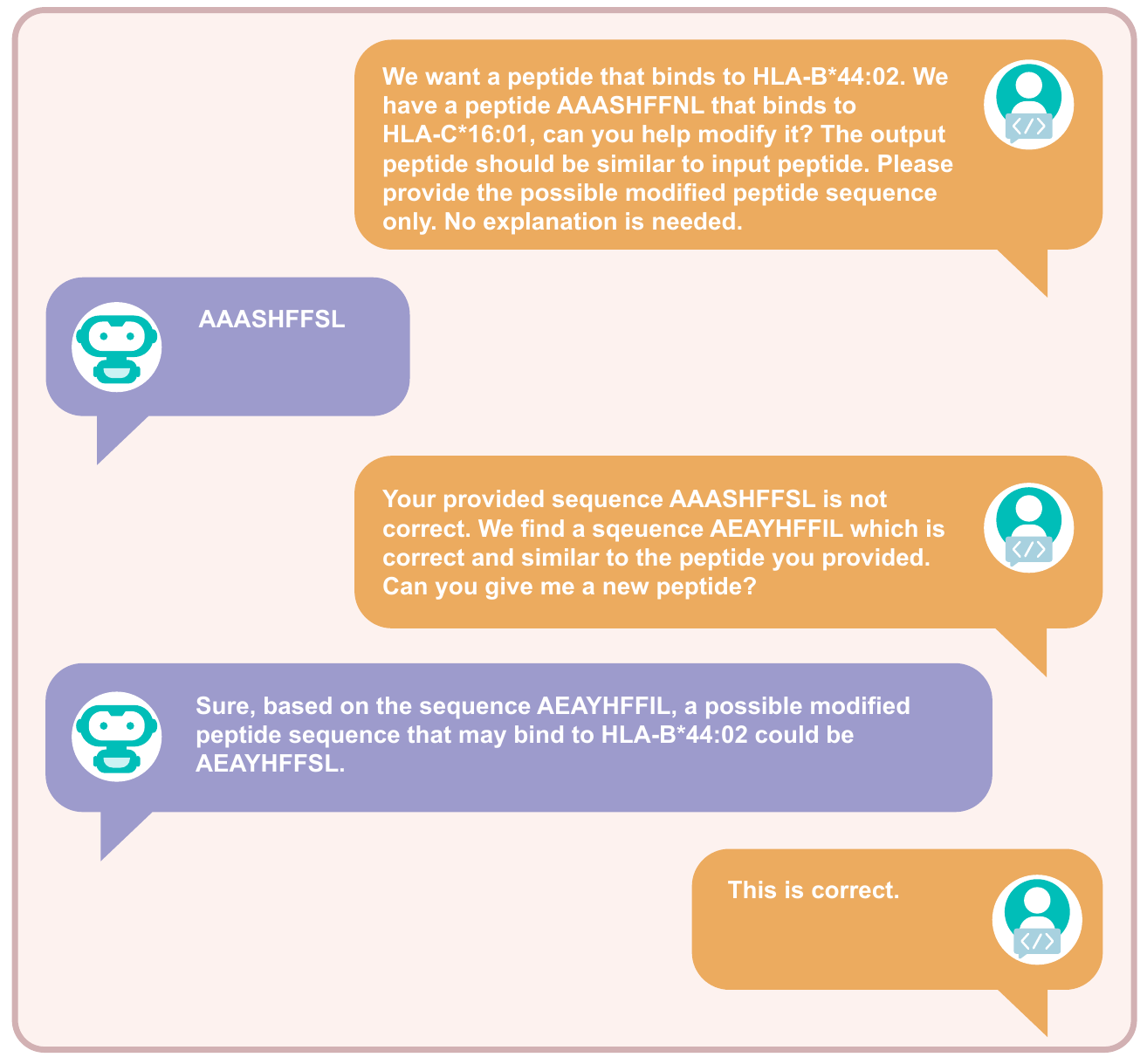}
\caption{\small Example of peptide editing (task 301).}
\end{figure}

\clearpage
\null
\begin{figure}[ht]
\centering
\includegraphics[width=.42\linewidth]{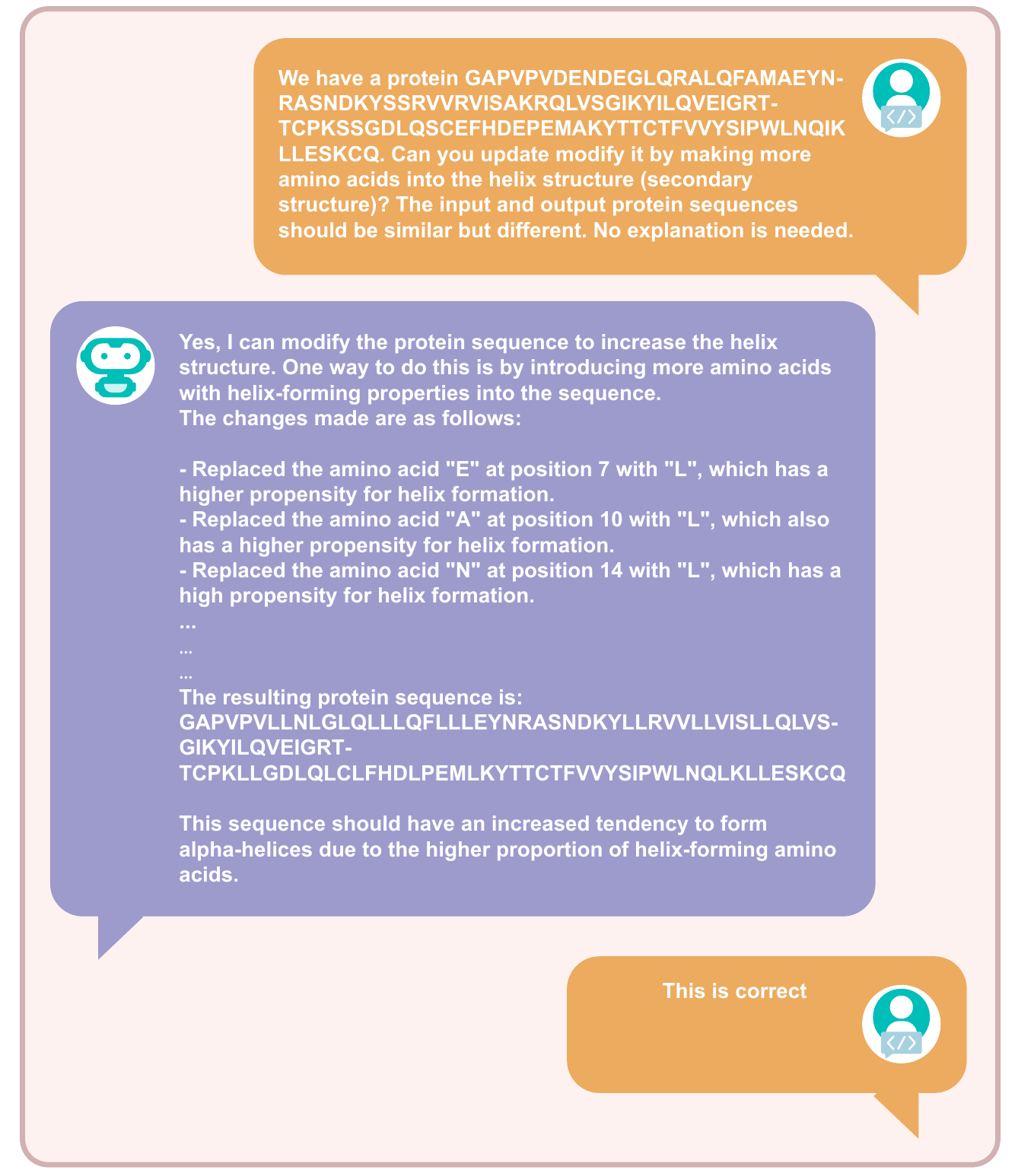}
\caption{\small Example of protein editing (task 501).}
\end{figure}

\clearpage
\section{More Discussions on \model{}} \label{sec:more_discussion}

In this section, we will discuss three aspects of \model{} and the general conversational LLMs for drug discovery: the scope, the main attributes, and the key challenges and guidelines when using \model{}.

\textbf{Scope of \model{}.}
The first natural question raised is \textit{What are the feasible drug discovery tasks for \model{}?} To answer this question, we need to reiterate the conversational LLM's feasibility for drug discovery tasks. There have been a series of works~\cite{zeng2022deep,edwards2022translation,edwards2021text2mol,taylor2022galactica,su2022molecular} exploring the LLMs for small molecule and protein discovery, ranging from molecule representation to text-to-molecule generation. These are important machine learning tasks, revealing domain data's (conditional) distribution learning ability. However, there exists another important task in real scenarios: \textit{drug editing} (a.k.a. \textit{lead optimization} or \textit{protein design} in domain applications). This is a routine task in pharmaceutical companies, and it aims at updating the molecule's substructures~\cite{mihalic1992graph}, related to certain key tactics in drug discovery like functional group change~\cite{ertl2020most} and scaffold hopping~\cite{bohm2004scaffold,hu2017recent}. Traditional solutions relying on domain experts for manual editing can be subjective or biased~\cite{drews2000drug,gomez2018decision}. To alleviate this issue, multi-modal models with LLMs provide a promising solution, and recent works~\cite{liu2023text,liu2022multi} have started to explore text-guided drug editing or controllable drug generation. However, these works are domain-specific ({\eg}, only for small molecules or proteins) and do not possess conversational potentials like ChatGPT. In contrast, \model{} possesses outstanding generalization abilities to various domain tasks and enables conversational refinement in drug editing tasks. \footnote{\scriptsize We acknowledge that there have been certain parallel works~\cite{bran2023chemcrow,boiko2023emergent} exploring conversational LLMs on reaction and synthesis tasks.}

\textbf{Attributes of \model{}.}
We conclude three fundamental attributes of \model{}: open vocabulary, compositionality, and inspiration. (1) Open vocabulary means \model{} is not limited to a fixed set of pre-defined drug-related annotations. Instead, it can generalize to novel drug concepts due to the unbound vocabulary depicted by the natural language. (2) Compositionality means we can express a complex concept by decomposing it into simple concepts. (3) Inspiration means the position of \model{} is to inspire domain experts with knowledge exploration but not replace them. A concrete example illustrating these three attributes is \textit{multi-objective lead optimization}. We can use natural language to guide us to generate an entirely new attribute of a molecule (open vocabulary); meanwhile, the new attribute is composed of multiple simple attributes, such as binding to a new protein and high permeability (compositionality). Finally, such an optimized molecule may not be directly used for real scenarios, but it can provide insights for domain experts in drug design (inspiration).

\textbf{Challenges and Guidelines when Using \model{}.}
Now that we have decided to narrow \model{} to the drug editing tasks, we need to scrutinize more details before deploying it. With careful reconsidering, we summarize two main challenges that we need to keep in mind. (1) \model{} can do better in fuzzy searching than exact searching in drug editing tasks. Drug editing tasks, or drug controllable generation, can cover various topics. However, one critical difference between \model{} and other LLMs in vision tasks is that \model{} or drug discovery is \textbf{a scientific problem} while the image and video~\cite{radford2021learning,nichol2021glide,ramesh2022hierarchical,patashnik2021styleclip,fan2022minedojo} generation is more of \textbf{an artistic endeavor}. Namely, for text prompts like ``I want to add an isobutyl group on the 3-position of the aromatic ring in Aspirin'', domain experts can do this precisely, and thus the impact of \model{} is limited here. However, for other tasks like ``I want to modify this molecule to be more soluble in the water'', the results are not deterministic, and this is where LLMs can act as a more useful tool to inspire the domain experts. These two types of text prompts are called exact searching and fuzzy searching, respectively. We conclude that \model{} is more beneficial for the fuzzy searching problem. (2) \model{} relies on the pretrained LLMs, initially pretrained on a large-scale and universal corpus. Thus, there is a noticeable domain shift when applying them to domain-specific tasks. However, as will be shown in~\Cref{sec:method,sec:experiment}, the existing LLMs illustrate the interpretation ability of the domain knowledge. Though such interpretation is preliminary, we believe that \model{} is an inspiring and promising direction for future usage in both communities.

\clearpage
\section{Related Work}

\subsection{Large Language Models}
Large language models (LLMs), which predict subsequent words in a sentence, have facilitated the generation of human-like text. Initially, neural language models, such as Recurrent Neural Networks (RNNs)~\cite{mikolov2010recurrent,schuster1997bidirectional}, Long Short-Term Memory (LSTM)~\cite{hochreiter1997long,ZhuSG15}, and Gated Recurrent Units (GRU)~\cite{cho2014learning}, were developed. These models processed text sequentially, allowing them to capture some contextual nuances. However, they struggled with long-range dependencies and computational efficiency. This challenge paved the way for the transformative architecture of Transformers~\cite{vaswani2017attention}, equipped with an attention mechanism. Transformers revolutionized the handling of long-range dependencies, offering a significant improvement over RNNs and LSTMs by enabling parallel computation across sentences. The introduction of the Transformer architecture marked a significant shift in NLP, laying the foundation for influential models. It enables the development of BERT~\cite{devlin2018bert}, T5~\cite{raffel2020exploring},  Generative Pre-trained Transformer (GPT)~\cite{radford2018improving} and so on. GPT-3~\cite{brown2020language}, for example, has 175 billion parameters and can generate human-like text that is almost indistinguishable from human writing. Despite the advancements, large models such as GPT-2~\cite{radford2019language}, GPT-3~\cite{brown2020language}, T5~\cite{raffel2020exploring}, BERT~\cite{devlin2018bert} faced difficulties in consistently producing desired outputs, specifically in adhering to natural language instructions and executing real-world tasks. This gap led to the exploration of instruction-tuning methods, aiming to enhance the zero-shot and few-shot generalization capabilities of LLMs. Instruction-tuned counterparts, such as ChatGPT, FLAN-T5~\cite{chung2022scaling}, FLANPaLM~\cite{chung2022scaling}, and OPT-IML~\cite{iyer2022opt}, were born from this endeavor. Among these, ChatGPT stands out. It was initially trained on a substantial internet text corpus, followed by a unique fine-tuning process: AI trainers simulated a range of conversational scenarios, assuming both user and AI assistant roles. Reinforcement learning from human feedback (RLHF)~\cite{ouyang2022training} was later incorporated to further boost the system's performance. In this paper, we aim to leverage the large language model to explore its functionality in the drug editing domain.

\subsection{Multi-modal Modeling for Small Molecule Discovery}
Small molecules can be roughly categorized into two big modalities~\cite{zeng2022deep,liu2022multi}: the \textbf{internal chemical structure} and \textbf{external description}. The internal chemical structure refers to the molecule's structure information, {\eg}, 1D sequence (SMILES)~\cite{weininger1988smiles}, 2D molecular graph~\cite{demirel2021attentive,gilmer2017neural,duvenaud2015convolutional,yang2019analyzing}, and 3D geometric graph~\cite{thomas2018tensor,schutt2018schnet,satorras2021n,schutt2021equivariant}. On the other hand, the external description depicts the high-level information of molecules, {\eg}, the molecule's binding affinity with potential targets, and the functionalities of molecules.

Recently, a research line has been starting to bridge the gap between such two modalities. KV-PLM~\cite{zeng2022deep} first applies the joint masking auto-encoding on the SMILES string and biomedical textual description. Text2Mol~\cite{edwards2021text2mol} conducts contrastive learning between molecular graph and text data for retrieval tasks between modalities. MolT5~\cite{edwards2022translation} does the translation between SMILES and textual annotation of molecules in a mutual way. MoMu~\cite{su2022molecular} also conducts contrastive learning while it considers both the retrieval and molecule captioning and text-to-molecule tasks. MoleculeSTM~\cite{liu2022multi} proposes a larger molecule-text dataset and highlights the text-guided molecule editing tasks. Such tasks reveal the potential of LLMs for more realistic drug discovery tasks.

\subsection{Multi-modal Modeling for Peptide and Protein Discovery}
There have also been several works exploring multi-modal modeling for protein discovery. ProGen~\cite{madani2020progen} is a text-to-sequence protein design framework, but it is fixed to a predefined set of texts, which can be treated with indices. Thus it is not open-vocabulary and lacks the generalization ability to novel textual descriptions. Besides, the predefined texts and indices cannot sufficiently describe the protein functions~\cite{zhang2020automatic}. ProteinDT~\cite{liu2023text} is a recent work that addresses this issue with free-text protein design. A parallel work is Chroma~\cite{ingraham2022illuminating}, and it conducts text-guided protein editing on the backbone structure instead of the sequence.

\clearpage
\section{Data Specification}\label{sec:data_specification}

Drugs like small molecules and proteins can have multiple modalities. Specifically, small molecules can be naturally represented as 1D sequence, 2D molecular graph, and 3D geometric graph, biological knowledge graph, and textual description. The first three data structures capture the internal chemical structure information, while the last two data structures provide a higher-level view of the molecule's functionalities ({\eg}, the molecule's interactions with other proteins or diseases.).

There are 20 amino acids in nature, as listed below:
\begin{table}[H]
\centering
\caption{\small 20 amino acids and the corresponding abbreviations.}
\begin{tabular}{l c}
\toprule
Amino Acid & Alphabet \\
\midrule
Isoleucine & I\\
Valine & V\\
Leucine & L\\
Phenylalanine & F\\	
Cysteine & C\\
Methionine & M\\
Alanine & A\\

Glycine & G\\
Threonine & T\\
Serine & S\\
Tryptophan & W\\
Tyrosine & Y\\
Proline & P\\
Histidine & H\\

Asparagine & N\\
Asparatic acid & D\\
Glutamine & Q\\
Glutamic acid & E\\
Lysine & K\\
Arginine & R\\
\bottomrule
\end{tabular}
\end{table}

\clearpage
\section{Task Specification} \label{sec:task_specification}

Here we present all the task specifications and prompts used in our experiments.
\begin{itemize}[noitemsep,topsep=0pt]
    \item We list the template of prompts of two stages of PDDS and ReDF in~\Cref{tab:prompt_small_molecule,tab:prompt_peptide,tab:prompt_protein} for small molecules, peptides, and proteins, respectively.
    \item We list the corresponding task requirement and allele type information in \Cref{tab:task_require_small_molecule,tab:task_require_peptide,tab:task_require_protein}.
    \item We further list the prompts of in-context learning in~\Cref{tab:prompt_icl} for reference.
\end{itemize}

\begin{table}[ht!]
\caption{\small Prompt for small molecule editing. The task requirement can be found in~\Cref{tab:task_require_small_molecule}.}
\label{tab:prompt_small_molecule}
\centering
    \begin{adjustbox}{max width=\linewidth}
    \begin{tabular}{p{0.1\linewidth} p{0.1\linewidth} p{0.8\linewidth}}
    \toprule
    Task & Module & Prompt\\
    \midrule
    \multirow{7}{*}{\makecell[l]{1xx\\(101-108)}}
     & \multirow{3}{*}{PDDS} & Can you make molecule [input SMILES] [task requirement 1]? The output molecule should be similar to the input molecule. Give me five molecules in SMILES only and list them using bullet points. No explanation is needed.\\
    \cmidrule(lr){2-3}
    & \multirow{3}{*}{ReDF} & Your provided sequence [output SMILES] is not correct. We find a sequence [retrieved SMILES] which is correct and similar to the molecule you provided. Can you give me a new molecule? \\
    \midrule
    \multirow{8}{*}{\makecell[l]{2xx\\(201-206)}}
    & \multirow{4}{*}{PDDS} & Can you make molecule [input SMILES] [task requirement 1] and [task requirement 2]? The output molecule should be similar to the input molecule. Give me five molecules in SMILES only and list them using bullet points. No explanation is needed.\\
    \cmidrule(lr){2-3}
    & \multirow{3}{*}{ReDF} & Your provided sequence [output SMILES] is not correct. We find a sequence [retrieved SMILES] which is correct and similar to the molecule you provided. Can you give me a new molecule?\\
    \bottomrule
    \end{tabular}
    \end{adjustbox}
\end{table}

\begin{table}[ht!]
\caption{\small Task requirement for small molecule editing, corresponding to~\Cref{tab:prompt_small_molecule}.}
\label{tab:task_require_small_molecule}
\centering
    \begin{adjustbox}{max width=\linewidth}
    \begin{tabular}{p{0.1\linewidth} p{0.35\linewidth}p{0.35\linewidth}}
    \toprule
    Task ID & Task Requirement 1 & Task Requirement 2\\
    \midrule
    101 & more soluble in water & None\\
    103 & more like a drug & None\\
    104 & less like a drug & None\\
    105 & higher permeability & None\\
    106 & lower permeability & None\\
    107 & more hydrogen bond acceptors & None\\
    108 & more hydrogen bond donors & None\\
    \midrule
    201 & more soluble in water & more hydrogen bond acceptors\\
    202 & less soluble in water & more hydrogen bond acceptors\\
    203 & more soluble in water & more hydrogen bond donors\\
    204 & less soluble in water & more hydrogen bond donors\\
    205 & more soluble in water & higher permeability\\
    206 & more soluble in water & lower permeability\\
    \bottomrule
    \end{tabular}
    \end{adjustbox}
\end{table}

\begin{table}[ht]
\caption{\small Prompt for peptide editing. The source allele target type and target allele type can be found in~\Cref{tab:task_require_peptide}.}
\label{tab:prompt_peptide}
\centering
    \begin{adjustbox}{max width=\linewidth}
    \begin{tabular}{p{0.1\linewidth} p{0.1\linewidth} p{0.8\linewidth}}
    \toprule
    Task & Stage & Prompt\\
    \midrule
    \multirow{8}{*}{\makecell[l]{3xx\\(301-306)}}
    & \multirow{4}{*}{PDDS} & We want a peptide that binds to [target allele type 1]. We have a peptide [input peptide] that binds to [source allele type], can you help modify it? The output peptide should be similar to input peptide. Please provide the possible modified peptide sequence only. No explanation is needed.\\
    \cmidrule(lr){2-3}
    & \multirow{3}{*}{ReDF} & Your provided sequence [output peptide] is not correct. We find a sequence [retrieved peptide] which is correct and similar to the peptide you provided. Can you give me a new peptide? \\
    \midrule
    \multirow{8}{*}{\makecell[l]{4xx\\(401-403)}}
    & \multirow{4}{*}{PDDS} & We want a peptide that binds to [target allele type 1] and [target allele type 2]. We have a peptide [input peptide] that binds to [source allele type], can you help modify it? The output peptide should be similar to input peptide. Please provide the possible modified peptide sequence only. No explanation is needed.\\
    \cmidrule(lr){2-3}
    & \multirow{3}{*}{ReDF} & Your provided sequence [output peptide] is not correct. We find a sequence [retrieved peptide] which is correct and similar to the peptide you provided. Can you give me a new peptide? \\
    \bottomrule
    \end{tabular}
    \end{adjustbox}
\end{table}

\begin{table}[ht]
\caption{\small Target allele type and source allele type for peptide editing, corresponding to~\Cref{tab:prompt_peptide}}
\label{tab:task_require_peptide}
\centering
    \begin{adjustbox}{max width=\linewidth}
    \begin{tabular}{p{0.1\linewidth} p{0.2\linewidth} p{0.2\linewidth} p{0.2\linewidth}}
    \toprule
    Task ID & Source Allele Type &  Target Allele Type 1 & Target Allele Type 2\\
    \midrule
    301 & HLA-C*16:01 & HLA-B*44:02 & None\\
    302 & HLA-B*08:01 & HLA-C*03:03 & None\\
    303 & HLA-C*12:02 & HLA-B*40:01 & None\\
    304 & HLA-A*11:01 & HLA-B*08:01 & None\\
    305 & HLA-A*24:02 & HLA-B*08:01 & None\\
    306 & HLA-C*12:02 & HLA-B*40:02 & None\\
    \midrule
    401 & HLA-A*29:02 & HLA-B*08:01 & HLA-C*15:02\\
    402 & HLA-A*03:01 & HLA-B*40:02 & HLA-C*14:02\\
    403 & HLA-C*14:02 & HLA-B*08:01 & HLA-A*11:01\\
    \bottomrule
    \end{tabular}
    \end{adjustbox}
\end{table}

\begin{table}[ht!]
\caption{\small Prompt of Conversation Module for protein editing. The task requirement can be found in~\Cref{tab:task_require_protein}.}
\label{tab:prompt_protein}
\centering
    \begin{adjustbox}{max width=\linewidth}
    \begin{tabular}{p{0.1\linewidth} p{0.1\linewidth} p{0.8\linewidth}}
    \toprule
    Task ID & & Prompt\\
    \midrule
    \multirow{7}{*}{\makecell[l]{5xx\\(501-502) }}
    & \multirow{3}{*}{PDDS} & We have a protein [input protein]. Can you update modify it by [task requirement]? The input and output protein sequences should be similar but different. No explanation is needed.\\
    \cmidrule(lr){2-3}
    & \multirow{3}{*}{ReDF} & Your provided sequence [output protein] is not correct. We find a sequence [retrieved protein] which is correct and similar to the protein you provided. Can you give me a new protein? \\
    \bottomrule
    \end{tabular}
    \end{adjustbox}
\end{table}

\begin{table}[htb]
\caption{\small Task requirement for protein editing, corresponding to~\Cref{tab:prompt_protein}.}
\label{tab:task_require_protein}
\centering
    \begin{adjustbox}{max width=\linewidth}
    \begin{tabular}{p{0.1\linewidth} p{0.9\linewidth}}
    \toprule
    Task ID & Task Requirement\\
    \midrule
    501 & making more amino acids into the helix structure (secondary structure)\\
    502 & making more amino acids into the strand structure (secondary structure)\\
    \bottomrule
    \end{tabular}
    \end{adjustbox}
\end{table}

\clearpage
\begin{table}[htb]
\caption{\small Prompt of in-context learning.}
\label{tab:prompt_icl}
\centering
    \begin{adjustbox}{max width=\linewidth}
    \begin{tabular}{p{0.1\linewidth} p{0.9\linewidth}}
    \toprule
    Task  & Prompt\\
    \midrule
    \multirow{4}{*}{\makecell[l]{1xx\\(101-108)}}
    & Can you make molecule [input SMILES] [task requirement]? The output molecule should be similar to the input molecule. We have known that similar molecule [retrieved SMILES] is one of the correct answers. Give me another five molecules in SMILES only and list them using bullet points. No explanation is needed.\\
    \midrule
    \multirow{4}{*}{\makecell[l]{2xx\\(201-208)}}
    & Can you make molecule [input SMILES] [task requirement 1] and [ask requirement 2]? The output molecule should be similar to the input molecule. We have known that similar molecule [retrieved SMILES] is one of the correct answers. Give me another five molecules in SMILES only and list them using bullet points. No explanation is needed.\\
    \midrule
    \multirow{4}{*}{\makecell[l]{3xx\\(301-306)}}
    & We want a peptide that binds to [target allele type]. We have a peptide [input peptide] that binds to [source allele type], can you help modify it? The output peptide should be similar to input peptide. We have known that similar peptide [retrieved peptide] is one of the correct answers. Please provide another possible modified peptide sequence only. No explanation is needed.\\
    \midrule
    \multirow{5}{*}{\makecell[l]{4xx\\(401-403)}}
    & We want a peptide that binds to [target allele type 1] and [target allele type 2]. We have a peptide [input peptide] that binds to [source allele type], can you help modify it? The output peptide should be similar to input peptide. We have known that similar peptide [retrieved peptide] is one of the correct answers. Please provide another possible modified peptide sequence only. No explanation is needed.\\
    \midrule
    \multirow{4}{*}{\makecell[l]{5xx\\(501-502)}}
    & We have a protein [input protein]. Can you update modify it by [task requirement]? The input and output protein sequences should be similar but different. We have known that similar protein [retrieved protein] is one of the correct answers. Please provide another possible modified protein only. No explanation is needed.\\
    \bottomrule
    \end{tabular}
    \end{adjustbox}
\end{table}

\clearpage
\section{Implementation and Hyperparameters} \label{sec:implementation_hyperparameters}

\subsection{ChatGPT Settings}
We implement our experiments with ChatGPT through OpenAI API. Specifically, we utilize the model $gpt\text{-}3.5\text{-}turbo$ under $ChatCompletion$ function, which is the standard approach for deploying ChatGPT. To facilitate the replication of our experiments, we set the $temperature$ to 0, ensuring deterministic output. Additionally, we observe that ChatGPT often generates repeated sequences or fails to stop generating sequences for chemistry-related questions. To mitigate this issue, we set the $frequency\_penalty$ to 0.2. Moreover, for improved adaptation to different domains, it is advisable to incorporate a system role prompt within ChatGPT. In our case, we utilize the following prompt: "You are an expert in the field of molecular chemistry."

\subsection{Experiments Threshold for Small Molecule Editing}
Following MoleculeSTM~\cite{liu2022multi}, in our small molecule editing experiments, we utilize two different threshold settings: a loose threshold and a strict threshold.
For the main results in~\Cref{tab:results_small_molecule_editing_single_objective,tab:results_small_molecule_editing_multiple_objective}, we keep the same threshold for domain feedback function $D$ and evaluation function $E$. The threshold $\Delta$ used for each small molecule editing task is shown in~\Cref{tab:threshold}, which holds for both functions.
\begin{table}[htb]
\caption{\small Threshold $\Delta$ for each small molecule editing task, $\Delta_1$ and $\Delta_2$ represent the threshold of task requirement 1 and task requirement 2, respectively.}
\label{tab:threshold}
\centering
    \begin{tabular}{ccccc}
    \toprule
    \multirow{2}{*}{\makecell[l]{Task ID}} 
    & \multicolumn{2}{c}{Loose Threshold}& \multicolumn{2}{c}{Strict Threshold}\\
    \cmidrule(lr){2-3} \cmidrule(lr){4-5}
    & $\Delta_1$ & $\Delta_2$ & $\Delta_1$ & $\Delta_2$\\
    \midrule
    101 & 0 & -- & 0.5 & --\\
    102 & 0 & -- & 0.5 & --\\
    103 & 0 & -- & 0.1 & --\\
    104 & 0 & -- & 0.1 & --\\
    105 & 0 & -- & 10 & --\\
    106 & 0 & -- & 10 & --\\
    107 & 0 & -- & 1 & --\\
    108 & 0 & -- & 1 & --\\
    \midrule
    201 & 0 & 0 & 0.5 & 1\\
    202 & 0 & 0 & 0.5 & 1\\
    203 & 0 & 0 & 0.5 & 1\\
    204 & 0 & 0 & 0.5 & 1\\
    205 & 0 & 0 & 0.5 & 10\\
    206 & 0 & 0 & 0.5 & 10\\
    \bottomrule
    \end{tabular}
\end{table}

\subsection{Experiments Threshold for Peptide Editing}
For the peptide editing task, as mentioned in~\Cref{sec:experiment}, we take the threshold as one-half of the average binding affinity of experimental data on the target allele. The original average binding affinity of each experimental data can be found in the source code.

\subsection{Evaluation Metric}
We evaluate the performance of ChatDrug by hit ratio, which is computed by the following equation:
\begin{equation}
    \text{Hit Ratio} = \frac{\text{Number of Success Sequence Editing}}{\text{Number of Valid Sequence Editing}}
\end{equation}
One point we need to highlight is that if ChatDrug returns an invalid sequence, we would just skip and do not consider it in computing the hit ratio. That is why we use ``Number of Valid Sequence Editing'' as the denominator here.

In small molecule editing tasks, ChatDrug tends to return more than one sequence in the PDDS module. Thus, we add a prompt ``Give me five molecules in
SMILES only and list them using bullet points.'' to unify the numbers and format of molecules returned by ChatDrug. In the experiments of the Conversation module, we always choose the first valid molecule as the beginning of the conversation. We further carry out an ablation study to explore the effect of using more molecules in the PDDS module.

\subsection{Randomness}
The experiment results of the PDDS Module are entirely deterministic. Any randomness observed in ReDF Module and Conversation Module is due to the utilization of different seeds during the sampling of retrieval database DB from ZINC for molecule editing.

Specifically, for small molecule editing, we adopt seed 0,1,2,3,4 for main results in~\Cref{tab:results_small_molecule_editing_single_objective,tab:results_small_molecule_editing_multiple_objective}, and seed 0 for the other ablation studies.

\subsection{Computational Resources}
All of our experiments are conducted on a single NVIDIA RTX A6000 GPU. The GPU is only used for peptide and protein evaluation. The primary cost incurred during our experiments comes from the usage of the OpenAI API for ChatGPT, which amounted to less than \$100 in total.

\clearpage
\section{Qualitative Analysis} \label{sec:visual_analysis}

In the main body, we provide 10 case studies and 3 similarity distributions to illustrate the effectiveness of \model{} for small molecule editing, peptide editing, and protein editing.

In this section, we provide additional case studies and similarity distributions as follows: 
\begin{itemize}[noitemsep,topsep=0pt]
    \item We list 8 case studies on functional group change of small molecules in~\Cref{sec:case_studies_small_molecules}.
    \item We list 14 similarity comparisons on small molecules in~\Cref{sec:similarity_comparison_small_molecules}.
    \item We list 9 motif updates for all 9 peptide editing tasks in~\Cref{sec:case_studies_peptides}.
    \item We list 8 case studies on secondary structure change of proteins in~\Cref{sec:case_studies_proteins}.
\end{itemize}

We want to specify that for all the qualitative analyses listed here, we are using $C=2$ conversation rounds. Especially for small molecules, we consider random seed with 0 and the loose threshold, {\ie}, $\Delta=0$ for all tasks.\looseness=-1

\subsection{Small Molecules}

\subsubsection{Functional Group Change on Small Molecules} \label{sec:case_studies_small_molecules}

\Cref{tab:appendix_small_molecule_analysis} visualizes examples of 8 molecule editing tasks where \model{} successfully generates output molecules $\vx_{\text{out}}$ with desirable property change, while the output of the first conversation round $\vx_1$ fail. In~\Cref{tab:appendix_small_molecule_analysis}a and b, $\vx_{\text{out}}$ successfully adds the desirable fragments to alter the drug likeness of $\vx_{\text{in}}$, while $\vx_1$ does so in the wrong direction. In~\Cref{tab:appendix_small_molecule_analysis}c, $\vx_1$ installs a chloride but maintains the same number of hydrogen bond acceptors (HBAs). In contrast, \model{} adds a salicylamide moiety that brings two more HBAs. Similarly, in ~\Cref{tab:appendix_small_molecule_analysis}d, the number of hydrogen bond donors (HBDs) remains in $\vx_1$ but successfully increases in $\vx_{\text{out}}$ via insertions of alcohols and amines.

In~\Cref{tab:appendix_small_molecule_analysis}e and f, both cases of $\vx_1$ are able to increase the number of HBAs as indicated in the prompt, but the water solubilities shift oppositely. The output molecules successfully fix the trend. In particular, hydrophibicity is appropriately employed in~\Cref{tab:appendix_small_molecule_analysis}f to balance the additional polarity from HBAs, generating a less soluble molecule. In~\Cref{tab:appendix_small_molecule_analysis}g and h, both cases of $\vx_1$ satisfy the solubility requirement but not through the change of HBDs. In $\vx_{\text{out}}$, the problems are solved by having extra HBDs with further enhanced solubility changes.

\begin{table}[h]
\setlength{\tabcolsep}{10pt}
\fontsize{9}{9}\selectfont
\centering
\caption{\small
Visualization of additional eight small molecule editing cases. The \colorbox{SmallMoleculeBlue}{blue regions}, \colorbox{SmallMoleculeRed}{red regions}, and \colorbox{SmallMoleculeGreen}{green regions} correspond to the edited substructures in the input molecule $\vx_{\text{in}}$, intermediate molecule $\vx_1$ in the 1st conversation round, and the output molecule $\vx_{\text{out}}$, respectively.
}
\label{tab:appendix_small_molecule_analysis}
\vspace{+1ex}
    \begin{adjustbox}{max width=\linewidth}
    \small
    \begin{tabular}{cccccc}
    \toprule
    \multicolumn{3}{c}{(a) Prompt for 103: more like a drug} &  \multicolumn{3}{c}{(b) Prompt for 104: less like a drug}\\
    \cmidrule(lr){1-3}\cmidrule(lr){4-6}
    {Input Molecule $\vx_{\text{in}}$} & {Intermediate Molecule $\vx_{1}$} & {Output Molecule $\vx_{\text{out}}$} & {Input Molecule $\vx_{\text{in}}$} & {Intermediate Molecule $\vx_{1}$} & {Output Molecule $\vx_{\text{out}}$}\\
    \cmidrule(lr){1-1}\cmidrule(lr){2-2}\cmidrule(lr){3-3}\cmidrule(lr){4-4}\cmidrule(lr){5-5}\cmidrule(lr){6-6}
    \adjustbox{valign=c}{\includegraphics[width=0.25\linewidth]{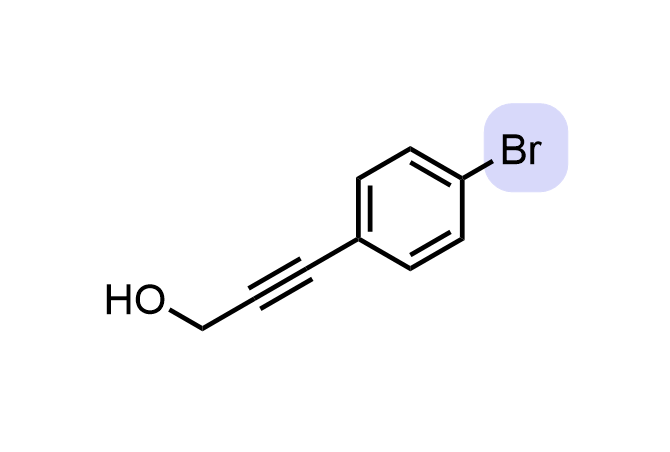}} &
    \adjustbox{valign=c}{\includegraphics[width=0.25\linewidth]{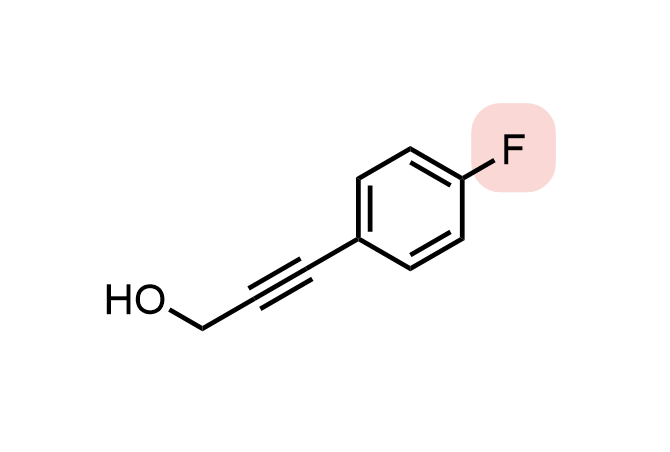}} &
    \adjustbox{valign=c}{\includegraphics[width=0.25\linewidth]{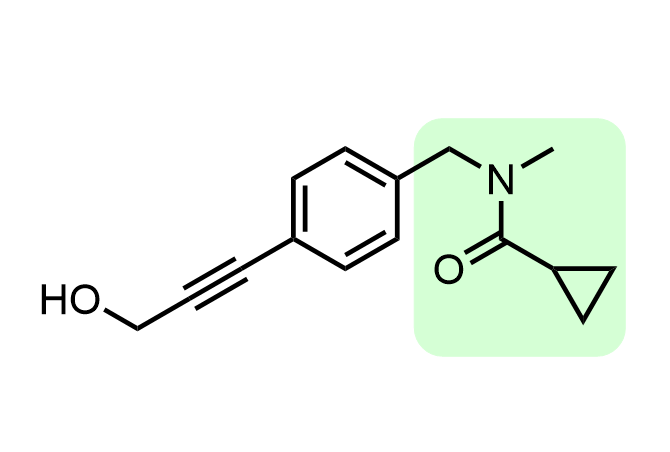}} &
    \adjustbox{valign=c}{\includegraphics[width=0.25\linewidth]{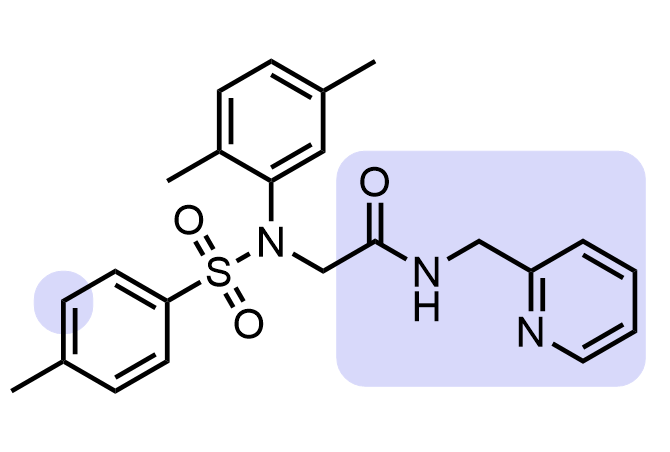}} &
    \adjustbox{valign=c}{\includegraphics[width=0.25\linewidth]{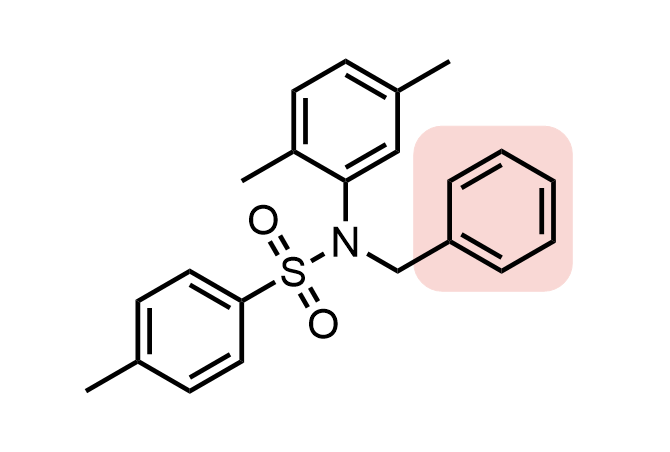}} &
    \adjustbox{valign=c}{\includegraphics[width=0.25\linewidth]{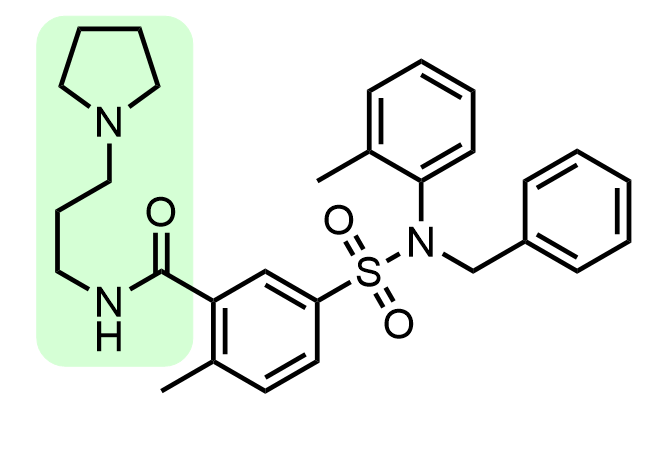}}
    \\
    {\small QED: 0.648} & {\small QED: 0.550} & {\small QED: 0.815} & {\small QED: 0.631} & {\small QED: 0.637} & {\small QED: 0.397}\\[5pt]
    \toprule
    \multicolumn{3}{c}{(c) Prompt for 107: more hydrogen bond acceptors} &  \multicolumn{3}{c}{(d) Prompt for 108: more hydrogen bond donors}\\
    \cmidrule(lr){1-3}\cmidrule(lr){4-6}
    {Input Molecule $\vx_{\text{in}}$} & {Intermediate Molecule $\vx_{1}$} & {Output Molecule $\vx_{\text{out}}$} & {Input Molecule $\vx_{\text{in}}$} & {Intermediate Molecule $\vx_{1}$} & {Output Molecule $\vx_{\text{out}}$}\\
    \cmidrule(lr){1-1}\cmidrule(lr){2-2}\cmidrule(lr){3-3}\cmidrule(lr){4-4}\cmidrule(lr){5-5}\cmidrule(lr){6-6}
    \adjustbox{valign=c}{\includegraphics[width=0.25\linewidth]{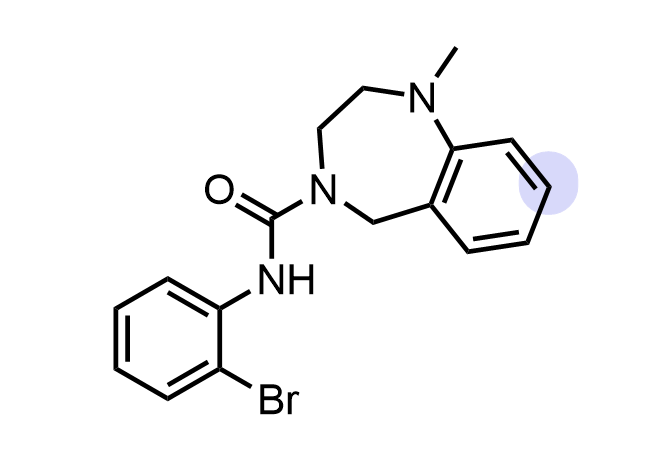}} &
    \adjustbox{valign=c}{\includegraphics[width=0.25\linewidth]{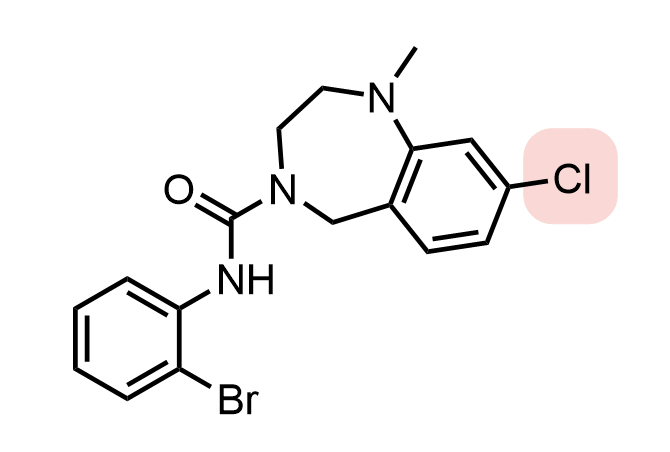}} &
    \adjustbox{valign=c}{\includegraphics[width=0.25\linewidth]{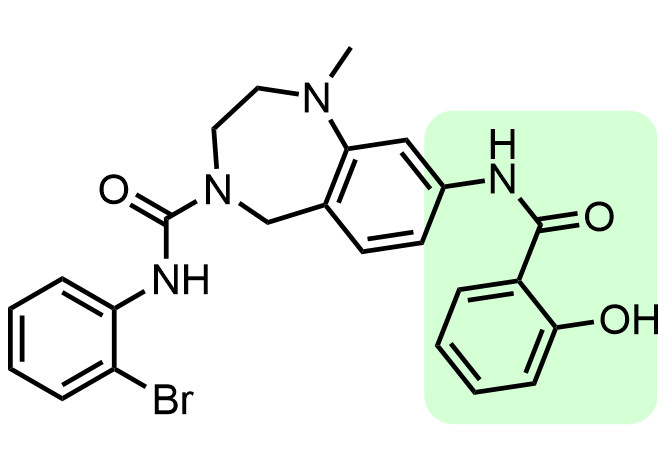}} &
    \adjustbox{valign=c}{\includegraphics[width=0.25\linewidth]{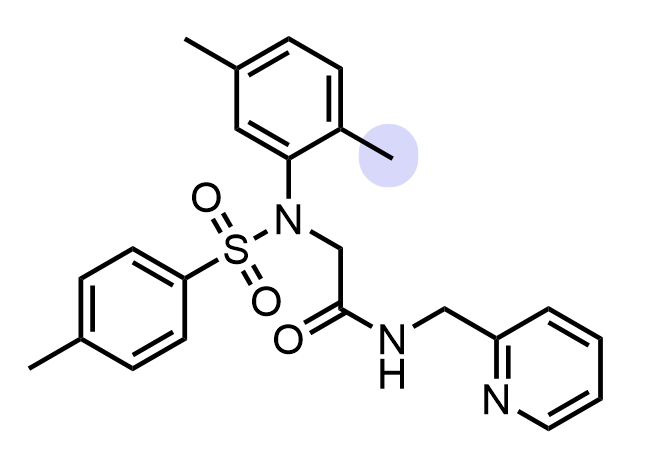}} &
    \adjustbox{valign=c}{\includegraphics[width=0.25\linewidth]{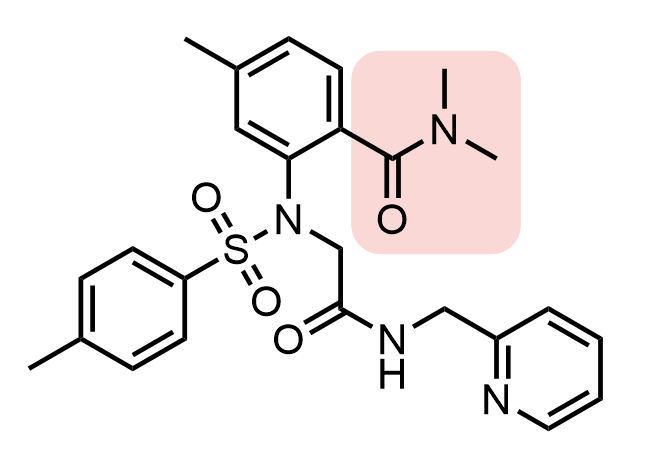}} &
    \adjustbox{valign=c}{\includegraphics[width=0.25\linewidth]{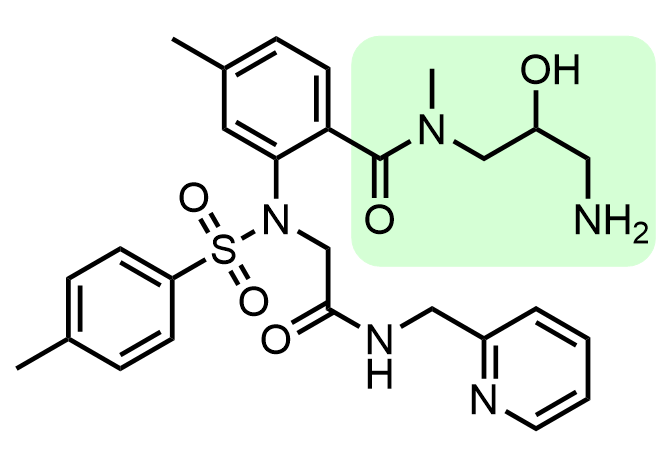}}
    \\
    {\small HBA: 2} & {\small HBA: 2} & {\small HBA: 4} & {\small HBD: 1} & {\small HBD: 1} & {\small HBD: 3}\\[5pt]
    \toprule
    \multicolumn{3}{c}{(e) Prompt for 201: more soluble in water and more hydrogen bond acceptors} &  \multicolumn{3}{c}{(f) Prompt for 202: less soluble in water and more hydrogen bond acceptors}\\
    \cmidrule(lr){1-3}\cmidrule(lr){4-6}
    {Input Molecule $\vx_{\text{in}}$} & {Intermediate Molecule $\vx_{1}$} & {Output Molecule $\vx_{\text{out}}$} & {Input Molecule $\vx_{\text{in}}$} & {Intermediate Molecule $\vx_{1}$} & {Output Molecule $\vx_{\text{out}}$}\\
    \cmidrule(lr){1-1}\cmidrule(lr){2-2}\cmidrule(lr){3-3}\cmidrule(lr){4-4}\cmidrule(lr){5-5}\cmidrule(lr){6-6}
    \adjustbox{valign=c}{\includegraphics[width=0.25\linewidth]{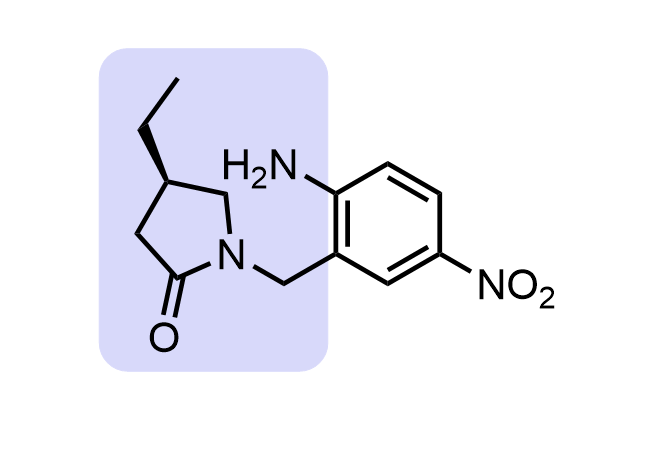}} &
    \adjustbox{valign=c}{\includegraphics[width=0.25\linewidth]{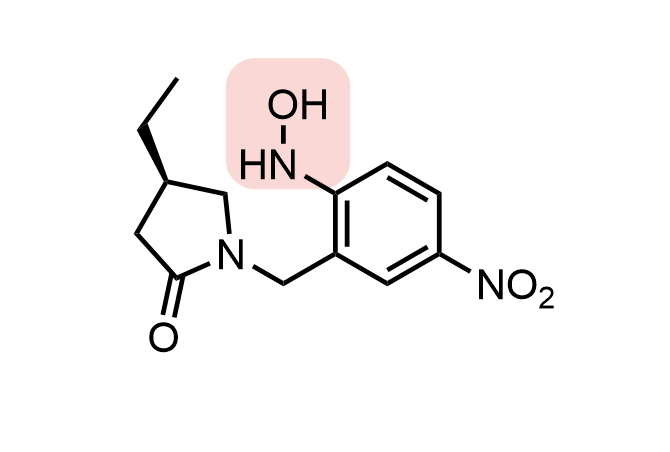}} &
    \adjustbox{valign=c}{\includegraphics[width=0.25\linewidth]{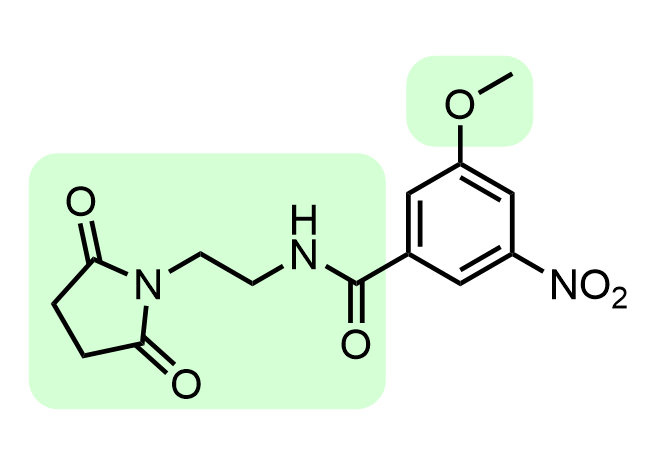}} &
    \adjustbox{valign=c}{\includegraphics[width=0.25\linewidth]{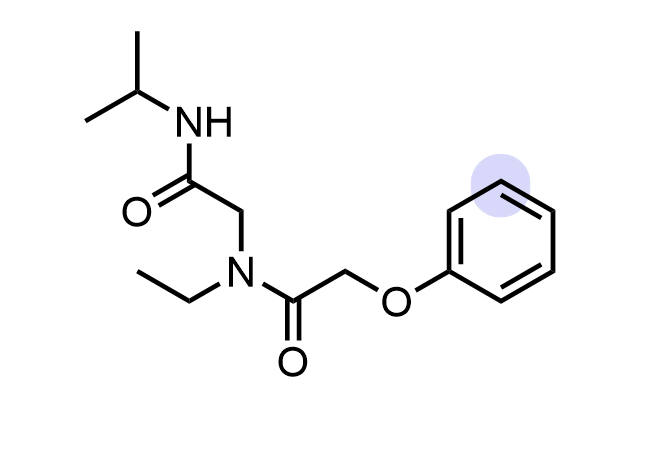}} &
    \adjustbox{valign=c}{\includegraphics[width=0.25\linewidth]{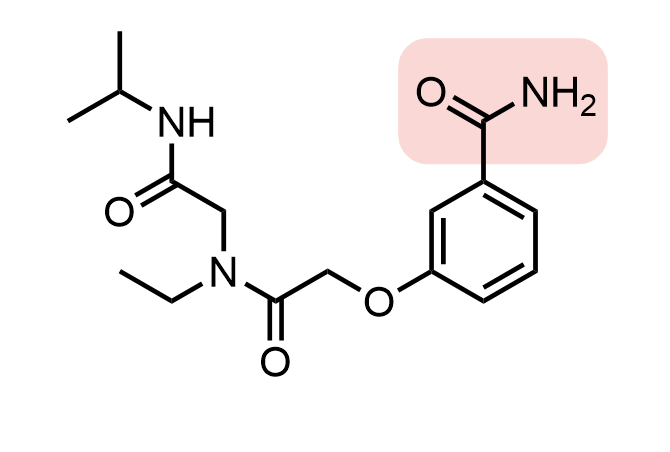}} &
    \adjustbox{valign=c}{\includegraphics[width=0.25\linewidth]{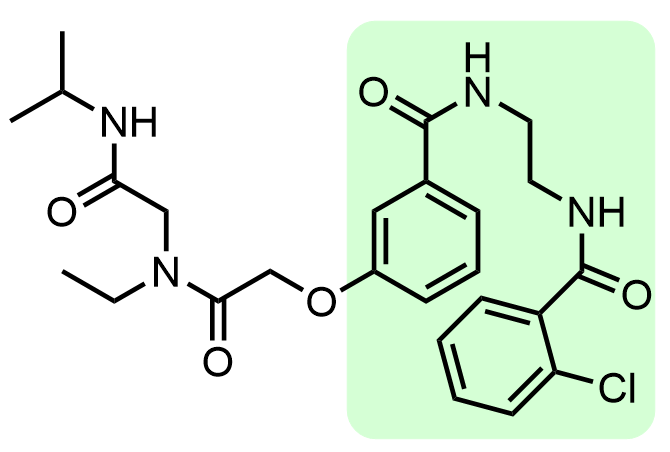}}
    \\
    {\small LogP: 1.12, HBA: 4} & {\small LogP: 1.34, HBA: 5} & {\small LogP: -0.42, HBA: 6} & {\small LogP: 0.98, HBA: 3} & {\small LogP: 0.02, HBA: 4} & {\small LogP: 1.68, HBA: 5}\\[5pt]
    \toprule
    \multicolumn{3}{c}{(g) Prompt for 203: more soluble in water and more hydrogen bond donors} &  \multicolumn{3}{c}{(h) Prompt for 204: less soluble in water and more hydrogen bond donors}\\
    \cmidrule(lr){1-3}\cmidrule(lr){4-6}
    {Input Molecule $\vx_{\text{in}}$} & {Intermediate Molecule $\vx_{1}$} & {Output Molecule $\vx_{\text{out}}$} & {Input Molecule $\vx_{\text{in}}$} & {Intermediate Molecule $\vx_{1}$} & {Output Molecule $\vx_{\text{out}}$}\\
    \cmidrule(lr){1-1}\cmidrule(lr){2-2}\cmidrule(lr){3-3}\cmidrule(lr){4-4}\cmidrule(lr){5-5}\cmidrule(lr){6-6}
    \adjustbox{valign=c}{\includegraphics[width=0.25\linewidth]{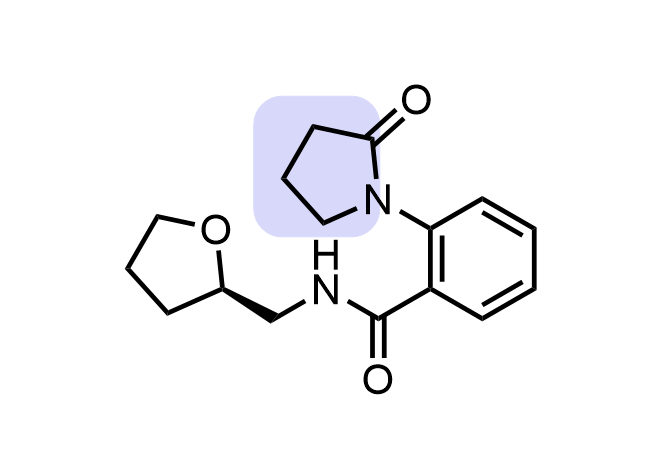}} &
    \adjustbox{valign=c}{\includegraphics[width=0.25\linewidth]{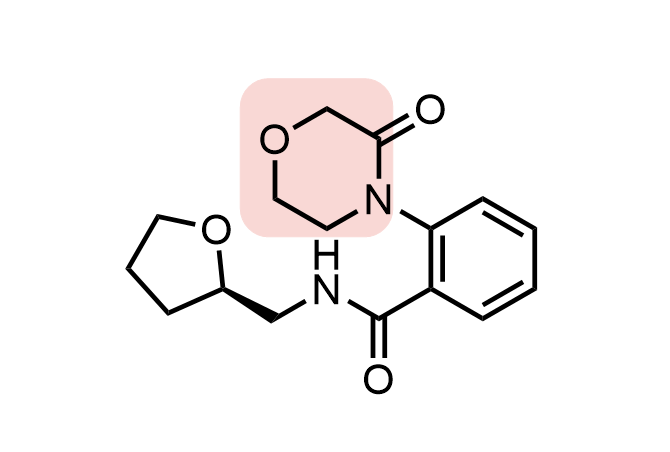}} &
    \adjustbox{valign=c}{\includegraphics[width=0.25\linewidth]{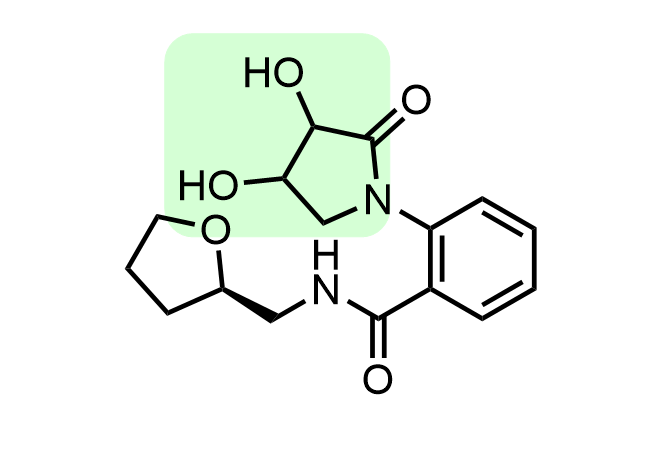}} &
    \adjustbox{valign=c}{\includegraphics[width=0.25\linewidth]{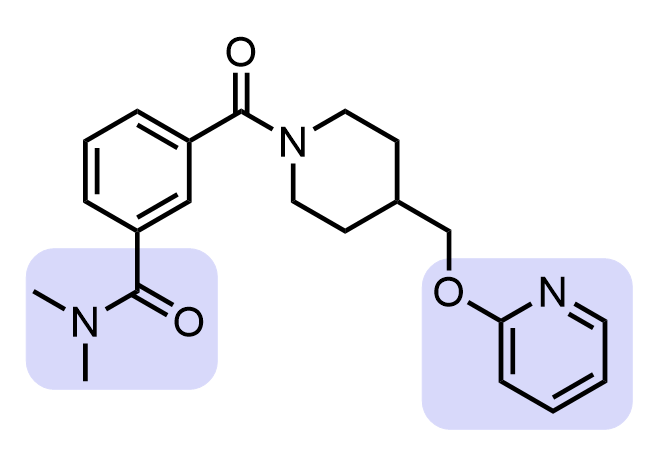}} &
    \adjustbox{valign=c}{\includegraphics[width=0.25\linewidth]{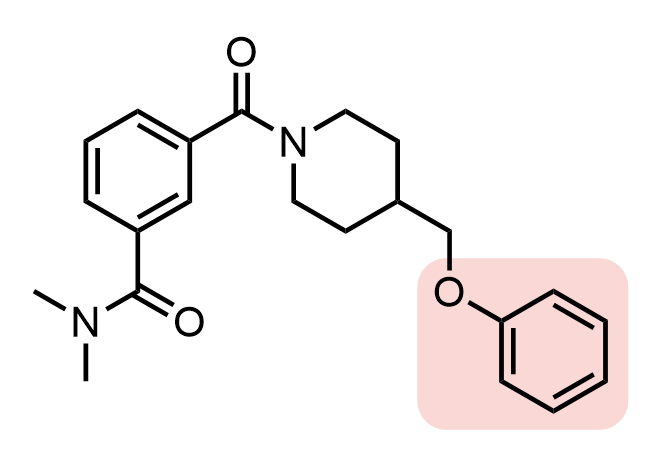}} &
    \adjustbox{valign=c}{\includegraphics[width=0.25\linewidth]{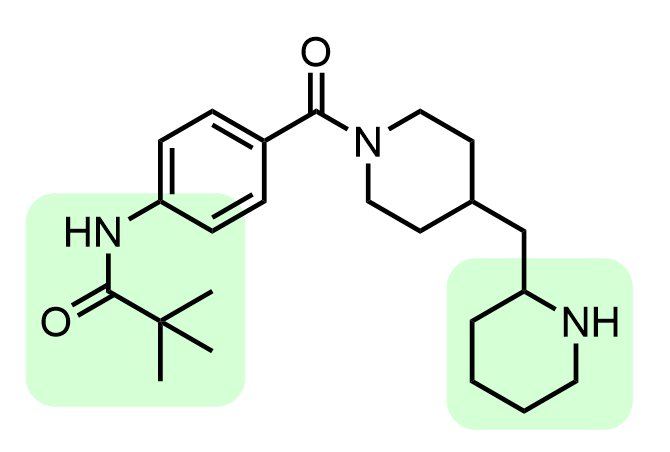}}
    \\
    {\small LogP: 1.29, HBD: 1} & {\small LogP: 0.52, HBD: 1} & {\small LogP: -0.77, HBD: 3} & {\small LogP: 2.14, HBD: 0} & {\small LogP: 2.75, HBD: 0} & {\small LogP: 3.62, HBD: 2}\\[5pt]
    \bottomrule
    \end{tabular}
    \end{adjustbox}
\vspace{-3ex}
\end{table}

\newpage
\subsubsection{Similarity Between Input, Intermediate, Retrieved, and Output Molecules} \label{sec:similarity_comparison_small_molecules}

\begin{figure}[h]
\centering
\begin{subfigure}[b]{0.32\linewidth}
    \centering
    \includegraphics[width=\linewidth]{figure/small_molecule_similarity/101.png}
    \caption{\small Task 101.}
\end{subfigure}
\begin{subfigure}[b]{0.32\linewidth}
    \centering
    \includegraphics[width=\linewidth]{figure/small_molecule_similarity/102.png}
    \caption{\small Task 102.}
\end{subfigure}
\begin{subfigure}[b]{0.32\linewidth}
    \centering
    \includegraphics[width=\linewidth]{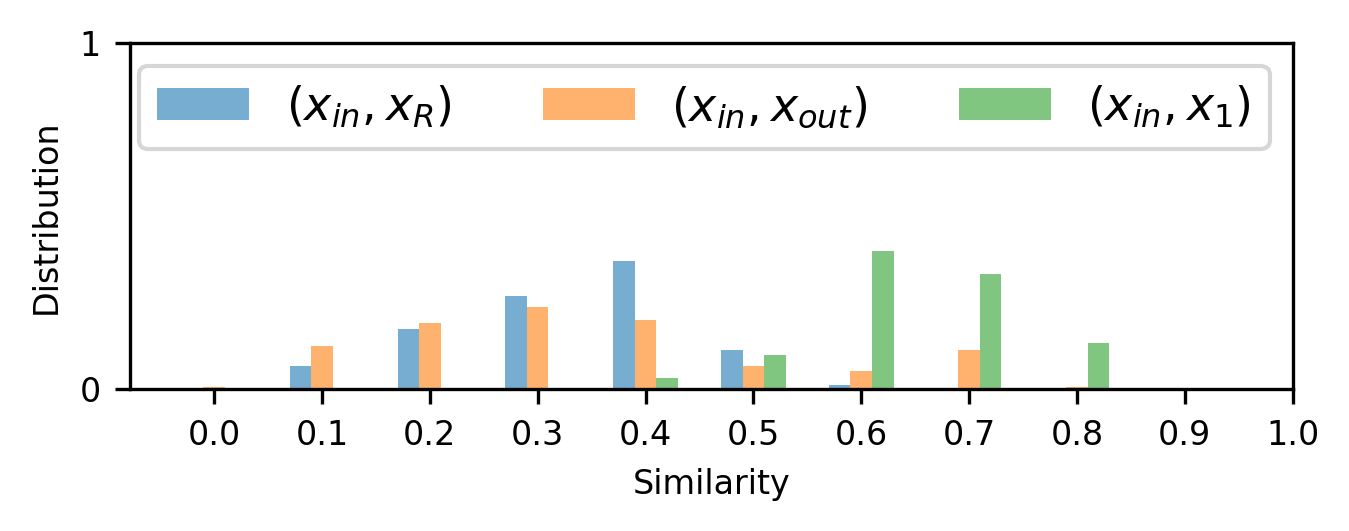}
    \caption{\small Task 103.}
\end{subfigure}
\begin{subfigure}[b]{0.32\linewidth}
    \centering
    \includegraphics[width=\linewidth]{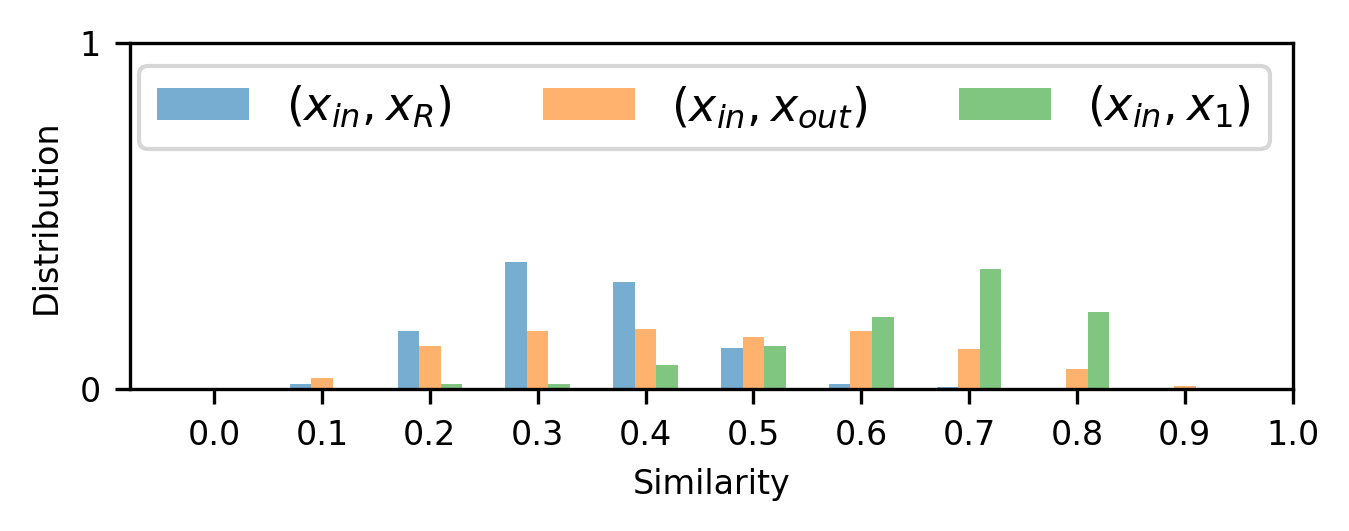}
    \caption{\small Task 104.}
\end{subfigure}
\begin{subfigure}[b]{0.32\linewidth}
    \centering
    \includegraphics[width=\linewidth]{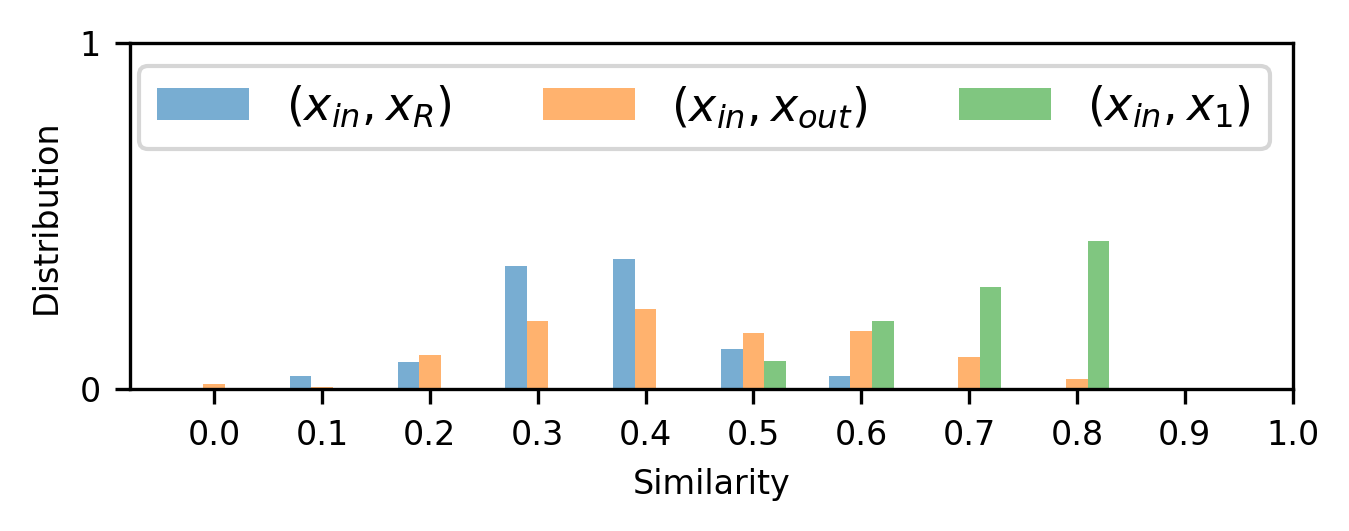}
    \caption{\small Task 105.}
\end{subfigure}
\begin{subfigure}[b]{0.32\linewidth}
    \centering
    \includegraphics[width=\linewidth]{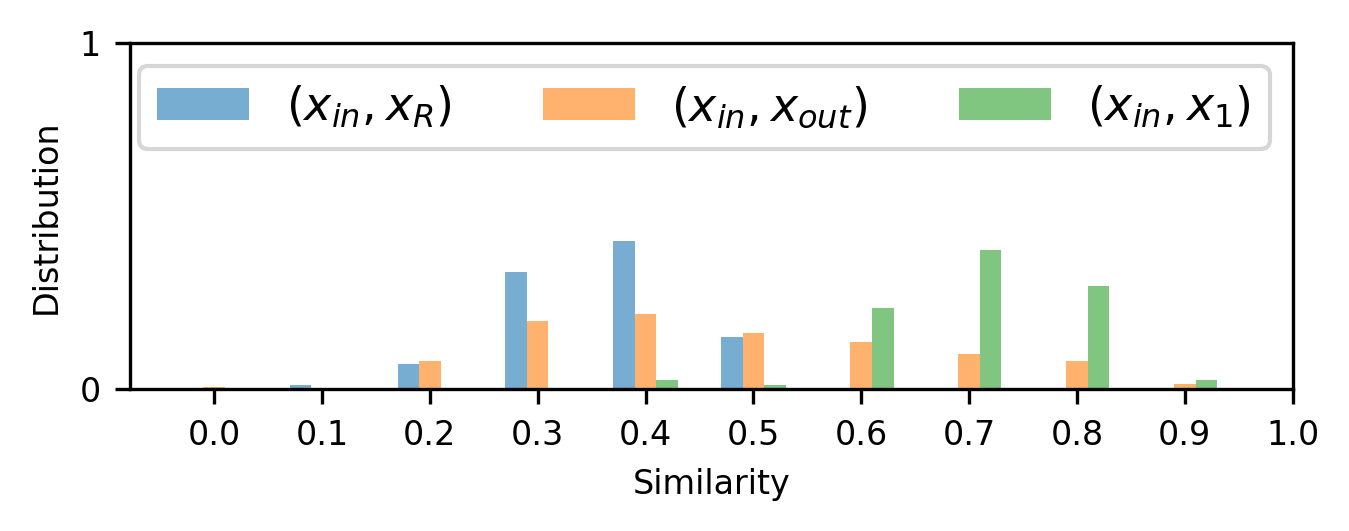}
    \caption{\small Task 106.}
\end{subfigure}
\begin{subfigure}[b]{0.32\linewidth}
    \centering
    \includegraphics[width=\linewidth]{figure/small_molecule_similarity/107.png}
    \caption{\small Task 107.}
\end{subfigure}
\begin{subfigure}[b]{0.32\linewidth}
    \centering
    \includegraphics[width=\linewidth]{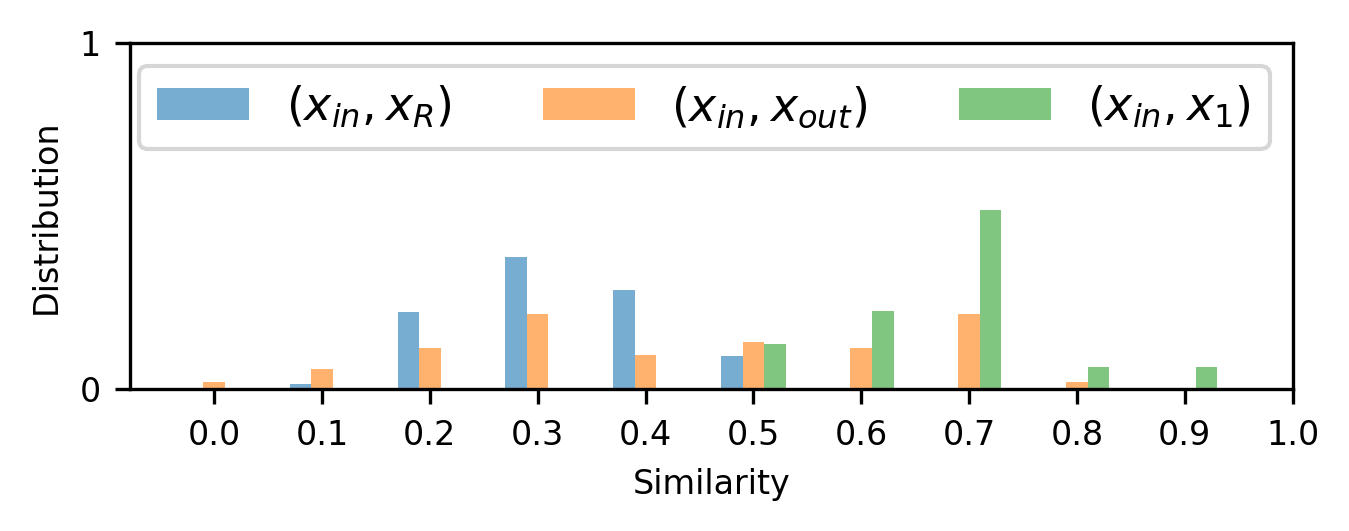}
    \caption{\small Task 108.}
\end{subfigure}
\vspace{-1.5ex}
\caption{\small Similarity distribution between input molecules $\vx_{\text{in}}$ and retrieval $\vx_R$, intermediate $\vx_1$, and output molecules $\vx_{\text{out}}$. Here we show the distribution of 8 single-objective editing tasks.}
\label{fig:similarity_distribution_shift_small_molecules_single}
\vspace{+2ex}
\end{figure}

\begin{figure}[h]
\centering
\begin{subfigure}[b]{0.32\linewidth}
    \centering
    \includegraphics[width=\linewidth]{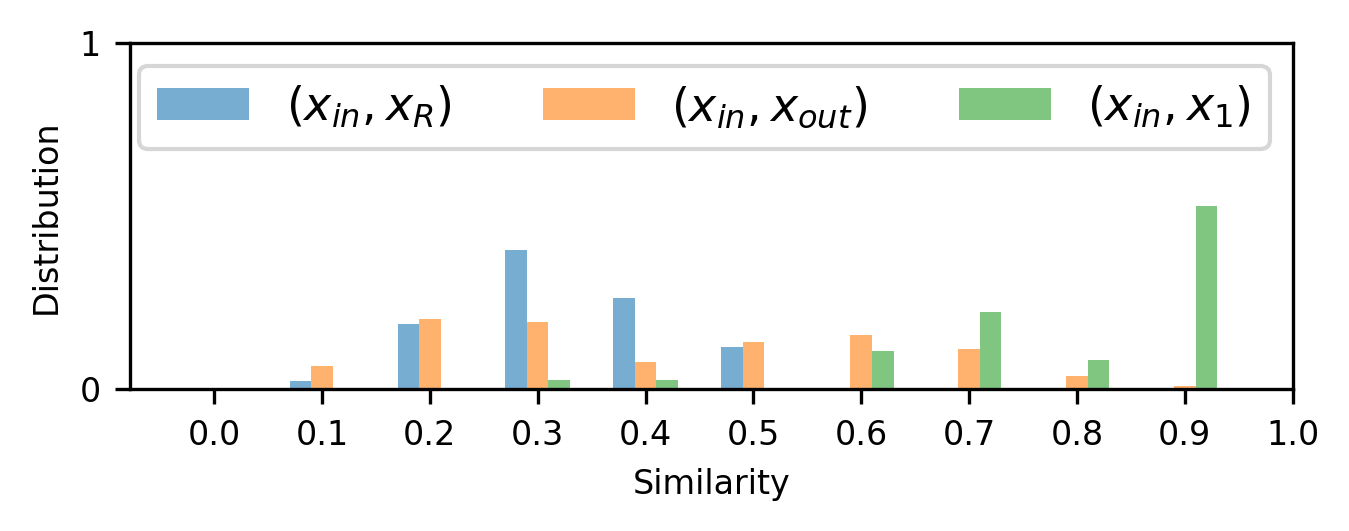}
    \caption{\small Task 201.}
\end{subfigure}
\begin{subfigure}[b]{0.32\linewidth}
    \centering
    \includegraphics[width=\linewidth]{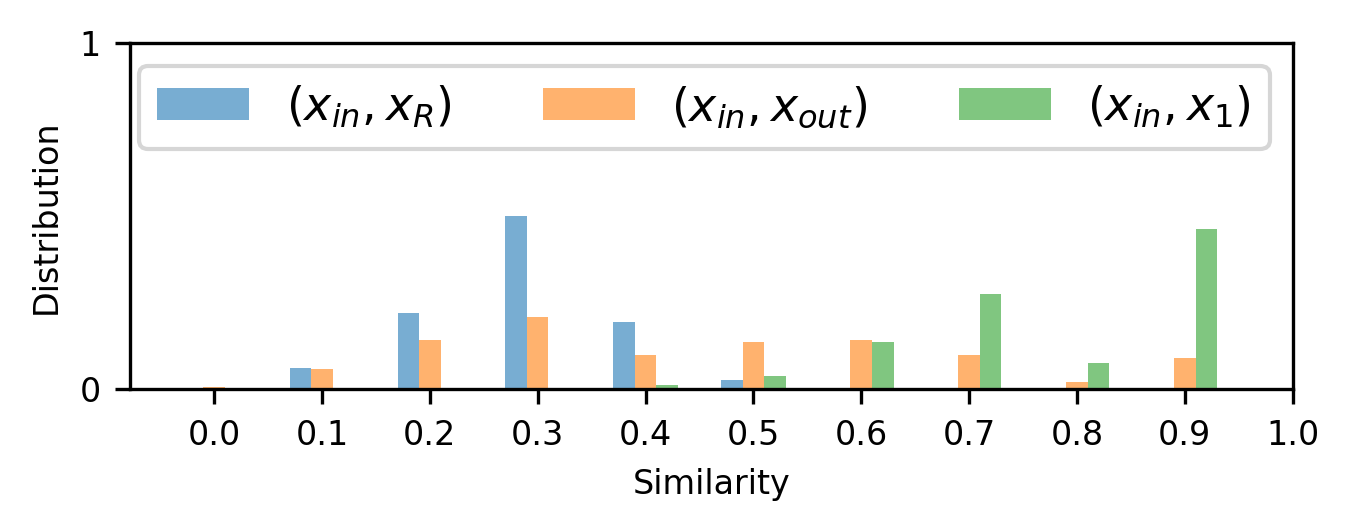}
    \caption{\small Task 202.}
\end{subfigure}
\begin{subfigure}[b]{0.32\linewidth}
    \centering
    \includegraphics[width=\linewidth]{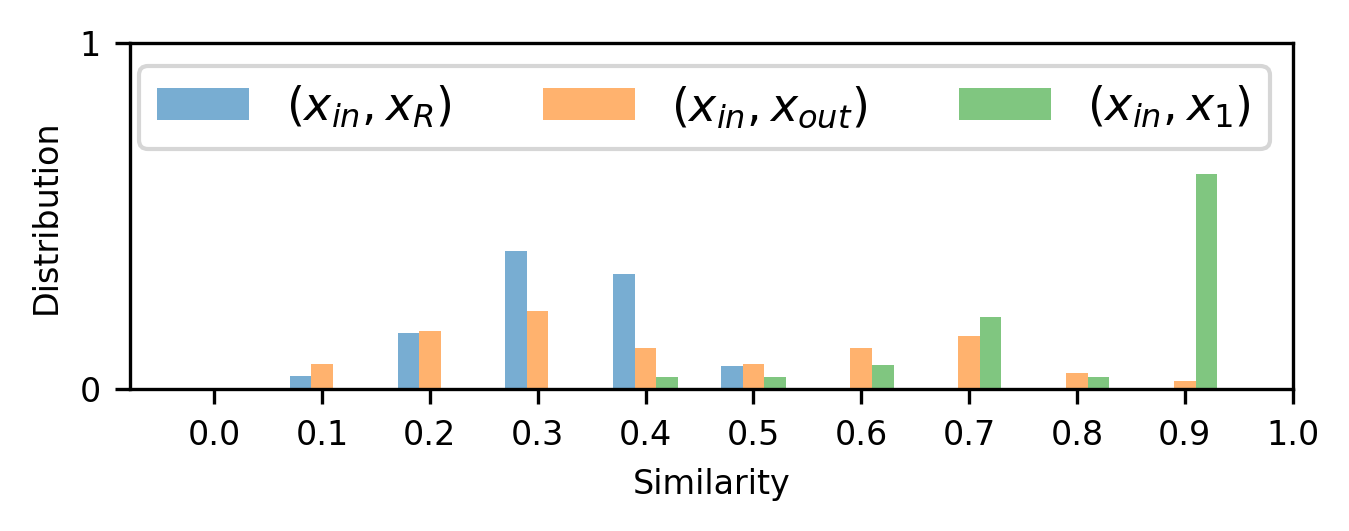}
    \caption{\small Task 203.}
\end{subfigure}
\begin{subfigure}[b]{0.32\linewidth}
    \centering
    \includegraphics[width=\linewidth]{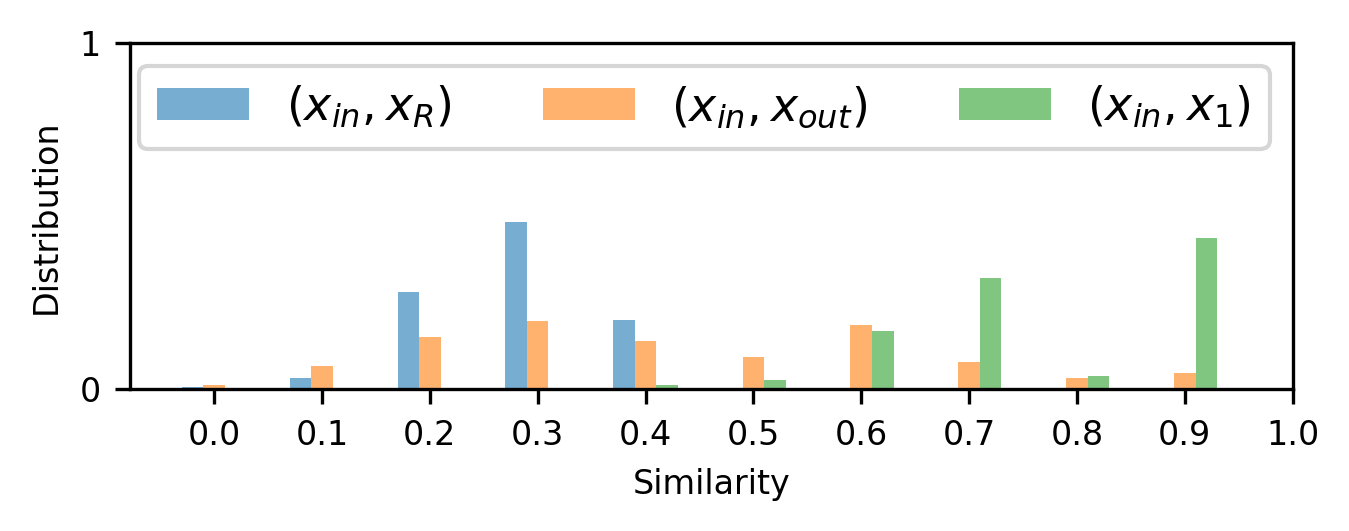}
    \caption{\small Task 204.}
\end{subfigure}
\begin{subfigure}[b]{0.32\linewidth}
    \centering
    \includegraphics[width=\linewidth]{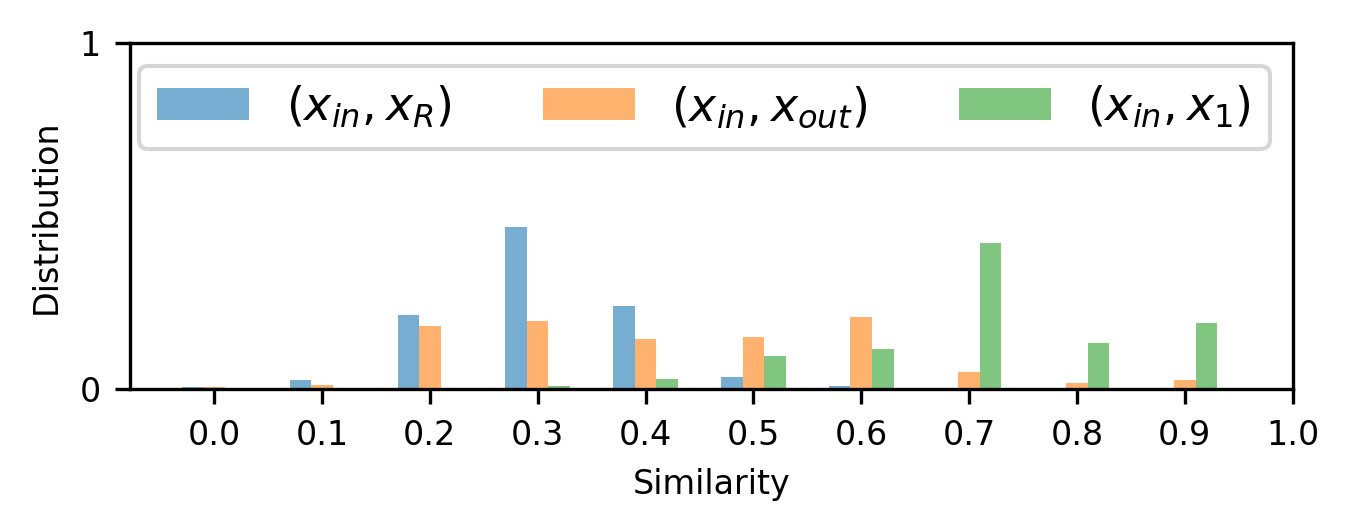}
    \caption{\small Task 205.}
\end{subfigure}
\begin{subfigure}[b]{0.32\linewidth}
    \centering
    \includegraphics[width=\linewidth]{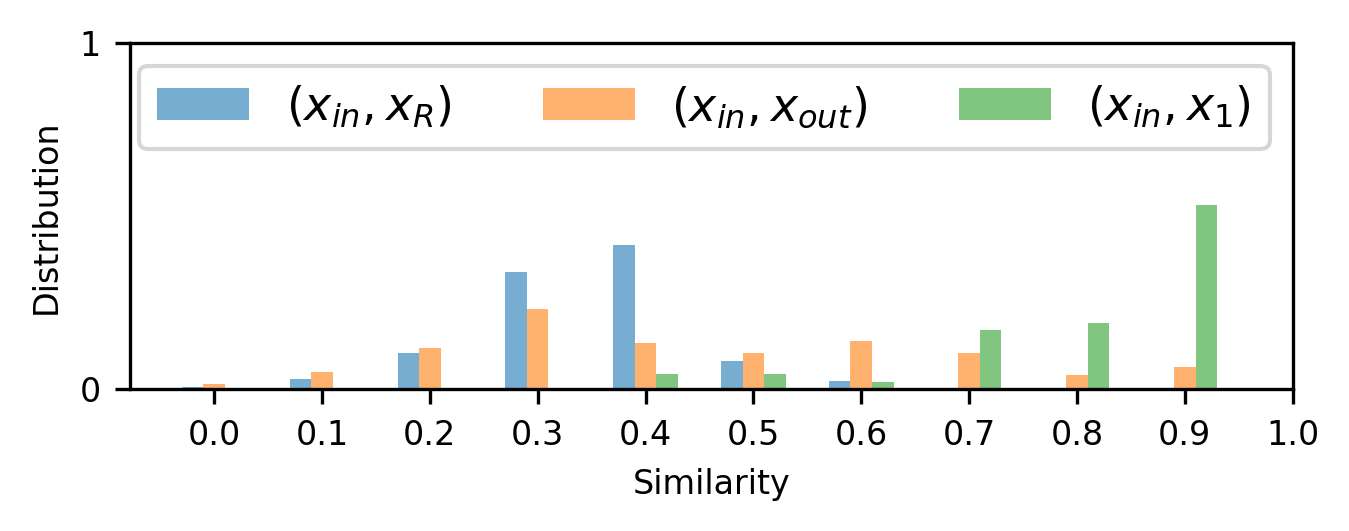}
    \caption{\small Task 206.}
\end{subfigure}
\vspace{-1.5ex}
\caption{\small Similarity distribution between input molecules $\vx_{\text{in}}$ and retrieval $\vx_R$, intermediate $\vx_1$, and output molecules $\vx_{\text{out}}$. Here we show the distribution of 6 multi-objective editing tasks.}
\label{fig:similarity_distribution_shift_small_molecules_multi}
\end{figure}

In~\Cref{fig:similarity_distribution_shift}, we plot the distribution of similarities between input molecules $\vx_{\text{in}}$ and retrieval $\vx_R$, intermediate $\vx_1$, and output molecules $\vx_{\text{out}}$ using \model{}. Here we provide more results. The distributions of 8 single-objective small molecule editing tasks can be found in~\Cref{fig:similarity_distribution_shift_small_molecules_single}, and 6 multi-objective small molecule editing tasks can be found in~\Cref{fig:similarity_distribution_shift_small_molecules_multi}.

As shown in~\Cref{fig:similarity_distribution_shift_small_molecules_single,fig:similarity_distribution_shift_small_molecules_multi}, the observation of similarity distribution satisfies for all 8 single-objective and 6 multi-objective tasks.

\newpage
\subsection{Peptide} \label{sec:case_studies_peptides}

\begin{figure}[ht]
\centering
\begin{subfigure}[b]{0.32\linewidth}
    \centering
    \includegraphics[width=\linewidth]{figure/peptide/501_0_input.png}
    \caption{\fontsize{7.5}{6}\selectfont Motifs of input peptides.}
\end{subfigure}
\hfill
\begin{subfigure}[b]{0.32\linewidth}
    \centering
    \includegraphics[width=\linewidth]{figure/peptide/501_1_edit.png}
    \caption{\fontsize{7.5}{6}\selectfont Motifs of edited peptides.}
\end{subfigure}
\hfill
\begin{subfigure}[b]{0.32\linewidth}
    \centering
    \includegraphics[width=\linewidth]{figure/peptide/501_2_target.png}
    \caption{\fontsize{7.5}{6}\selectfont Motifs of experimental peptides.}
\end{subfigure}
\vspace{-1ex}
\caption{\small Visualization for peptide editing for task 301, higher binding affinity to HLA-B*44:02.}
\label{fig:peptide_motif_analysis_501}
\vspace{+1ex}
\centering
\begin{subfigure}[b]{0.32\linewidth}
    \centering
    \includegraphics[width=\linewidth]{figure/peptide/502_0_input.png}
    \caption{\fontsize{7.5}{6}\selectfont Motifs of input peptides.}
\end{subfigure}
\hfill
\begin{subfigure}[b]{0.32\linewidth}
    \centering
    \includegraphics[width=\linewidth]{figure/peptide/502_1_edit.png}
    \caption{\fontsize{7.5}{6}\selectfont Motifs of edited peptides.}
\end{subfigure}
\hfill
\begin{subfigure}[b]{0.32\linewidth}
    \centering
    \includegraphics[width=\linewidth]{figure/peptide/502_2_target.png}
    \caption{\fontsize{7.5}{6}\selectfont Motifs of experimental peptides.}
\end{subfigure}
\vspace{-1ex}
\caption{\small Visualization for peptide editing for task 302, higher binding affinity to HLA-C*03:03.}
\label{fig:peptide_motif_analysis_502}
\vspace{+1ex}
\centering
\begin{subfigure}[b]{0.32\linewidth}
    \centering
    \includegraphics[width=\linewidth]{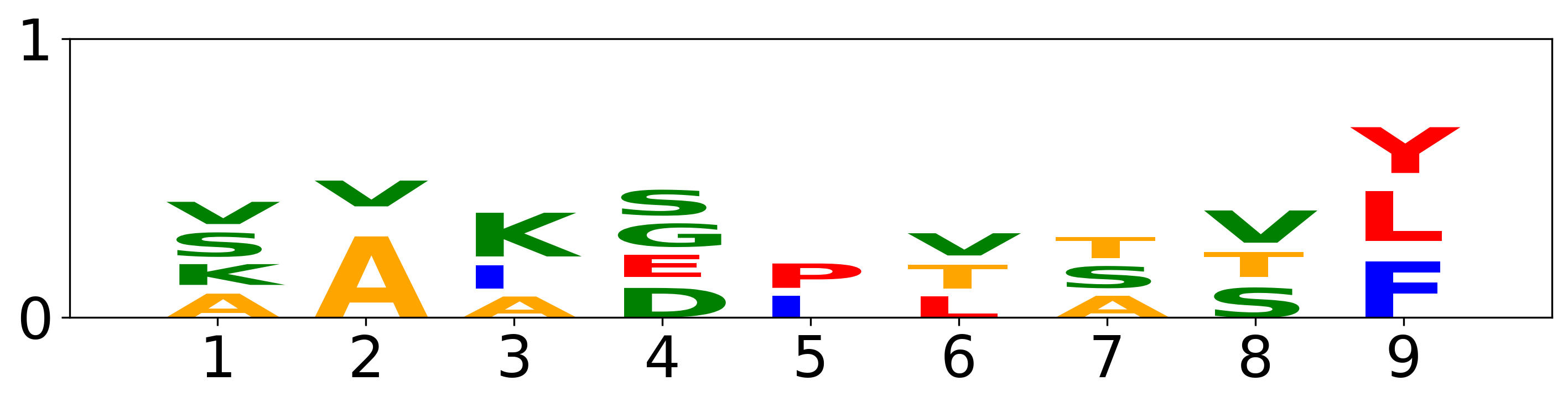}
    \caption{\fontsize{7.5}{6}\selectfont Motifs of input peptides.}
\end{subfigure}
\hfill
\begin{subfigure}[b]{0.32\linewidth}
    \centering
    \includegraphics[width=\linewidth]{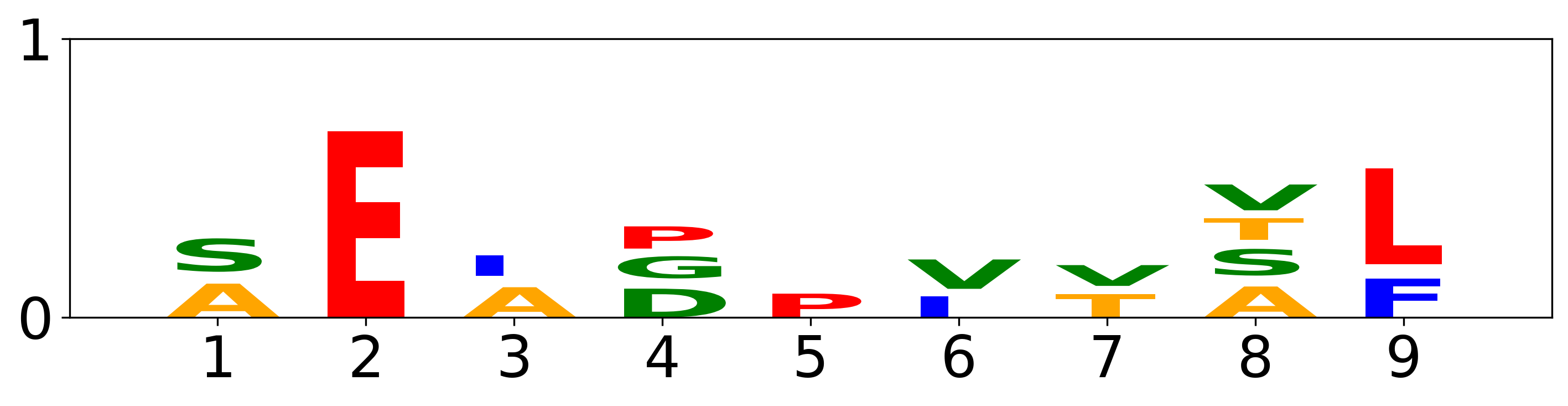}
    \caption{\fontsize{7.5}{6}\selectfont Motifs of edited peptides.}
\end{subfigure}
\hfill
\begin{subfigure}[b]{0.32\linewidth}
    \centering
    \includegraphics[width=\linewidth]{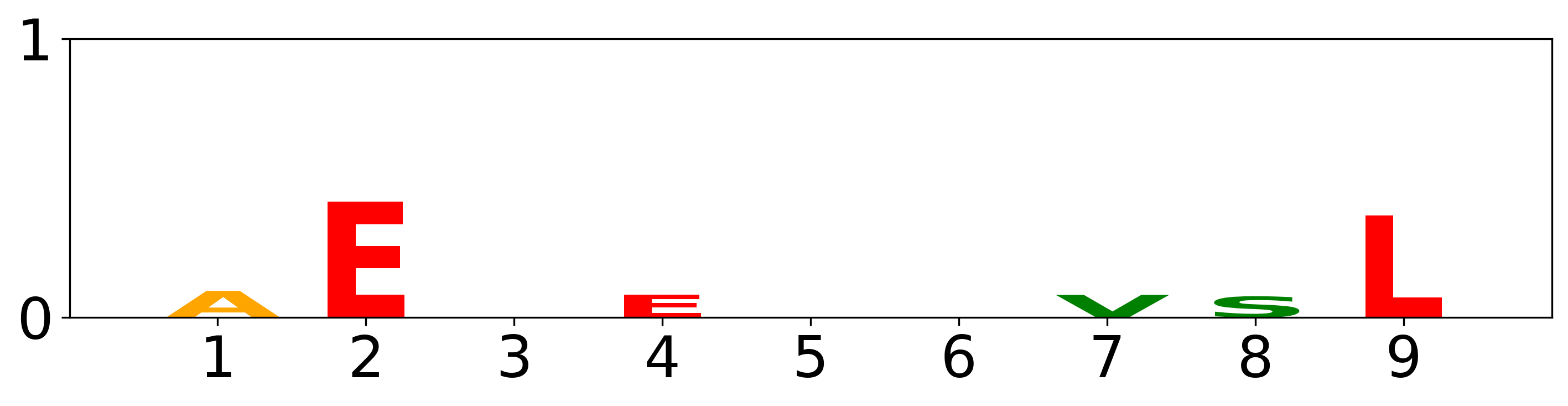}
    \caption{\fontsize{7.5}{6}\selectfont Motifs of experimental peptides.}
\end{subfigure}
\vspace{-1ex}
\caption{\small Visualization for peptide editing for task 303, higher binding affinity to HLA-B*40:01.}
\label{fig:peptide_motif_analysis_503}
\vspace{+1ex}
\centering
\begin{subfigure}[b]{0.32\linewidth}
    \centering
    \includegraphics[width=\linewidth]{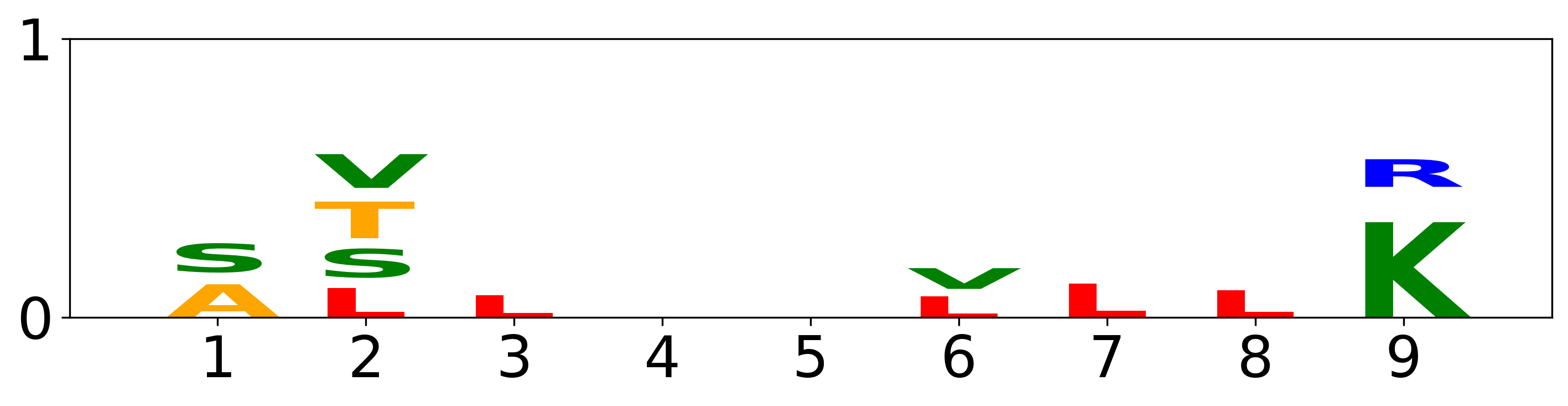}
    \caption{\fontsize{7.5}{6}\selectfont Motifs of input peptides.}
\end{subfigure}
\hfill
\begin{subfigure}[b]{0.32\linewidth}
    \centering
    \includegraphics[width=\linewidth]{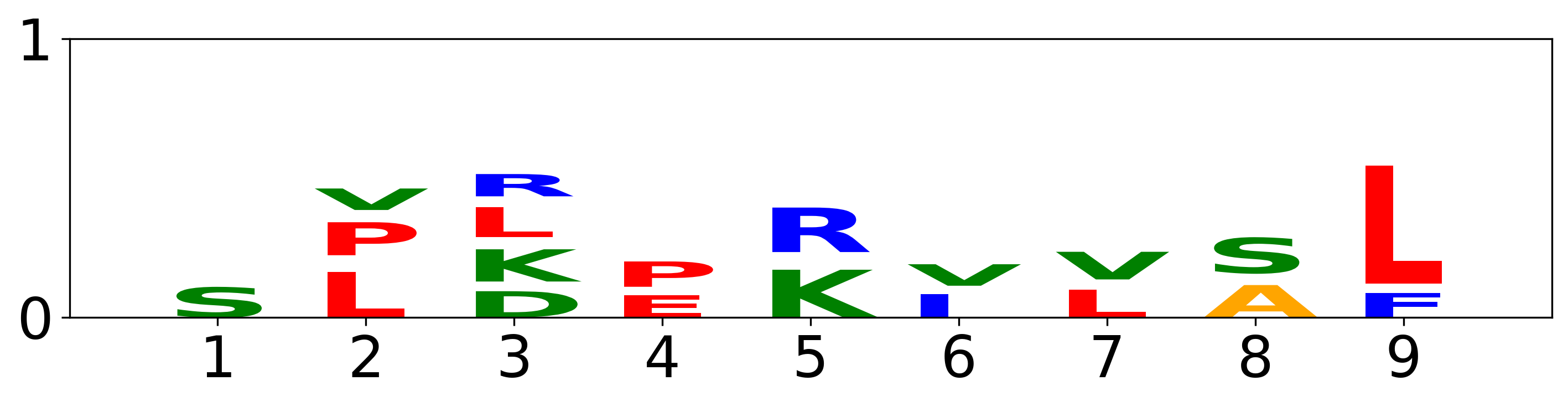}
    \caption{\fontsize{7.5}{6}\selectfont Motifs of edited peptides.}
\end{subfigure}
\hfill
\begin{subfigure}[b]{0.32\linewidth}
    \centering
    \includegraphics[width=\linewidth]{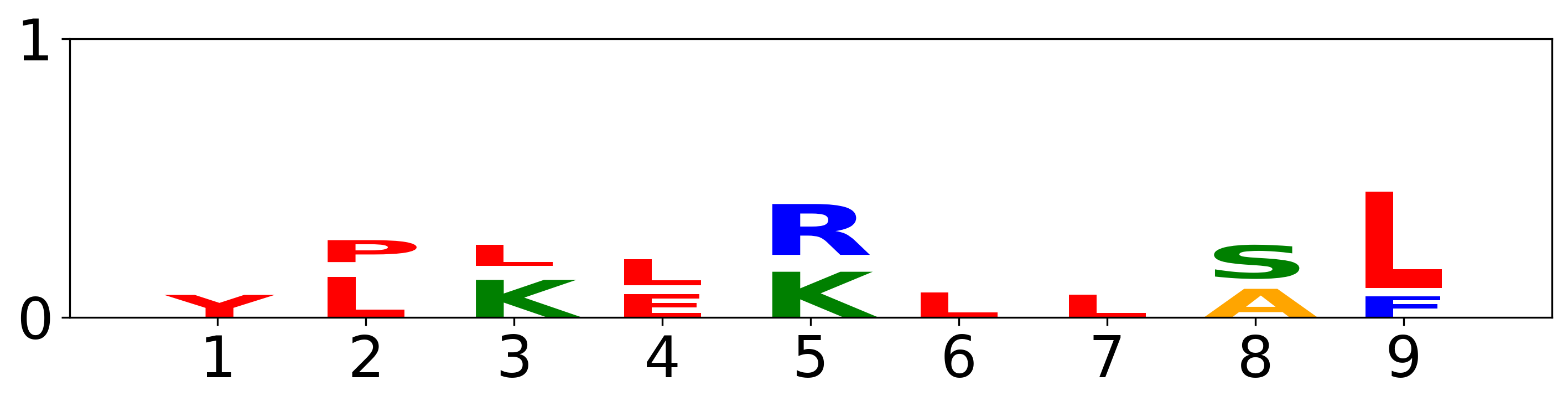}
    \caption{\fontsize{7.5}{6}\selectfont Motifs of experimental peptides.}
\end{subfigure}
\vspace{-1ex}
\caption{\small Visualization for peptide editing for task 304, higher binding affinity to HLA-B*08:01.}
\label{fig:peptide_motif_analysis_504}
\vspace{+1ex}
\centering
\begin{subfigure}[b]{0.32\linewidth}
    \centering
    \includegraphics[width=\linewidth]{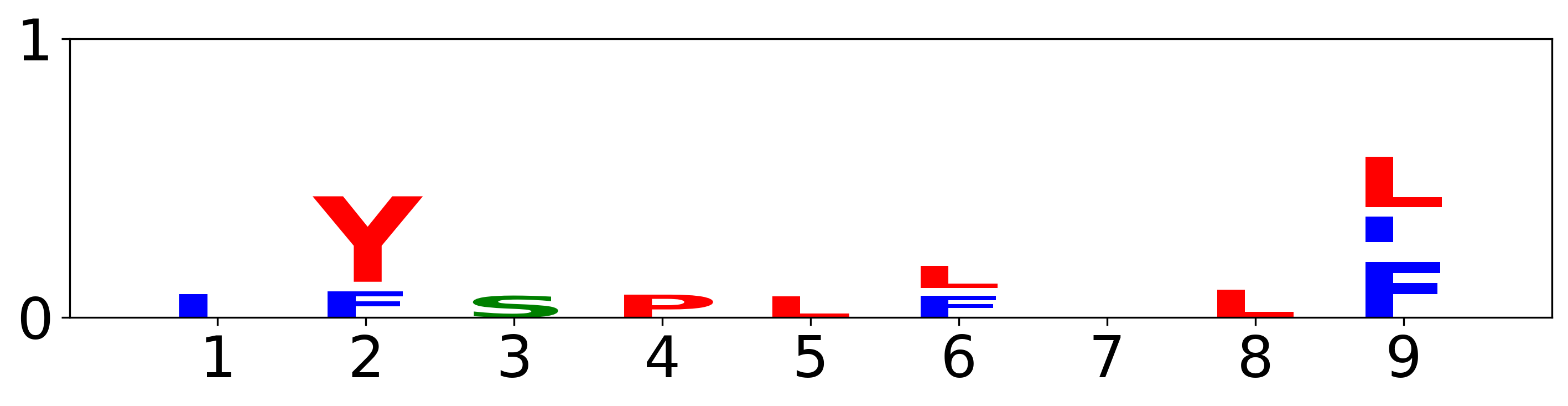}
    \caption{\fontsize{7.5}{6}\selectfont Motifs of input peptides.}
\end{subfigure}
\hfill
\begin{subfigure}[b]{0.32\linewidth}
    \centering
    \includegraphics[width=\linewidth]{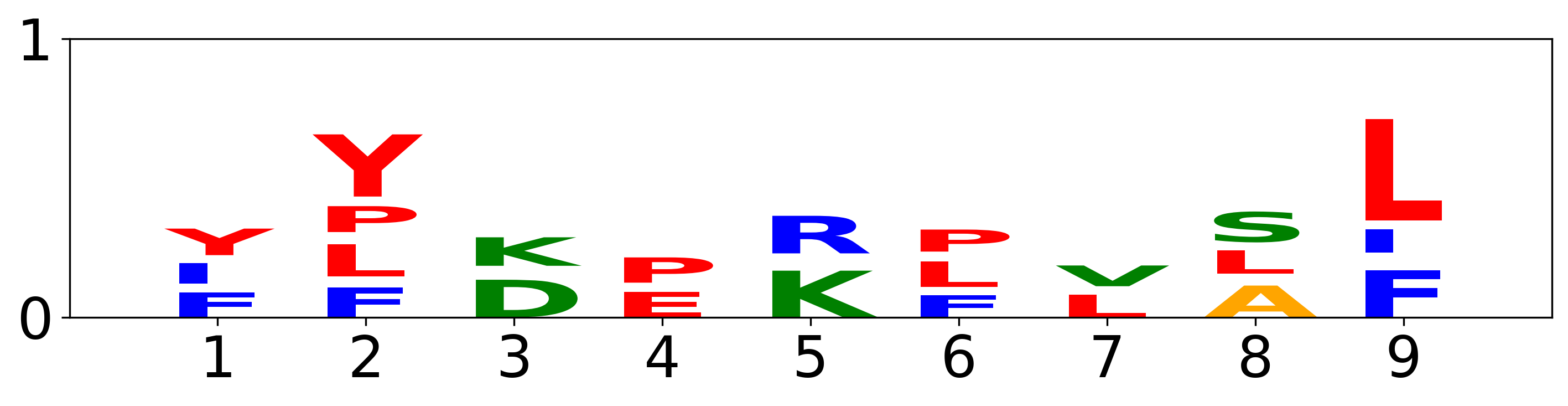}
    \caption{\fontsize{7.5}{6}\selectfont Motifs of edited peptides.}
\end{subfigure}
\hfill
\begin{subfigure}[b]{0.32\linewidth}
    \centering
    \includegraphics[width=\linewidth]{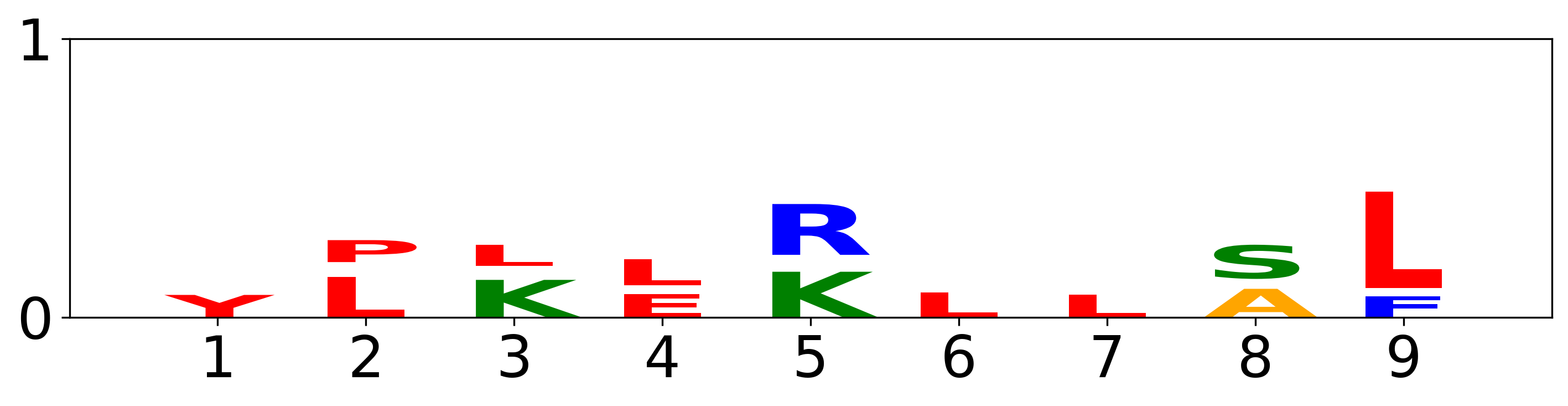}
    \caption{\fontsize{7.5}{6}\selectfont Motifs of experimental peptides.}
\end{subfigure}
\vspace{-1ex}
\caption{\small Visualization for peptide editing for task 305, higher binding affinity to HLA-B*08:01.}
\label{fig:peptide_motif_analysis_505}
\vspace{+1ex}
\begin{subfigure}[b]{0.32\linewidth}
    \centering
    \includegraphics[width=\linewidth]{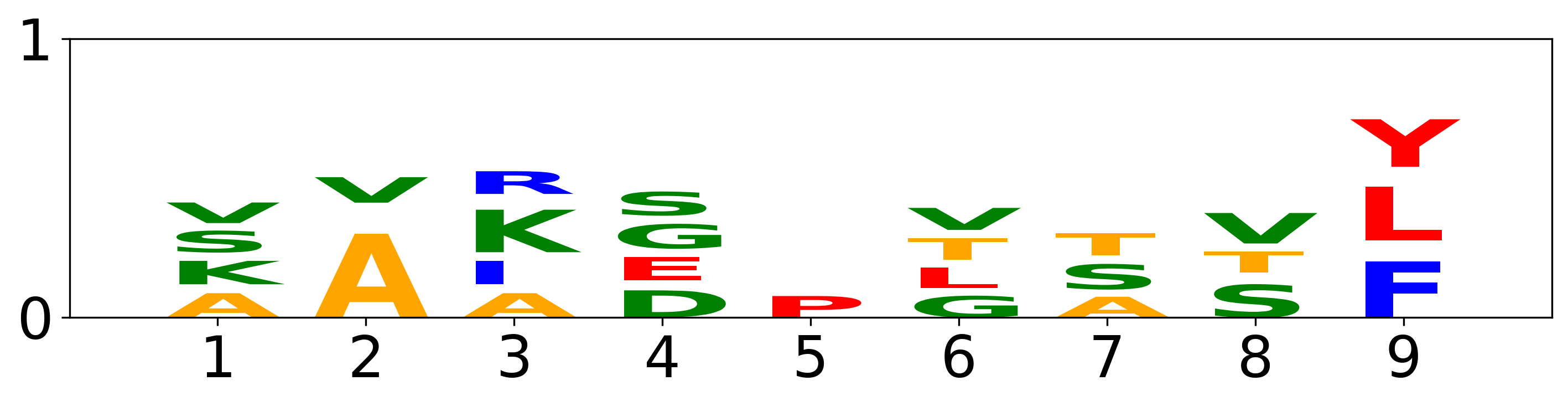}
    \caption{\fontsize{7.5}{6}\selectfont Motifs of input peptides.}
\end{subfigure}
\hfill
\begin{subfigure}[b]{0.32\linewidth}
    \centering
    \includegraphics[width=\linewidth]{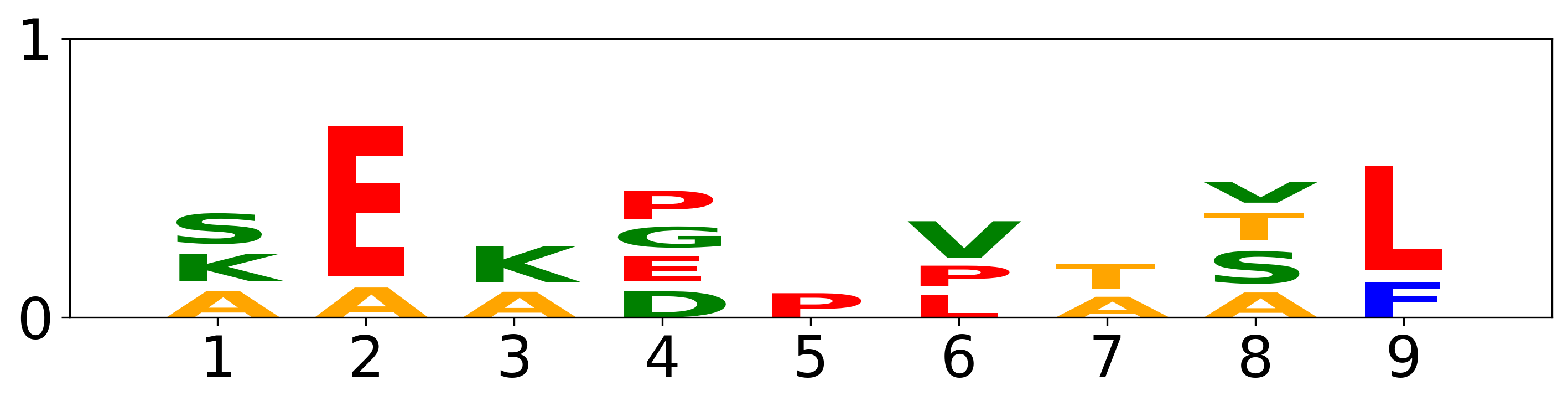}
    \caption{\fontsize{7.5}{6}\selectfont Motifs of edited peptides.}
\end{subfigure}
\hfill
\begin{subfigure}[b]{0.32\linewidth}
    \centering
    \includegraphics[width=\linewidth]{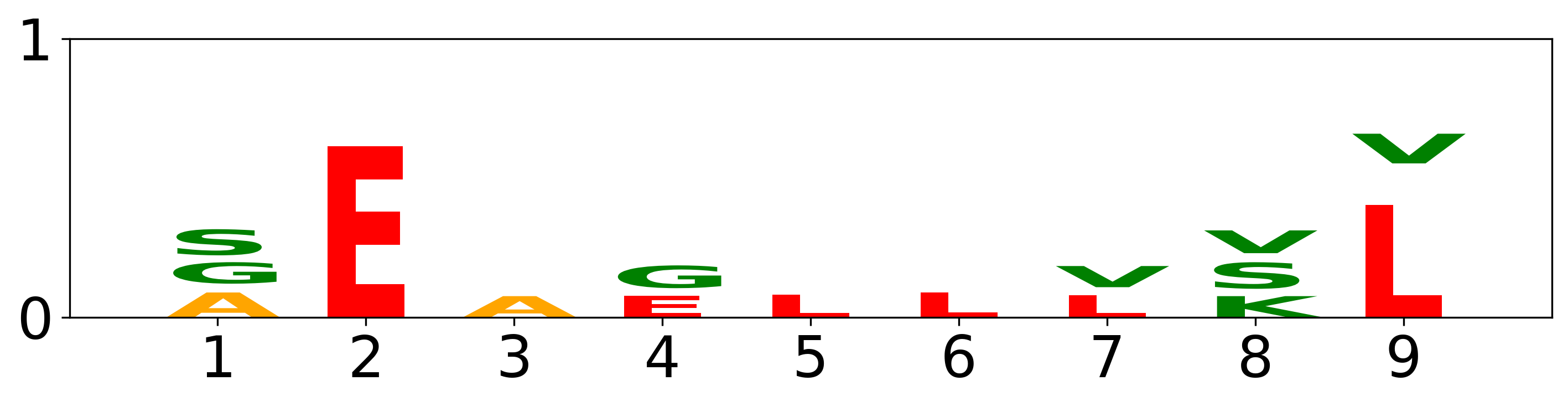}
    \caption{\fontsize{7.5}{6}\selectfont Motifs of experimental peptides.}
\end{subfigure}
\vspace{-1ex}
\caption{\small Visualization for peptide editing for task 306, higher binding affinity to HLA-B*40:02.}
\label{fig:peptide_motif_analysis_506}
\end{figure}

In the main body, we have illustrated how the motif of peptides changes for two peptide editing tasks. Here we show all 6 single-objective editing tasks in~\Cref{fig:peptide_motif_analysis_501,fig:peptide_motif_analysis_502,fig:peptide_motif_analysis_503,fig:peptide_motif_analysis_504,fig:peptide_motif_analysis_505,fig:peptide_motif_analysis_506}.
\begin{itemize}[noitemsep,topsep=0pt,itemindent=1em]
    \item For task 301 in~\Cref{fig:peptide_motif_analysis_501}, \model{} can successfully upweight E (Glutamic acid) at position 2.
    \item For task 302 in~\Cref{fig:peptide_motif_analysis_502}, \model{} can successfully upweight A (Alanine) at position 2, and L (Leucine) at position 9.
    \item For task 303 in~\Cref{fig:peptide_motif_analysis_503}, \model{} can successfully upweight E (Glutamic acid) at position 2, and L (Leucine) at position 9.
    \item For task 304 in~\Cref{fig:peptide_motif_analysis_504}, \model{} can successfully upweight R (Arginine) and K (Lysine) at position 5, and L (Leucine) at position 9.
    \item For task 305 in~\Cref{fig:peptide_motif_analysis_505}, \model{} can successfully upweight R (Arginine) and K (Lysine) at position 5, and L (Leucine) at position 9.
    \item For task 306 in~\Cref{fig:peptide_motif_analysis_506}, \model{} can successfully upweight E (Glutamic acid) at position 2, and L (Leucine) at position 9.
\end{itemize}

\begin{figure}[ht]
\centering
\begin{subfigure}[b]{0.4\linewidth}
    \centering
    \includegraphics[width=\linewidth]{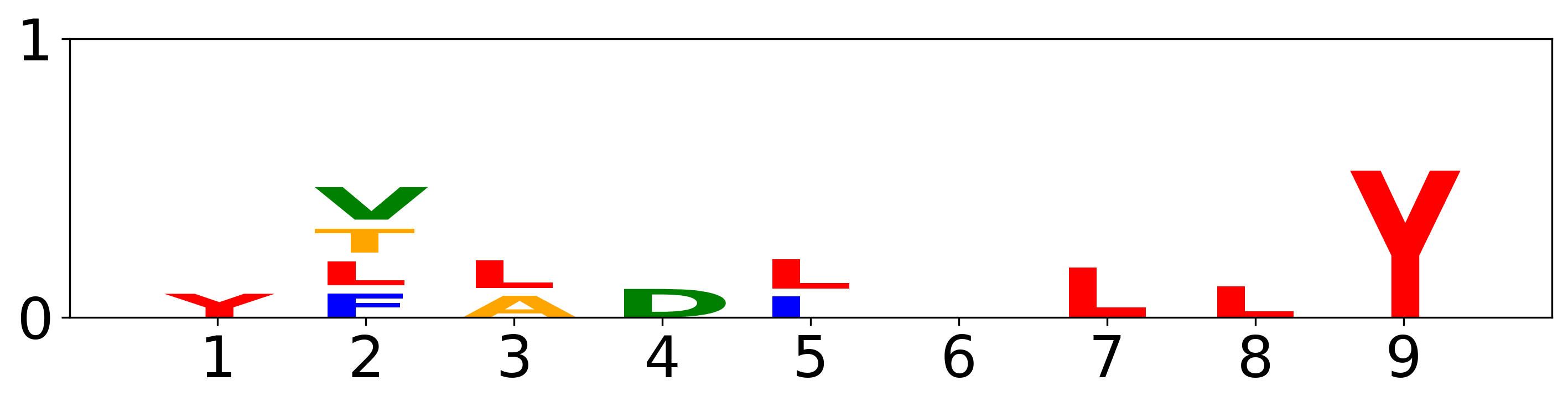}
    \caption{\fontsize{7.5}{6}\selectfont Motifs of input peptides.}
\end{subfigure}
\hfill
\begin{subfigure}[b]{0.4\linewidth}
    \centering
    \includegraphics[width=\linewidth]{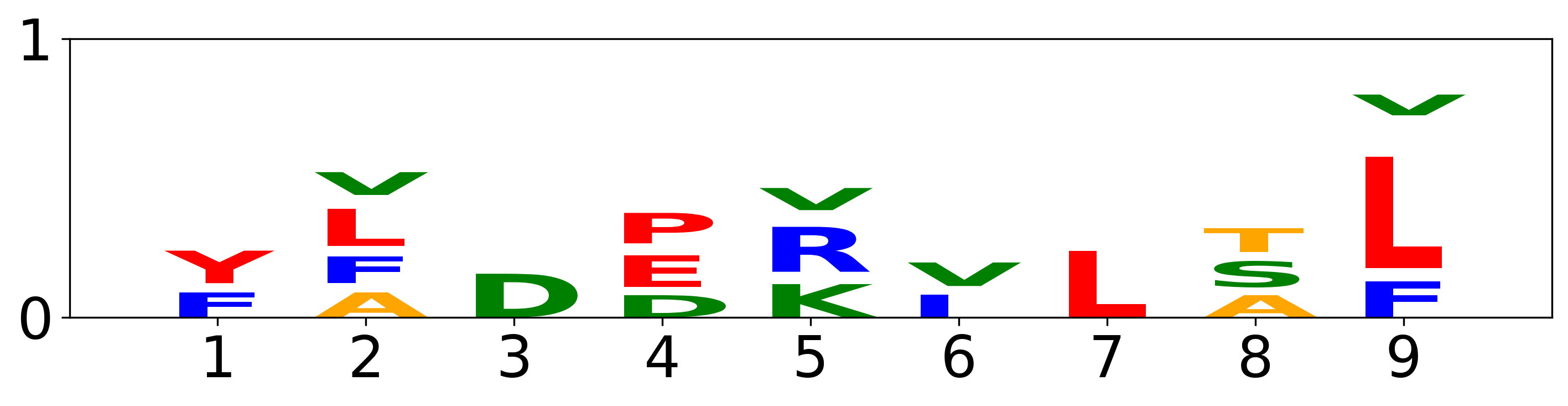}
    \caption{\fontsize{7.5}{6}\selectfont Motifs of edited peptides.}
\end{subfigure}
\\
\begin{subfigure}[b]{0.4\linewidth}
    \centering
    \includegraphics[width=\linewidth]{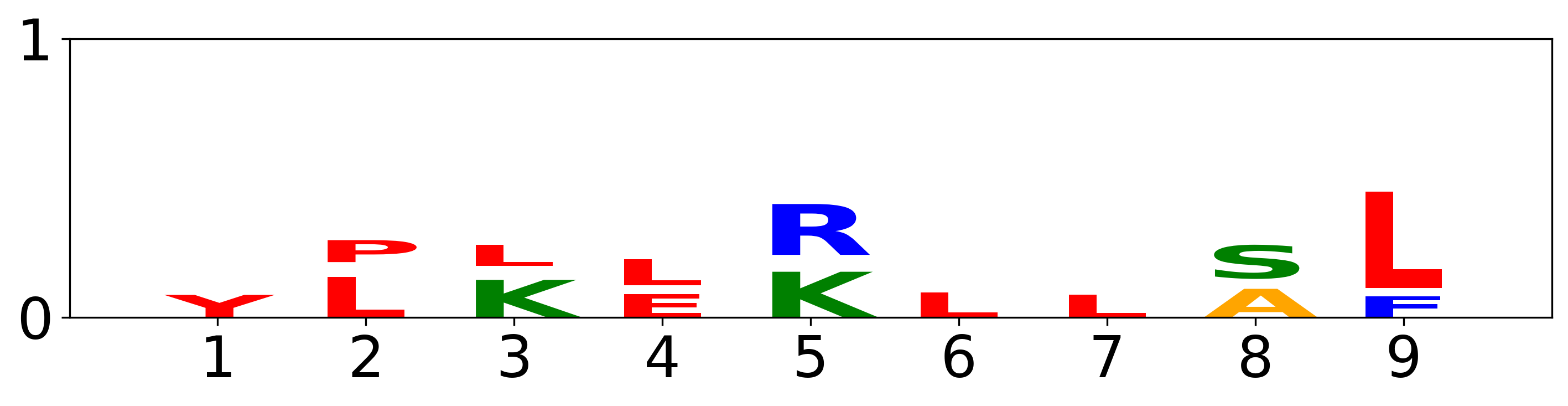}
    \caption{\fontsize{7.5}{6}\selectfont Motifs of experimental peptides (target allele 1).}
\end{subfigure}
\hfill
\begin{subfigure}[b]{0.4\linewidth}
    \centering
    \includegraphics[width=\linewidth]{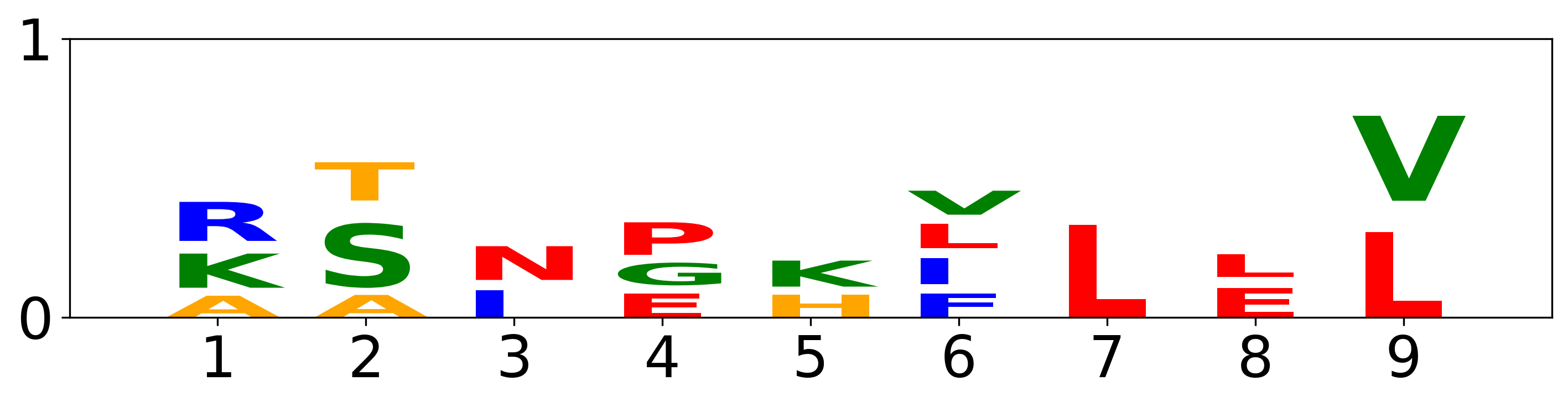}
    \caption{\fontsize{7.5}{6}\selectfont Motifs of experimental peptides (target allele 2).}
\end{subfigure}
\hfill
\caption{\small Visualization for peptide editing for task 401, higher binding affinity to HLA-B*08:01 and HLA-C*15:02.}
\label{fig:peptide_motif_analysis_601}
\vspace{+1ex}

\centering
\begin{subfigure}[b]{0.4\linewidth}
    \centering
    \includegraphics[width=\linewidth]{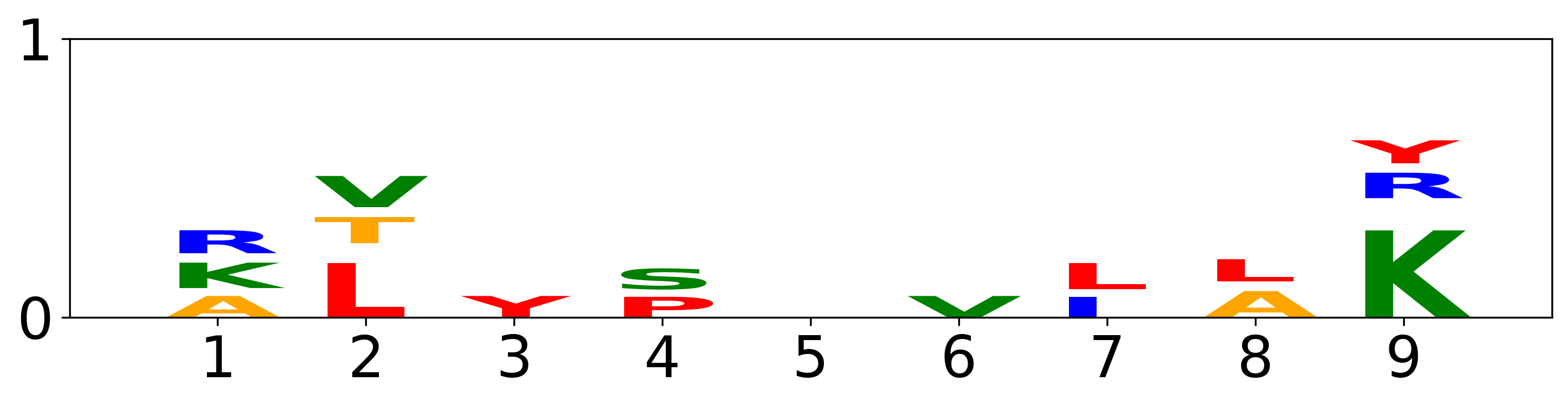}
    \caption{\fontsize{7.5}{6}\selectfont Motifs of input peptides.}
\end{subfigure}
\hfill
\begin{subfigure}[b]{0.4\linewidth}
    \centering
    \includegraphics[width=\linewidth]{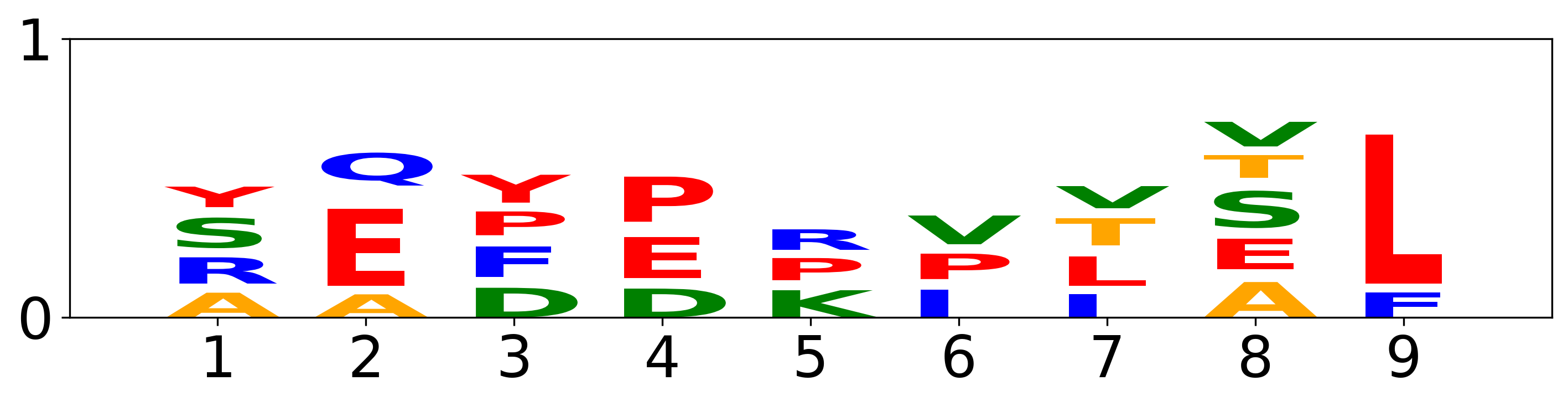}
    \caption{\fontsize{7.5}{6}\selectfont Motifs of edited peptides.}
\end{subfigure}
\\
\begin{subfigure}[b]{0.4\linewidth}
    \centering
    \includegraphics[width=\linewidth]{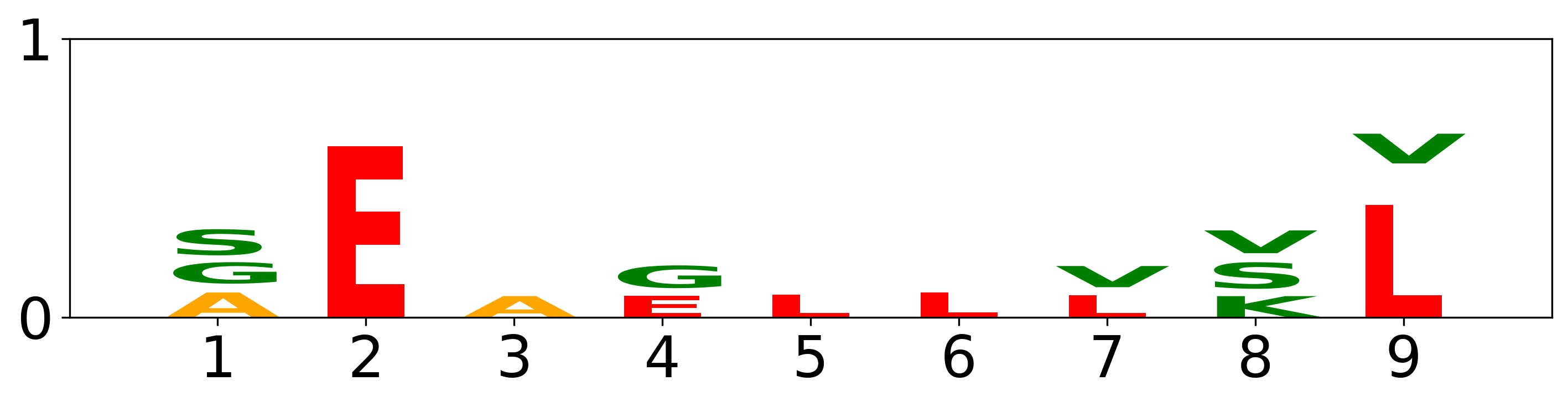}
    \caption{\fontsize{7.5}{6}\selectfont Motifs of experimental peptides (target allele 1).}
\end{subfigure}
\hfill
\begin{subfigure}[b]{0.4\linewidth}
    \centering
    \includegraphics[width=\linewidth]{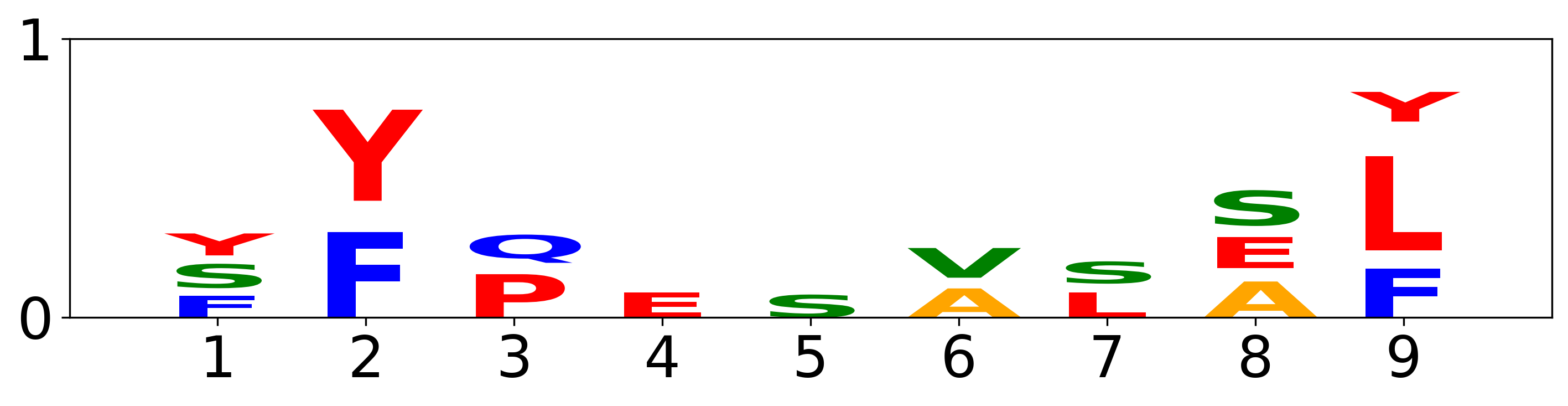}
    \caption{\fontsize{7.5}{6}\selectfont Motifs of experimental peptides (target allele 2).}
\end{subfigure}
\hfill
\caption{\small Visualization for peptide editing for task 402, higher binding affinity to HLA-B*40:02 and HLA-C*14:02.}
\label{fig:peptide_motif_analysis_602}
\vspace{+1ex}

\centering
\begin{subfigure}[b]{0.4\linewidth}
    \centering
    \includegraphics[width=\linewidth]{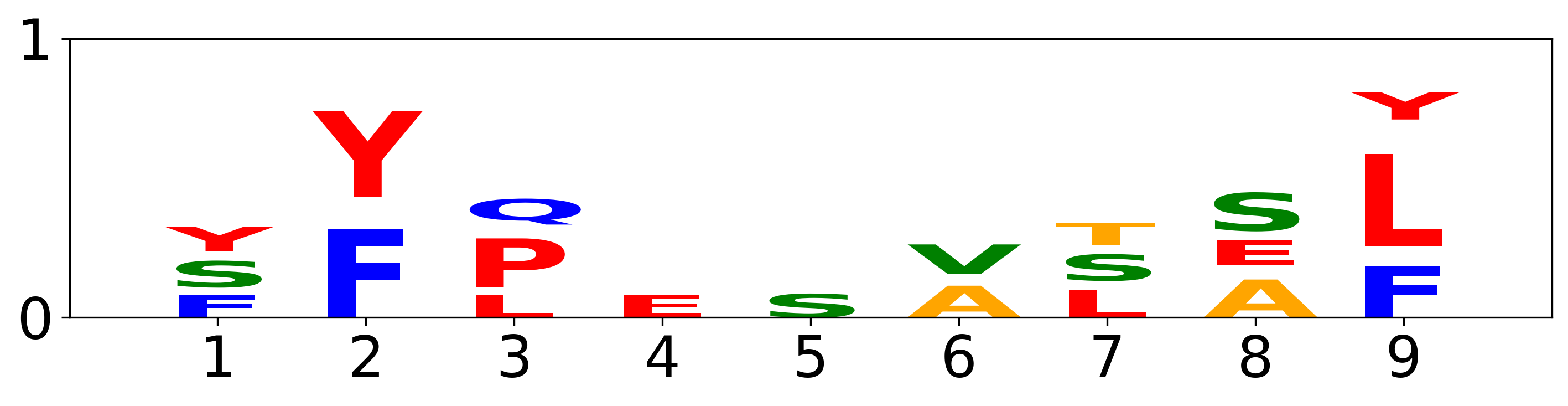}
    \caption{\fontsize{7.5}{6}\selectfont Motifs of input peptides.}
\end{subfigure}
\hfill
\begin{subfigure}[b]{0.4\linewidth}
    \centering
    \includegraphics[width=\linewidth]{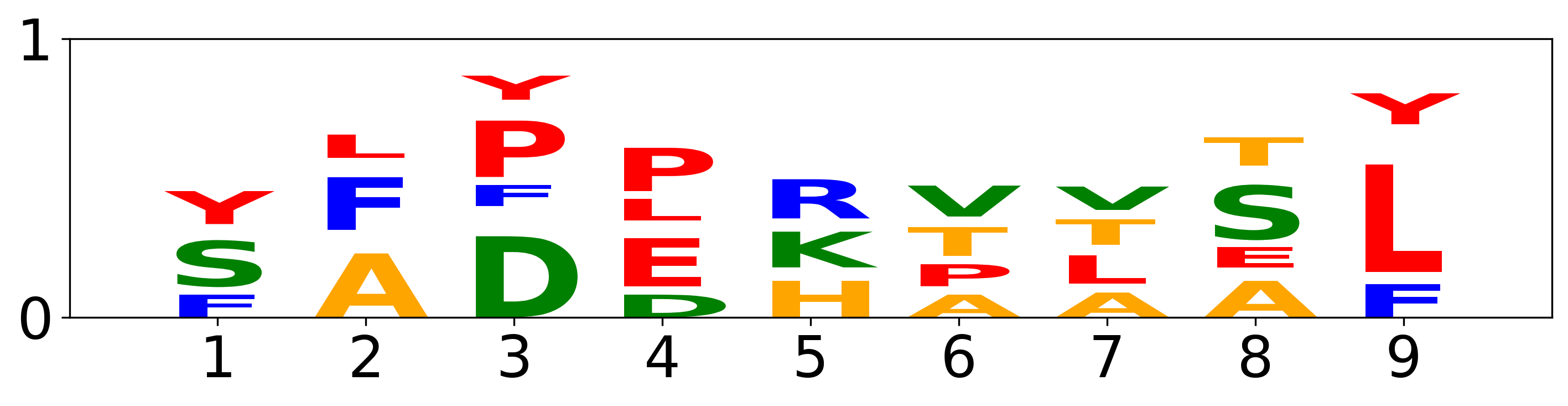}
    \caption{\fontsize{7.5}{6}\selectfont Motifs of edited peptides.}
\end{subfigure}
\\
\begin{subfigure}[b]{0.4\linewidth}
    \centering
    \includegraphics[width=\linewidth]{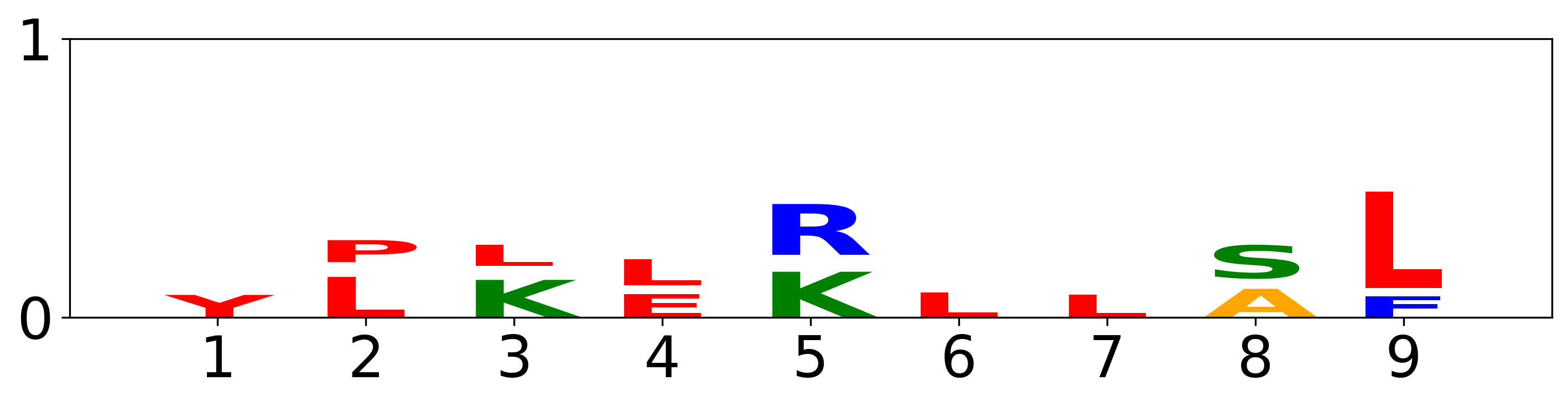}
    \caption{\fontsize{7.5}{6}\selectfont Motifs of experimental peptides (target allele 1).}
\end{subfigure}
\hfill
\begin{subfigure}[b]{0.4\linewidth}
    \centering
    \includegraphics[width=\linewidth]{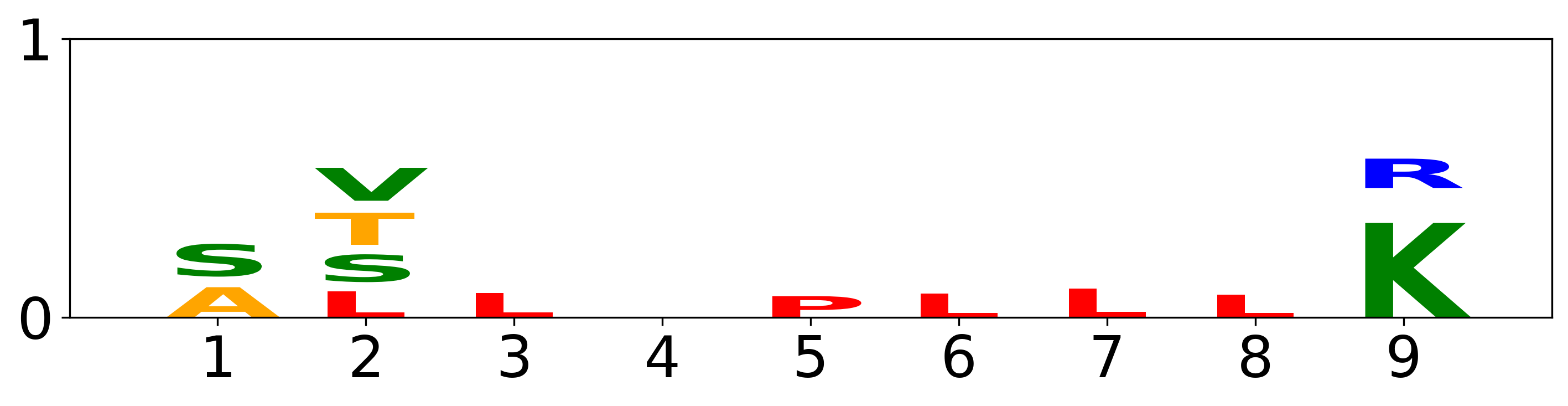}
    \caption{\fontsize{7.5}{6}\selectfont Motifs of experimental peptides (target allele 2).}
\end{subfigure}
\hfill
\caption{\small Visualization for peptide editing for task 403, higher binding affinity to HLA-B*08:01 and HLA-A*11:01.}
\label{fig:peptide_motif_analysis_603}
\vspace{+1ex}
\end{figure}

\clearpage
Here we show all 3 multi-objective editing tasks in~\Cref{fig:peptide_motif_analysis_601,fig:peptide_motif_analysis_602,fig:peptide_motif_analysis_603}. Notice that here there are two target allele types, and we mark them as ``target allele 1'' and ``target allele 2''.
\begin{itemize}[noitemsep,topsep=0pt,itemindent=1em]
    \item For task 401 in~\Cref{fig:peptide_motif_analysis_601}, \model{} can successfully upweight R (Arginine) and K (Lysine) at position 5, and L (Leucine) and F (Phenylalanine) at position 9 for target allele type 1. \model{} can also upweight L (Leucine) at position 7, and V (Valine) and L (Leucine) at position 9 for target allele type 2.
    \item For task 402 in~\Cref{fig:peptide_motif_analysis_602}, \model{} can successfully upweight E (Glutamic acid) at position 2, and L (Leucine) at position 9 for target allele type 1. \model{} can also upweight F (Phenylalanine) and L (Leucine) at position 9 for target allele type 2.
    \item For task 403 in~\Cref{fig:peptide_motif_analysis_603}, \model{} can successfully upweight R (Arginine) and K (Lysine) at position 5, and L (Leucine) at position 9 for target allele type 1.
\end{itemize}

\textbf{Potential issue on conflicts among target allele types.} One potential challenge is that for multi-objective editing, the target allele types could have conflicting motifs, like the two target alleles for task 403. We leave this for future exploration.

\clearpage
\subsection{Protein} \label{sec:case_studies_proteins}
Recall that we consider two types of secondary structures for protein editing tasks. Both the inputs and outputs are protein sequences. Then we use ESMFold~\cite{lin2022language} for protein folding (protein sequence to protein structure prediction) and then plot the protein structures using PyMOL~\cite{pymol}. For all the protein structure visualizations, we mark \textcolor{ProteinSecondaryStructureRed}{$\alpha$-helix structures} and \textcolor{ProteinSecondaryStructureYellow}{$\beta$-strand structures}. The edited regions are highlighted in the \textcolor{ProteinSecondaryStructureBlue}{blue circles}.

\textbf{Task 501: edit proteins with more helix structures.}
\begin{figure}[h]
\centering
\begin{subfigure}[b]{0.45\linewidth}
    \centering
    \includegraphics[width=\linewidth]{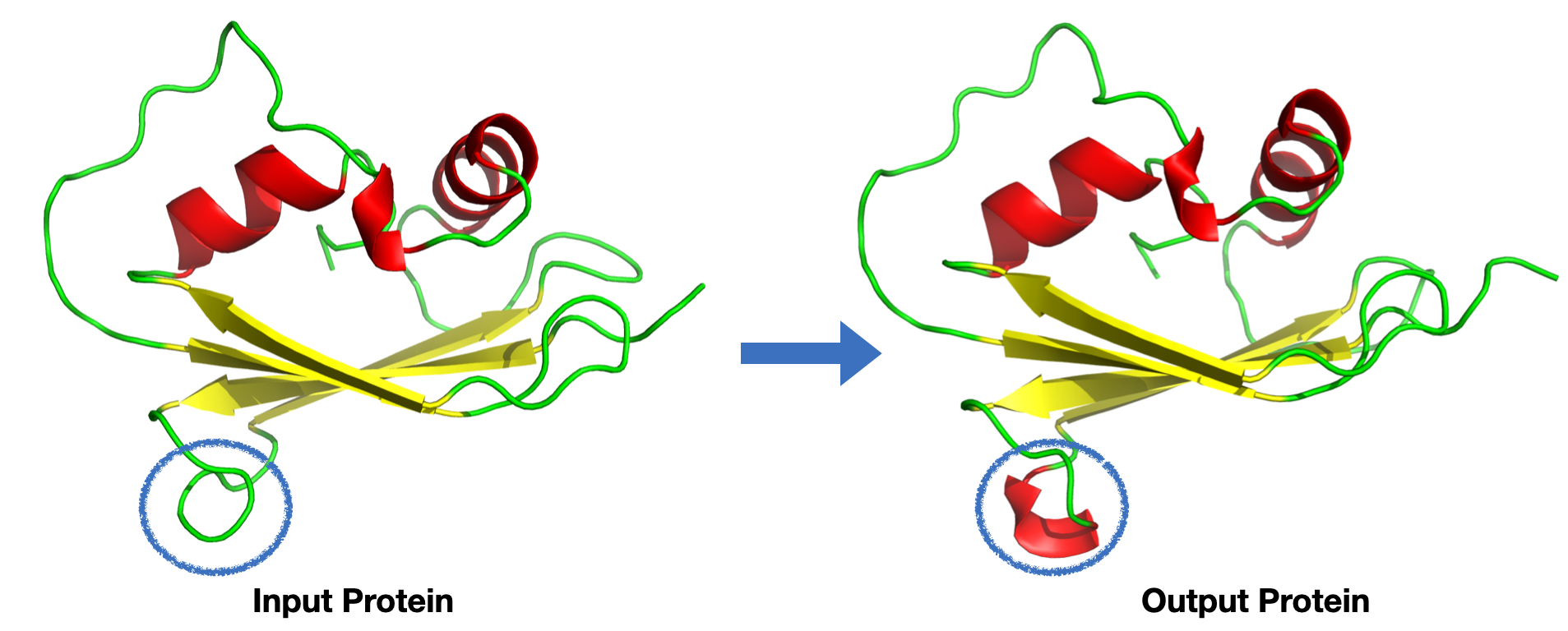}
    \caption{\small Protein editing with more $\alpha$-helix structures for data 1.}
\end{subfigure}
\hfill
\begin{subfigure}[b]{0.45\linewidth}
    \centering
    \includegraphics[width=\linewidth]{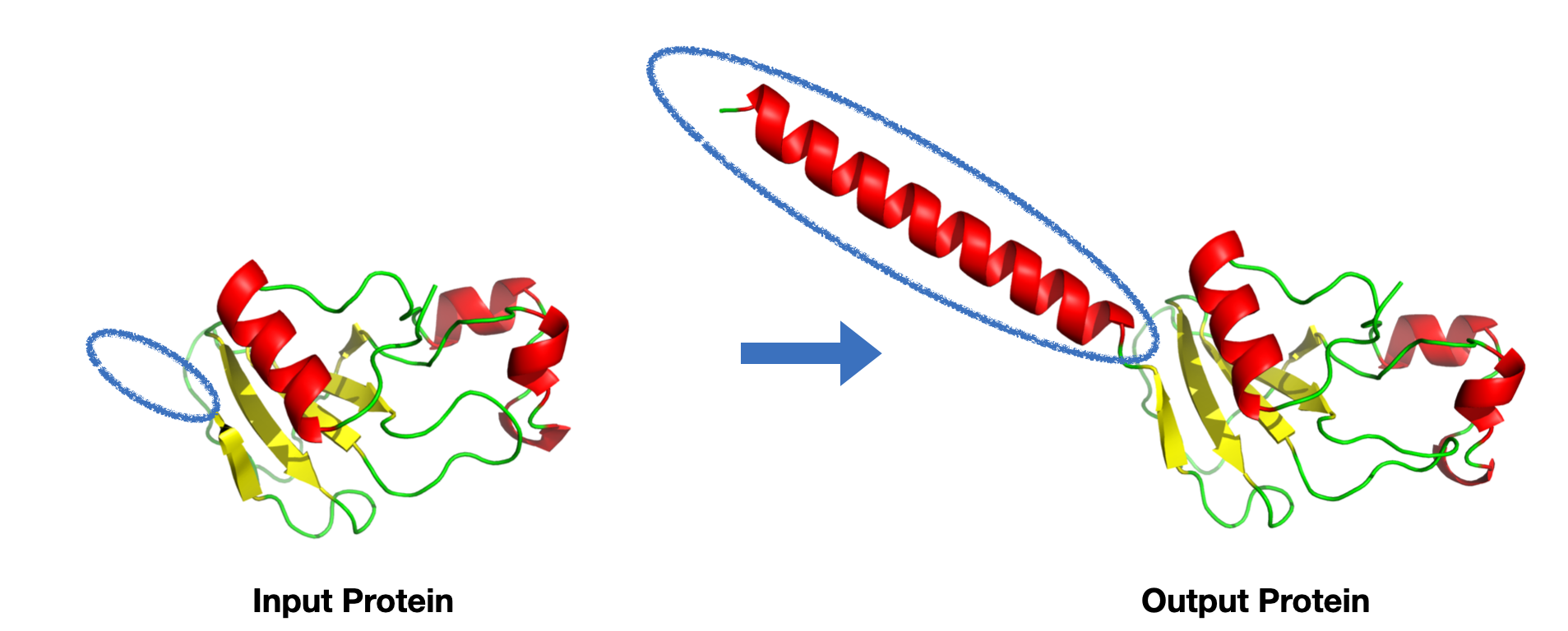}
    \caption{\small Protein editing with more $\alpha$-helix structures for data 2.}
\end{subfigure}
\begin{subfigure}[b]{0.45\linewidth}
    \centering
    \includegraphics[width=\linewidth]{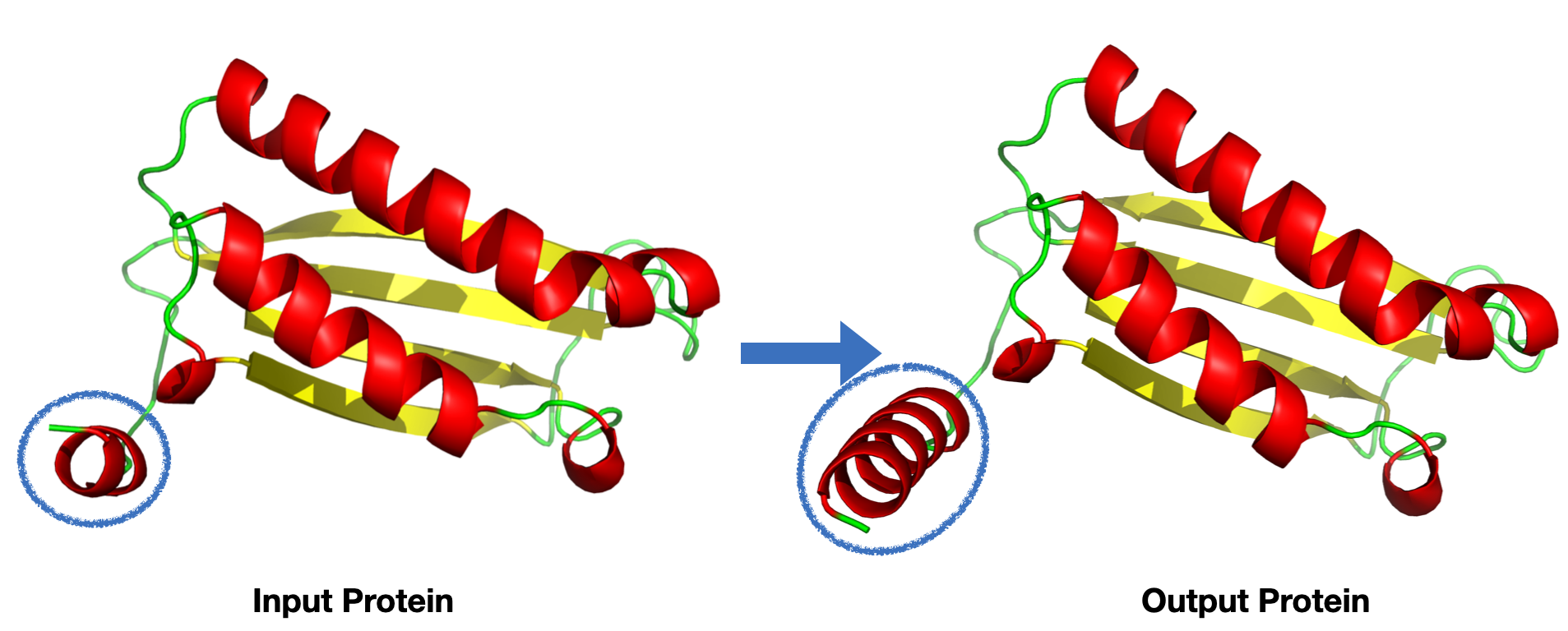}
    \caption{\small Protein editing with more $\alpha$-helix structures for data 3.}
\end{subfigure}
\hfill
\begin{subfigure}[b]{0.45\linewidth}
    \centering
    \includegraphics[width=\linewidth]{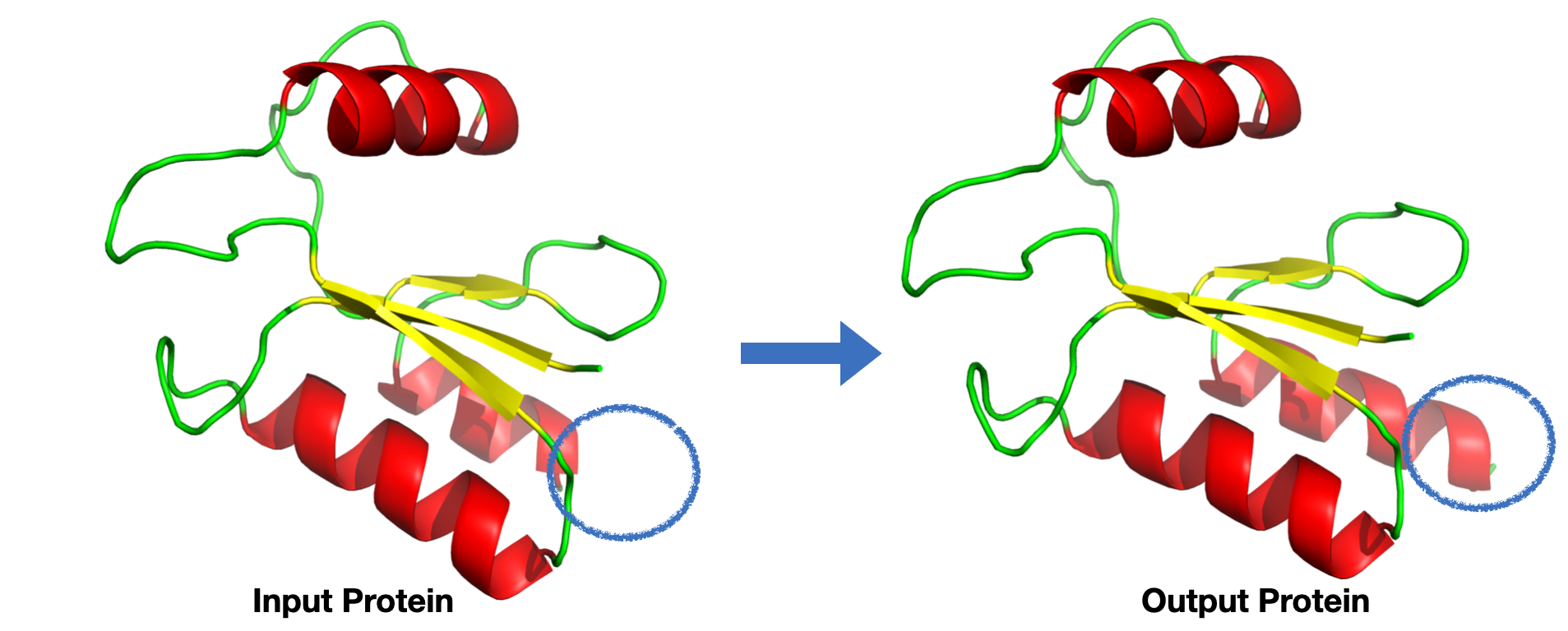}
    \caption{\small Protein editing with more $\alpha$-helix structures for data 4.}
\end{subfigure}
\caption{\small Protein editing with more $\alpha$-helix structures.}
\end{figure}

\textbf{Task 502: edit proteins with more strand structures.}
\begin{figure}[h]
\centering
\begin{subfigure}[b]{0.45\linewidth}
    \centering
    \includegraphics[width=\linewidth]{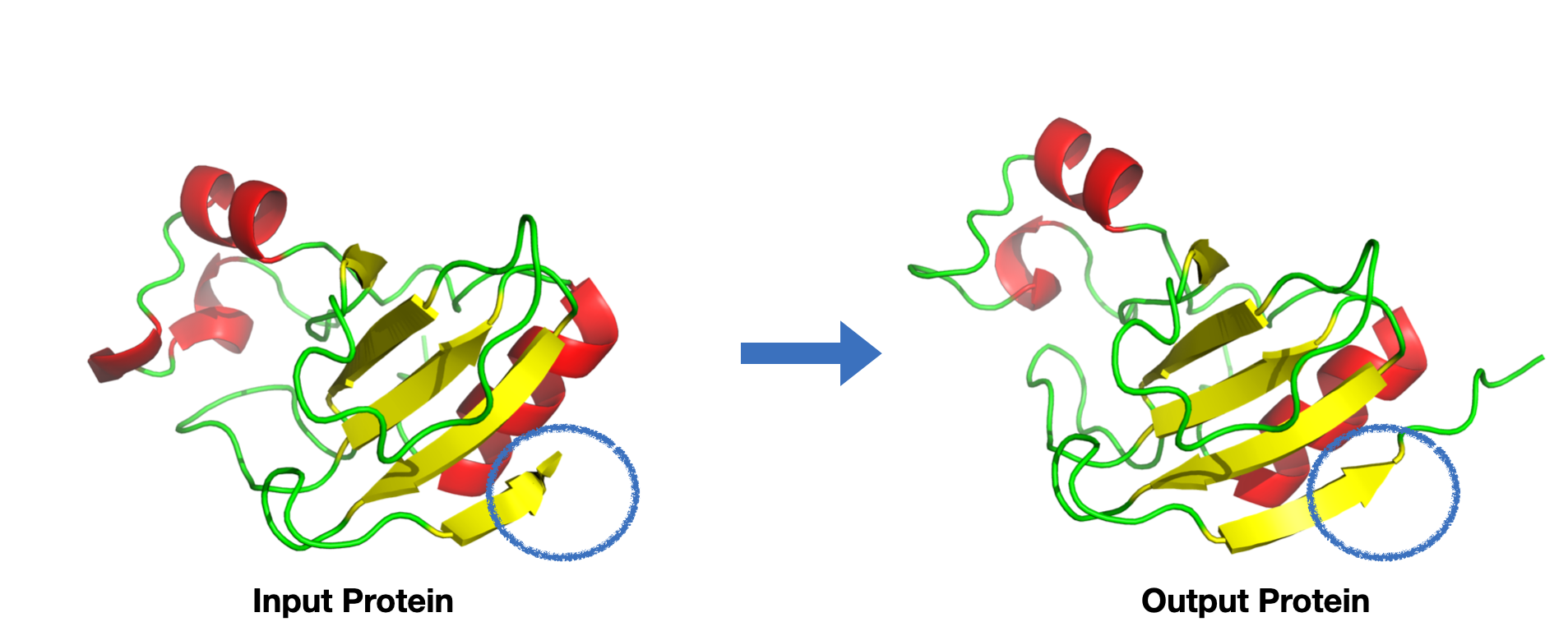}
    \caption{\small Protein editing with more $\beta$-strand structures for data 1.}
\end{subfigure}
\hfill
\begin{subfigure}[b]{0.45\linewidth}
    \centering
    \includegraphics[width=\linewidth]{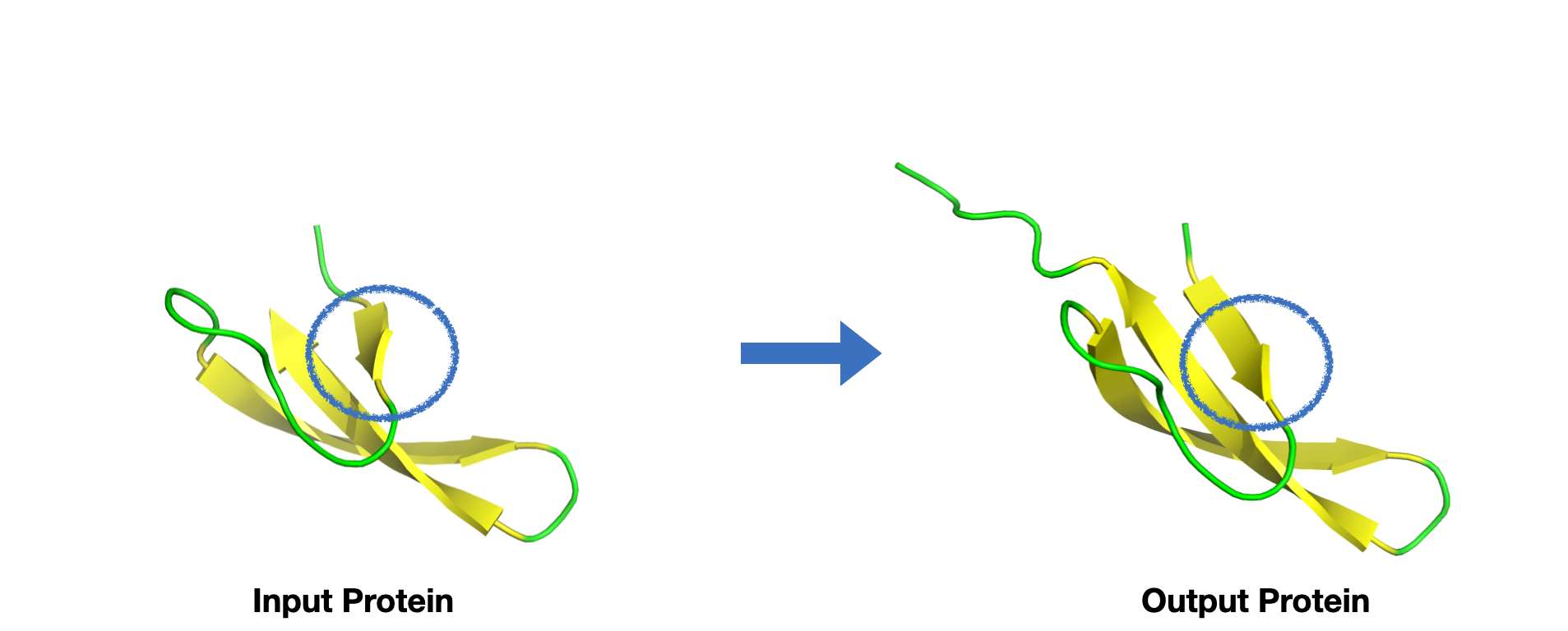}
    \caption{\small Protein editing with more $\beta$-strand structures for data 2.}
\end{subfigure}
\hfill
\begin{subfigure}[b]{0.45\linewidth}
    \centering
    \includegraphics[width=\linewidth]{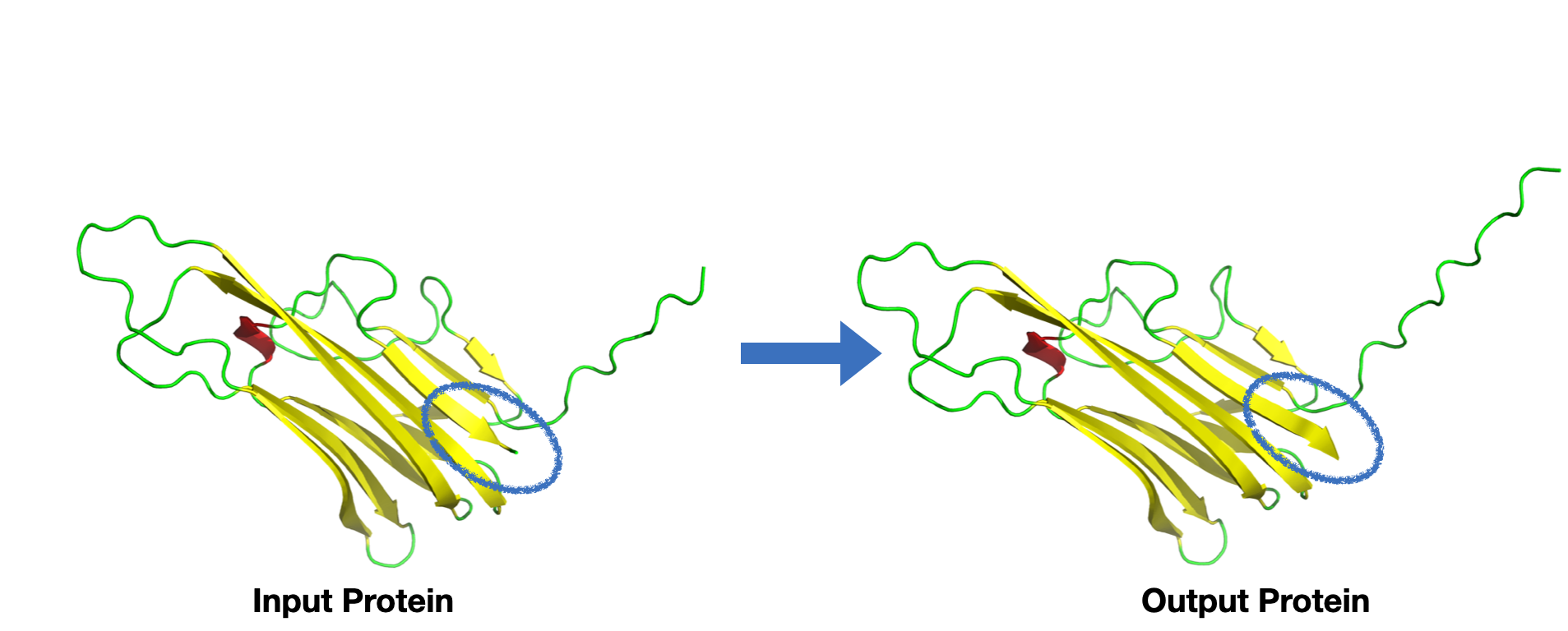}
    \caption{\small Protein editing with more $\beta$-strand structures for data 3.}
\end{subfigure}
\hfill
\begin{subfigure}[b]{0.45\linewidth}
    \centering
    \includegraphics[width=\linewidth]{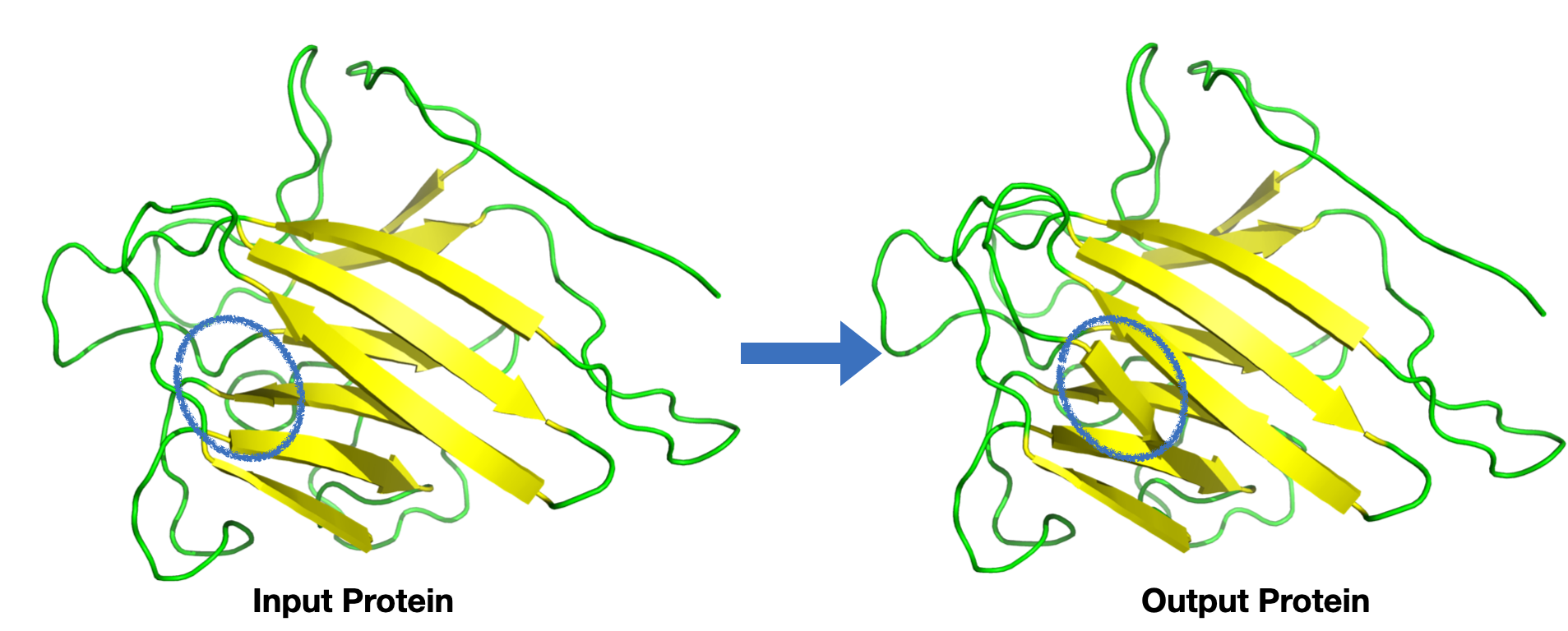}
    \caption{\small Protein editing with more $\beta$-strand structures for data 4.}
\end{subfigure}
\caption{\small Protein editing with more $\beta$-strand structures.}
\end{figure}

\clearpage
\section{Ablation Studies} \label{sec:ablation_studies}

\subsection{Zero-shot and In-context Learning for protein and peptide}
In~\Cref{sec:ablation_study_zero_shot_in_context_learning}, we conduct an ablation study on small molecules to show the comparison between the zero-shot, in-context learning, and \model{}. Here we conduct the same ablation study on peptides and proteins as follows.
\begin{table}[ht]
\centering
\caption{\small
Results on six single-objective and three multi-objective peptide editing tasks. Random Mutation-$R$ for $R$ mutated positions. The evaluation is the hit ratio of the increased binding affinity score. The best results are marked in \textbf{bold}. Due to the space limitation, please check~\Cref{sec:task_specification} for the text prompt of each task.
}
\setlength{\tabcolsep}{10pt}
\begin{adjustbox}{max width=\linewidth}
\begin{tabular}{l rrrrrr rrr}
\toprule
& \multicolumn{6}{c}{single-objective editing} & \multicolumn{3}{c}{multi-objective editing}\\
\cmidrule(lr){2-7} \cmidrule(lr){8-10}
& 301 & 302 & 303 & 304 & 305 & 306 & 401 & 402 & 403\\
\midrule
Random Mutation-1 & 1.80 & 14.40 & 1.80 & 1.80 & 12.00 & 5.60 & 3.20 & 0.80 & 0.40\\
Random Mutation-2 & 1.80 & 13.40 & 2.80 & 3.00 & 8.40 & 4.40 & 2.20 & 0.60 & 1.20\\
Random Mutation-3 & 1.80 & 9.40 & 2.40 & 4.20 & 9.00 & 3.80 & 3.00 & 0.60 & 0.80\\
In-context Learning (few-shot) & 24.05 & 38.40 & 27.40 & 32.00 & 45.50 & 32.80 & 29.20 & 17.47 & 14.40 \\
\model{} ($C=0$, zero-shot) & 1.60 & 16.80 & 2.40 & 8.22 & 15.00 & 8.02 & 5.41 & 2.00 & 1.20 \\
\model{} ($C=2$) & \textbf{58.60} & \textbf{69.34} & \textbf{58.52} & \textbf{55.11} & \textbf{64.40} & \textbf{62.73} & \textbf{53.71} & \textbf{41.45} & \textbf{54.71}\\
\bottomrule
\end{tabular}
\end{adjustbox}
\vspace{-1ex}
\end{table}

\begin{table}[ht]
\centering
\caption{\small 
Results on two protein editing tasks. Random Mutation-$R$ for $R$ mutated positions. The evaluation is the hit ratio of increased secondary structures accordingly. The best results are marked in \textbf{bold}.
}
\begin{adjustbox}{max width=0.53\linewidth}
\begin{tabular}{l rr}
\toprule
& 501 more helix & 502 more strand\\
\midrule
Random Mutation-1 & 18.32 & 17.35\\
Random Mutation-2 & 24.95 & 19.69\\
Random Mutation-3 & 26.90 & 21.44\\
In-context Learning (few-shot) &\textbf{36.64} &44.47 \\
\model{} ($C=0$, zero-shot) & 21.43 & 23.73\\
\model{} ($C=2$) & 34.79 & \textbf{51.38} \\
\bottomrule
\end{tabular}
\end{adjustbox}
\end{table}

\clearpage
\newpage
\subsection{Ablation Study on the Thresholds in Feedback Condition Function}
In the main body, we conduct an ablation study on the thresholds in the feedback condition function. Due to the space limitation, we only list the mean in~\Cref{tab:small_molecule_ablation_on_feedback_condition}. Here we list both the mean and standard deviation as follows.

\begin{table}[ht]
\centering
\setlength{\tabcolsep}{10pt}
\caption{\small Ablation studies on single-objective small molecule editing and feedback condition $D$ with five seeds and two conversational rounds. The evaluation metric $E$ uses the strict threshold for each task.}
\begin{adjustbox}{max width=\linewidth}
\begin{tabular}{l rrrrrrrr}
\toprule
& 101 & 102 & 103 & 104 & 105 & 106 & 107 & 108\\
\midrule
loose threshold & 80.73$\pm$1.32& 41.00$\pm$0.91& 11.23$\pm$2.70& 16.94$\pm$1.24& 33.16$\pm$2.22& 53.59$\pm$1.59& 14.96$\pm$1.96& 21.93$\pm$1.82 \\ 
strict threshold & 88.67$\pm$0.95& 70.08$\pm$3.44& 19.37$\pm$5.54& 30.99$\pm$2.66& 43.08$\pm$2.95& 66.69$\pm$2.74& 72.60$\pm$2.51& 76.43$\pm$3.32\\ 
\bottomrule
\end{tabular}
\end{adjustbox}
\end{table}

\begin{table}[ht]
\centering
\setlength{\tabcolsep}{20pt}
\caption{\small Ablation studies on multi-objective small molecule editing and feedback condition $D$ with five seeds and two conversational rounds. The evaluation metric $E$ uses the strict threshold for each task.}
\begin{adjustbox}{max width=\linewidth}
\begin{tabular}{l rrrrrr}
\toprule
& 201 & 202 & 203 & 204 & 205 & 206\\
\midrule
loose threshold & 20.14$\pm$0.86& 7.96$\pm$2.05 & 17.93$\pm$0.79& 5.79$\pm$1.38& 3.66$\pm$0.24& 41.04$\pm$1.66 \\ 
strict threshold & 49.64$\pm$2.66& 24.92$\pm$4.85& 53.64$\pm$5.81& 24.19$\pm$2.19& 10.44$\pm$5.75& 52.9$\pm$2.23\\ 
\bottomrule
\end{tabular}
\end{adjustbox}
\end{table}

\clearpage
\newpage
\subsection{Ablation Study on the Number of Request Answers in Zero-shot \model{}}
Notice that in~\Cref{tab:prompt_small_molecule}, we list five molecules (a.k.a. five trials) for each answer. In this subsection, we would like to conduct an ablation study to explore in the zero-shot setting of \model{}, {\ie}, with the conversation round $C=0$, if we can obtain higher performance using more trial numbers. This means that for each input small molecule, we have five edited small molecules; as long as one of them is a hit, then we say this is a successful hit. The results for 14 tasks with the loose threshold are listed below.

\begin{table}[htb!]
\centering
\caption{\small Ablation studies on different trial numbers on single-objective molecule editing, with $C=0$ and seed is 0.}
\begin{adjustbox}{max width=\linewidth}
\begin{tabular}{l rrrrrr}
\toprule
& \multicolumn{3}{c}{loose condition $\Delta=0$}
& \multicolumn{3}{c}{strict condition $\Delta>0$}\\
\cmidrule(lr){2-4} \cmidrule(lr){5-7}
& trial = 1 & trial = 3 & trial = 5 & trial = 1 & trial = 3 & trial = 5\\
\midrule
101 \textit{more soluble in water} & 78.26 & 88.77 & 93.05 & 68.48 & 80.21 & 85.03\\
102 \textit{less soluble in water} & 71.35 & 89.95 & 93.12 & 24.16 & 74.60 & 78.84\\
103 \textit{more like a drug} & 16.15 & 45.64 & 53.81 & 2.08 & 4.62 & 7.11\\
104 \textit{less like a drug} & 32.12 & 68.37 & 75.00 & 2.07 & 17.86 & 31.12\\
105 \textit{higher permeability} & 16.04 & 27.98 & 33.16 & 9.63 & 18.13 & 22.28\\
106 \textit{lower permeability} & 8.33 & 34.04 & 57.67 & 5.56 & 24.47 & 42.86\\
107 \textit{more hydrogen bond acceptors} & 59.41 & 76.57 & 83.15 & 1.76 & 18.29 & 33.71\\
108 \textit{more hydrogen bond donors} & 63.16 & 85.23 & 89.77 & 5.85 & 19.89 & 32.39\\
\bottomrule
\end{tabular}
\end{adjustbox}
\end{table}

\begin{table}[htb!]
\centering
\caption{\small Ablation studies on different trial numbers on multi-objective molecule editing, with $C=0$ and seed is 0.}
\begin{adjustbox}{max width=\linewidth}
\begin{tabular}{l rrrrrr}
\toprule
& \multicolumn{3}{c}{loose condition $\Delta=0$}
& \multicolumn{3}{c}{strict condition $\Delta>0$}\\
\cmidrule(lr){2-4} \cmidrule(lr){5-7}
& trial = 1 & trial = 3 & trial = 5 & trial = 1 & trial = 3 & trial = 5\\
\midrule
\makecell[l]{201 \textit{more soluble in water} and\\\textit{more hydrogen bond acceptors}} & 43.09 & 69.89 & 75.40 & 6.08 & 22.04 & 34.22\\
\midrule
\makecell[l]{202 \textit{less soluble in water} and\\\textit{more hydrogen bond acceptors}} & 0.52 & 13.47 & 31.44 & 0.00 & 0.52 & 2.06\\
\midrule
\makecell[l]{203 \textit{more soluble in water} and\\\textit{more hydrogen bond donors}} & 54.49 & 79.67 & 81.97 & 6.18 & 18.13 & 35.52\\
\midrule
\makecell[l]{204 \textit{less insoluble in water} and\\\textit{more hydrogen bond donors}} & 0.53 & 5.21 & 13.54 & 0.00 & 0.00 & 1.04 \\
\midrule
\makecell[l]{205 \textit{more soluble in water} and\\\textit{higher permeability}} & 2.11 & 4.15 & 5.18 & 1.05 & 1.04 & 1.04\\
\midrule
\makecell[l]{206 \textit{more soluble in water} and\\\textit{lower permeability}} & 22.22 & 59.07 & 69.43 & 16.40 & 48.70 & 55.44\\
\bottomrule
\end{tabular}
\end{adjustbox}
\end{table}

\clearpage
\newpage
\subsection{Ablation Study on Reasoning and Knowledge Extraction of \model{}}

\begin{figure}[ht]
\centering
\includegraphics[width=.4\linewidth]{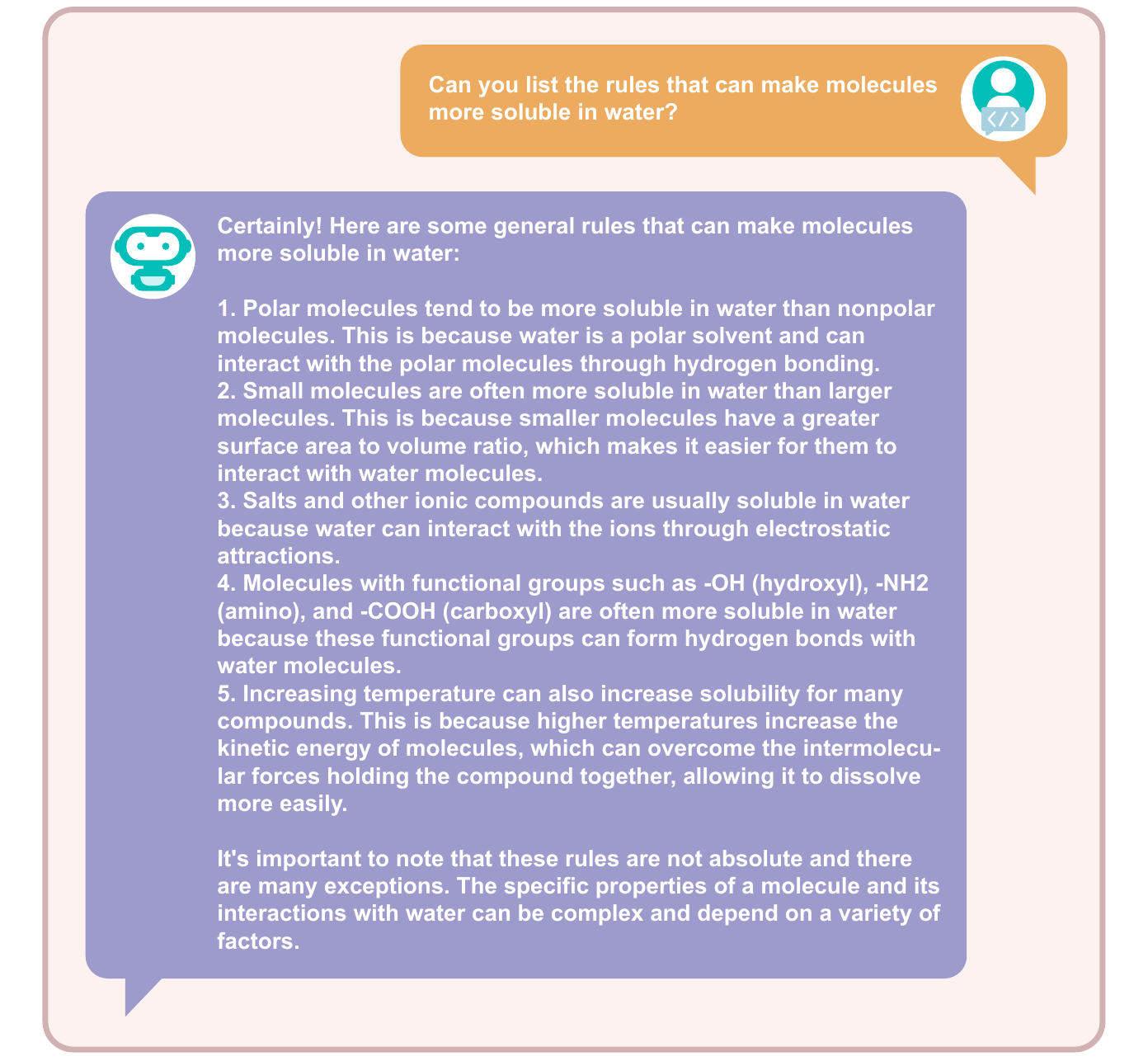}
\caption{\small Knowledge extraction directly without any conversations.}
\label{tab:conversation_knowledge_extraction_without_examples}
\end{figure}

In~\Cref{tab:conversation_knowledge_extraction_with_examples}, we show the knowledge extraction ability of \model{}. As a control experiment, here we test the knowledge extraction ability of \model{} without the examples illustrated in the conversation. As shown in~\Cref{tab:conversation_knowledge_extraction_without_examples}, the extracted knowledge is more general and less informative when we remove the examples.

Then as discussed previously, the extracted information in \model{} contains redundant information. For example, the one listed in~\Cref{tab:conversation_knowledge_extraction_without_examples} extracts three rules with overlap meanings. Then we conduct a further ablation study by forcing \model{} to extract three rules from the five original rules in~\Cref{tab:conversation_knowledge_extraction_with_examples}. We observe that \model{} successfully merges rule 1 and rule 5, both of which express the same concept that polar functional groups lead to good solubility. Interestingly, \model{} also merges rule 3 with rule 4. Though these 2 rules share the idea of ring structure, the essence of rule 3 is introducing polar substituents. Thus, rule 3 should be more relevant to rules 1 and 5. Nevertheless, it is clear that \model{} understands the domain knowledge and is capable of extracting and summarizing it. Overall, we would like to give positive feedback on \model{}.

\begin{figure}[ht]
\centering
\includegraphics[width=.4\linewidth]{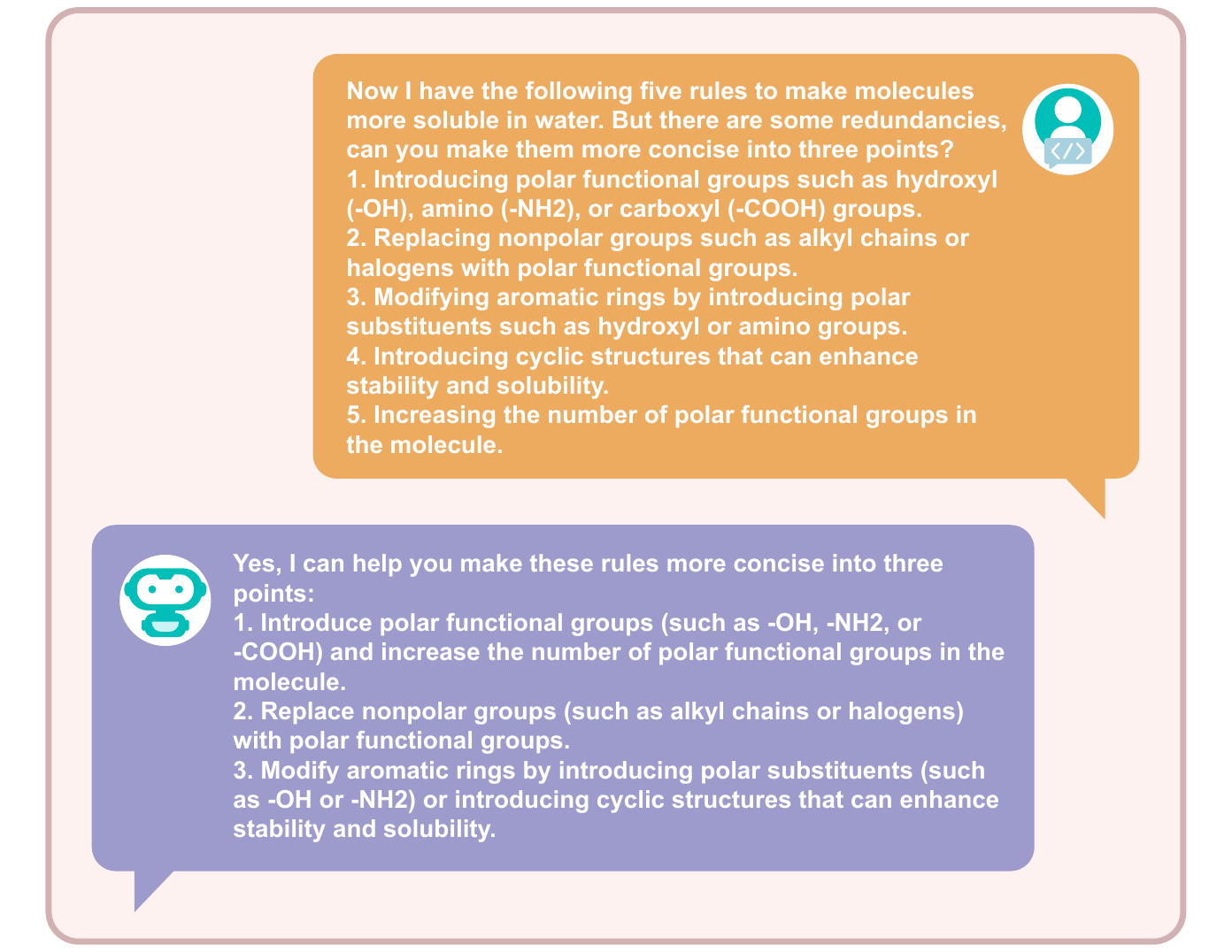}
\caption{\small Further knowledge extraction.}
\end{figure}

To sum up, we observe that \model{} can understand and extract the knowledge to some extent, though not perfectly. We believe this is a promising direction for future exploration.

\end{document}